\documentclass[preprint]{elsarticle}

\usepackage{amsmath,amssymb}
\usepackage{bm}
\usepackage{algorithm}
\usepackage[noend]{algpseudocode}
\algrenewcommand{\algorithmiccomment}[1]{\hspace{\fill}\{#1\}}
\usepackage{setspace}

\usepackage{color}

\biboptions{authoryear} 
\begin{document}

\begin{frontmatter}

\title{Coordinate descent heuristics for the irregular strip packing problem of rasterized shapes}

\author[osaka]{Shunji Umetani\corref{cor1}}
\ead{umetani@ist.osaka-u.ac.jp}
\author[osaka]{Shohei Murakami}
\cortext[cor1]{Corresponding author. Tel.: +81 (6) 6879 4793.}
\address[osaka]{Osaka University, 2-1 Yamadaoka, Suita, Osaka 565-0871, Japan}

\begin{abstract}
We consider the irregular strip packing problem of rasterized shapes, where a given set of pieces of irregular shapes represented in pixels should be placed into a rectangular container without overlap.
The rasterized shapes provide simple procedures of the intersection test without any exceptional handling due to geometric issues, while they often require much memory and computational effort in high-resolution.
To reduce the complexity of rasterized shapes, we propose a pair of scanlines representation called the double scanline representation that merges consecutive pixels in each row and column into strips with unit width, respectively.
Based on this, we develop coordinate descent heuristics for the raster model that repeat a line search in the horizontal and vertical directions alternately, where we also introduce a corner detection technique used in computer vision to reduce the search space.
Computational results for test instances show that the proposed algorithm obtains sufficiently dense layouts of rasterized shapes in high-resolution within a reasonable computation time.
\end{abstract}

\end{frontmatter}

\section{Introduction}\label{sec:introduction}
\begin{figure}[tb]
  \centering
  \includegraphics[width=0.95\textwidth]{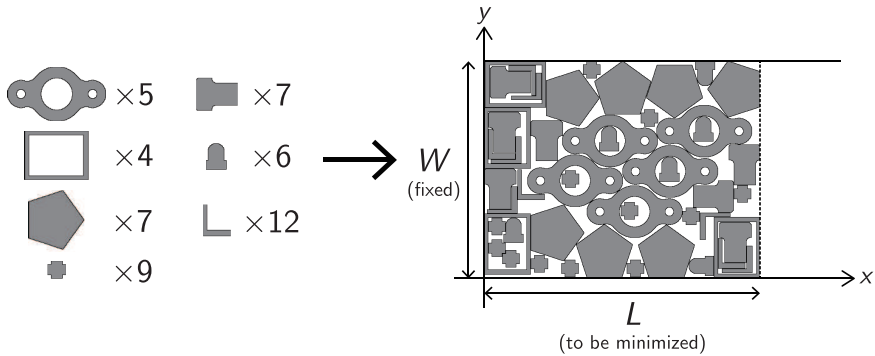}
  \caption{An instance of the irregular strip packing problem and its solution.\label{fig:isp_instance}}
\end{figure}

The \emph{irregular strip packing problem} (ISP), or often called the \emph{nesting problem}, is the one of the representative cutting and packing problems that emerges in a wide variety of industrial applications, such as garment manufacturing, sheet metal cutting, furniture making and shoe manufacturing~\citep{Alvarez-ValdesR2018,ScheithauerG2018}.
This problem is categorized as the two-dimensional, irregular open dimensional problem in \cite{DyckhoffH1990,WascherG2007}.
Given a set of pieces of irregular shapes and a rectangular container with a fixed width and a variable length, this problem asks a feasible layout of the pieces into the container such that no pair of pieces overlaps with each other and the container length is minimized.
We note that rotations of pieces are usually restricted to a few number of degrees (e.g., 0 or 180 degrees) in many industrial applications, because textiles have grain and may have a drawing pattern.
Figure~\ref{fig:isp_instance} shows an instance of ISP and a feasible solution.

\begin{figure}[tb]
  \centering
  \includegraphics[width=0.8\textwidth]{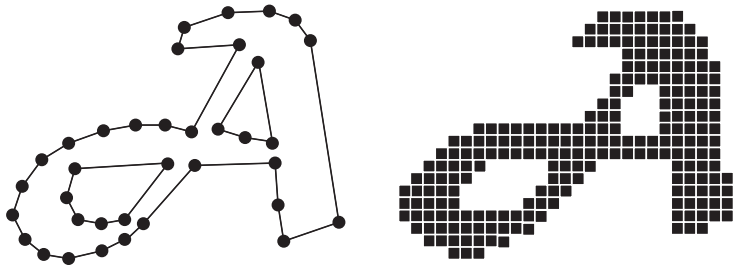}
  \caption{The vector and raster representations for an irregular shape.\label{fig:vector_raster}}
\end{figure}

The first issue encountered when handling ISP is how to represent the irregular shapes.
In computer graphics, the irregular shapes are often represented in two models as shown in Figure~\ref{fig:vector_raster}: the \emph{vector model} represents an irregular shape as a set of chained line and curve segments forming its outline, and the \emph{raster model} (also known as the \emph{bitmap model}) represents an irregular shape as a set of grid pixels forming its inside.
The vector model requires complicated trigonometric computations with many exception handling for the intersection test of the irregular shapes, while the raster model provides simple computations without any exception handling.
On the other hand, the vector model often consumes less memory usage and computation time for the intersection test than the raster model, because the number of line and curve segments in the vector model often becomes much smaller than that of grid pixels in the raster model.

For the vector model, in particular polygons, the recent development of computational geometry such as the no-fit polygon (NFP) enables us to compute their intersection test efficiently~\citep{BennellJA2008,LeaoAAS2020}.
Based on the efficient geometric computations, many efficient heuristic algorithms have been developed for ISP of the polygons (called the polygon packing problem in this paper)~\citep{BennellJA2009,HuY2018b}.
A standard approach for the polygon packing problem is to develop construction algorithms, e.g., the bottom-left (BL) algorithm and the bottom-left fill (BLF) algorithm that places the pieces one by one into the container based on a given order~\citep{AlbanoA1980,BlazewiczJ1993,OliveiraJF2000,DowslandKA2002,GomesAM2002,BennellJA2010}.
Another approach is to resort improvement algorithms that relocate pieces by solving the compaction problem and/or the separation problem~\citep{LiZ1995,BennellJA2001,GomesAM2006}.
The compaction problem relocates pieces from a given feasible placement so as to minimize the container length.
The separation problem relocates pieces from a given infeasible placement so as to make it feasible while minimizing the total amount of their translation.
The \emph{overlap minimization problem} (OMP) is a variant of the separation problem that places pieces within the container with given width and length so as to minimize the overlap penalty for all pairs of pieces~\citep{EgebladJ2007,ImamichiT2009,UmetaniS2009,LeungSCH2012,ElkeranA2013}.

\citet{MundimLR2017} and \citet{SatoAK2019} constrained the search space to place the pieces on a discrete set of positions (i.e., a grid), and developed a discretized NFP called the no-fit raster for the intersection test of the irregular shapes.
This constrained model of the search space on a grid is called the dotted-board model \citep{ToledoFMB2013}, which is similar to the raster model but different because the given pieces are represented in polygons (i.e., the vector model).

For the raster model, \citet{OliveiraJF1993}, \citet{SegenreichSA1986} and \citet{BabuAR2001} represented the pieces by matrices of different codes \citep{BennellJA2008}.
The raster representations are simple to code the irregular shapes and provide simple procedures for their intersection test, which enable us to develop practical software for a wide variety of free-form packing problems including many curve segments and holes only by converting input data to pixels.
However, their computational costs highly depend on the resolution of the raster representations; i.e., improving the resolution much increases the memory usage and computation time for the procedures.
To improve the computational efficiency of the intersection test and the overlap minimization for the raster model, \citet{OkanoH2002} and \citet{HuY2018a} proposed sophisticated data structures that merge the pixels of the irregular shapes into strips or rectangles.
Using these representations, several heuristic algorithms have been developed for ISP of the raster model~\citep{SegenreichSA1986,OliveiraJF1993,JainS1998,BabuAR2001,ChenP2004,WongWK2009}.
Despite these heuristic algorithms for the raster model, their computational results were still restricted for the instances of low-resolution and insufficient to convince their high performance for the instances of high-resolution.

In this paper, we develop an efficient line search algorithm for coordinate descent heuristics in the raster model, which makes it possible to high performance improvement algorithms for the instances of high-resolution.
The coordinate descent heuristics (CDH) is one of improvement algorithms that repeat a line search in the horizontal and vertical directions alternately, which have achieved high performance for ISP of the vector model~\citep{EgebladJ2007,UmetaniS2009}.
However, unlike the vector model, the conventional raster representations were computationally much expensive to develop any efficient line search algorithms especially in high-resolution.
To reduce the complexity of rasterized shapes, we propose a pair of scanlines representation called the \emph{double scanline representation} for NFPs of the rasterized shapes that reduces their complexity by merging consecutive pixels in each row and column into strips with unit width, respectively.
We also introduce a corner detection technique used in computer vision \citep{RostenE2006} to reduce the search space of the line search.
The coordinate descent heuristics are incorporated into a variant of the guided local search (GLS) for OMP based on \citet{UmetaniS2009}.
Using this as a main component, we then develop a heuristic algorithm for ISP, which we call the \emph{guided coordinate descent heuristics} (GCDH).

This paper is organized as follows.
We first formulate ISP and OMP in the raster model and illustrate the outline of the proposed algorithm GCDH for ISP in Section~\ref{sec:formulation}.
We then introduce an efficient intersection test for rasterized shapes in Section~\ref{sec:intersection}.
We explain the main component of the proposed algorithm CDH for OMP in Section~\ref{sec:coordinate_descent} and the construction algorithm for an initial solution of ISP in Section~\ref{sec:construction}.
Finally, we report computational results in Section~\ref{sec:computational_results} and make concluding remarks in Section~\ref{sec:conclusion}.

\section{Formulation and approach}\label{sec:formulation}

\subsection{Irregular strip packing problem of rasterized shapes}\label{sec:main_problem}
We are given a list of $n$ pieces $\mathcal{P} = \{ P_1, P_2, \dots, P_n \}$ of rasterized shape with a list of their possible orientations $\mathcal{O} = (O_1, O_2, \dots, O_n)$, where a piece $P_i$ can be rotated by $o$ degrees for each $o \in O_i$.
We assume without loss of generality that zero degree is always included in $O_i$.
We are also given a rectangular container $C = C(W,L)$ with a width $W$ and a length $L$, where $W$ is a non-negative constant and $L$ is a non-negative variable.
We assume that the container edges with the width $W$ and the length $L$ are parallel to the $y$-axis and the $x$-axis as shown in Figure~\ref{fig:isp_instance}, respectively, and the bottom-left corner of the container $C$ is the origin $(0,0)$.
We denote $P_i(o)$ as a piece $P_i$ rotated by $o \in O_i$ degrees, which may be written as $P_i$ for simplicity when its orientation is not specified or clear from the context.
We consider the bounding-box of a piece $P_i(o)$ as the smallest rectangle that encloses $P_i(o)$, and its width and length are denoted by $w_i(o)$ and $l_i(o)$, respectively.
We describe a position of a piece $P_i(o)$ by a coordinate $\bm{v}_i = (x_i,y_i)$ of the center of its bounding box called the reference point.
For convenience, we regard a piece $P_i$ as a set of grid pixels when its reference point is put at the origin $(0,0)$ and a rotated piece $P_i(o)$ is sometimes reshaped to fit grid pixels.
We then describe a piece $P_i$ placed at $\bm{v}_i$ by the Minkowski sum
\begin{equation}
P_i \oplus \bm{v}_i = \{ \bm{p} + \bm{v}_i \mid \bm{p} \in P_i \}.
\end{equation}

We describe a solution of ISP by lists of positions $\bm{v} = (\bm{v}_1, \bm{v}_2, \dots, \bm{v}_n)$ and orientations $\bm{o} = (o_1, o_2, \dots, o_n)$ of all pieces $P_i$ ($i=1,\dots,n$).
We note that a solution $(\bm{v}, \bm{o})$ uniquely determines a layout of the pieces.
The ISP is described as follows:
\begin{equation}
\begin{array}{llll}
  (\textnormal{ISP}) & \textnormal{minimize} & L\\
  & \textnormal{subject to} & \lfloor l_i(o_i) / 2 \rfloor \le x_i \le L - \lfloor l_i(o_i) / 2 \rfloor, & 1 \le i \le n,\\
  & & \lfloor w_i(o_i) / 2 \rfloor \le y_i \le W - \lfloor w_i(o_i) / 2 \rfloor, & 1 \le i \le n,\\
  & & (P_i(o_i) \oplus \bm{v}_i) \cap (P_j(o_j) \oplus \bm{v}_j) = \emptyset, & 1 \le i < j \le n,\\
  & & L \in \mathbb{Z}_+,\\
  & & \bm{v}_i = (x_i,y_i) \in \mathbb{Z}_+^2, & 1 \le i \le n,\\
  & & o_i \in O_i, & 1 \le i \le n,
\end{array}
\end{equation}
where $\mathbb{Z}_+$ is the set of nonnegative integer values.
We note that the coordinate $\bm{v}_i = (x_i,y_i)$ of a piece $P_i(o_i)$ takes $(\lfloor l_i(o_i) / 2 \rfloor, \lfloor w_i(o_i) / 2 \rfloor)$ and $(L - \lfloor l_i(o_i) / 2 \rfloor, W - \lfloor w_i(o_i) / 2 \rfloor)$ when it is placed at the bottom-left and top-right corners of the container $C$, respectively.
We also note that minimization of the length $L$ is equivalent to maximization of the density defined by $\sum_{1 \le i \le n} (\textnormal{area of} \; P_i) / WL$.

\subsection{Overlap minimization problem}\label{sec:overlap_minimization_problem}
We consider OMP as a sub-problem of ISP to find a feasible layout $(\bm{v},\bm{o})$ of given pieces for the container $C$ with a given length $L$.
A solution of OMP may have a number of overlapping pieces, and the total amount of overlap is penalized in such a way that a solution with no penalty gives a feasible layout for ISP.
Let $f_{ij}(\bm{v}_i,\bm{v}_j,o_i,o_j)$ be a function that measures the overlap amount for a pair of pieces $P_i(o_i) \oplus \bm{v}_i$ and $P_j(o_j) \oplus \bm{v}_j$.
The objective of OMP is to find a solution $(\bm{v},\bm{o})$ that minimizes the total amount of the overlap penalty $F(\bm{v},\bm{o}) = \sum_{1 \le i < j \le n} f_{ij}(\bm{v}_i,\bm{v}_j,o_i,o_j)$ under the constraint that all pieces $P_i$ ($1 \le i \le n$) are placed within the container $C(W,L)$.
\begin{equation}
  \begin{array}{llll}
    (\textnormal{OMP}) & \textnormal{minimize} & F(\bm{v},\bm{o}) = \displaystyle\sum_{1 \le i < j \le n} f_{ij}(\bm{v}_i,\bm{v}_j,o_i,o_j)\\
    & \textnormal{subject to} & \lfloor l_i(o_i) / 2 \rfloor \le x_i \le L - \lfloor l_i(o_i) / 2 \rfloor, & 1 \le i \le n,\\
    & & \lfloor w_i(o_i) / 2 \rfloor \le y_i \le W - \lfloor w_i(o_i) / 2 \rfloor, & 1 \le i \le n,\\
    & & \bm{v}_i = (x_i,y_i) \in \mathbb{Z}_+^2, & 1 \le i \le n,\\
    & & o_i \in O_i, & 1 \le i \le n.
  \end{array}
\end{equation}

We introduce the \emph{directional penetration depth} to define the overlap penalty function $f_{ij}(\bm{v}_i,\bm{v}_j,o_i,o_j)$ for a pair of pieces $P_i(o_i) \oplus \bm{v}_i$ and $P_j(o_j) \oplus \bm{v}_j$, which was originally defined as the minimum translational distance in a given direction to separate a pair of polygons~\citep{DobkinD1993}.
If they do not overlap, then their directional penetration depth is zero.
In the raster model, we define the directional penetration depth in the horizontal and vertical directions as follows:
\begin{equation}
  \label{eq:directional_penetration_depth}
  \delta(P_i, P_j,\bm{d}) = \min \{ |t| \mid P_i \cap (P_j \oplus t \bm{d}) = \emptyset, t \in \mathbb{Z} \}, \;\; \bm{d} \in \{ (1,0), (0,1) \}.
\end{equation}
We then define the overlap penalty $f_{ij}(\bm{v}_i,\bm{v}_j,o_i,o_j)$ for a pair of pieces $P_i(o_i) \oplus \bm{v}_i$ and $P_j(o_j) \oplus \bm{v}_j$ as the minimum within the horizontal and vertical penetration depths
\begin{equation}
  \label{eq:coordinate_penetration_depth}
    f_{ij}(\bm{v}_i,\bm{v}_j,o_i,o_j) = \min \{ \delta(P_i(o_i) \oplus \bm{v}_i, P_j(o_j) \oplus \bm{v}_j, \bm{d}) \mid \bm{d} \in \{ (1,0), (0,1) \} \}.
\end{equation}

\subsection{Entire algorithm for the irregular strip packing problem}\label{sec:entire_algorithm}
\begin{figure}[tb]
  \centering
  \includegraphics[height=0.3\textheight]{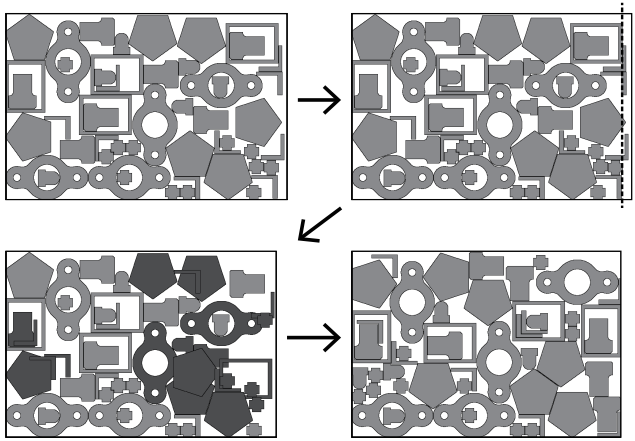}
  \caption{The outline of the proposed algorithm (GCDH).}\label{fig:outline}
\end{figure}

We give an entire description of the proposed algorithm GCDH for ISP.
The GCDH first generates an initial solution $(\bm{v},\bm{o})$ by a construction algorithm to be explained in Section~\ref{sec:construction}, and sets the minimum container length $L$ containing all pieces $P_i$ ($1 \le i \le n$).
It then searches the minimum feasible container length $L^{\ast}$ by shrinking or extending the right sides of the container $C$ until the time limit is reached, where the ratios of shrinking and extending container length are controlled by the parameters $r_{\mathrm{dec}}$ and $r_{\mathrm{inc}}$, respectively.
If the current solution $(\bm{v},\bm{o})$ is feasible, then it shrinks the container length $L$ to $\lfloor (1 - r_{\mathrm{dec}}) L \rfloor$ and relocates protruding pieces $P_i$ at random positions in the container $C(W,L)$; otherwise it extends the container length $L$ to $\lfloor (1 + r_{\mathrm{inc}}) L \rfloor$.
If there is at least one overlapping pair of pieces in the current solution, then it tries to resolve overlap by CDH for OMP explained in Section~\ref{sec:coordinate_descent}.
Figure~\ref{fig:outline} shows the outline of the proposed algorithm GCDH.
The algorithm is formally described in Algorithm~\ref{alg:GCDH}, where we omit the input data $W, \mathcal{P},\mathcal{O}$ commonly used in all algorithms in this paper.
\begin{algorithm}
  \caption{GCDH}\label{alg:GCDH}
  \begin{spacing}{1.0}
    \begin{small}
\begin{algorithmic}[1]
  \Statex
  \State $L, (\bm{v},\bm{o}) \leftarrow \textnormal{{\sc Construct}}$ \Comment{See Algorithm~\ref{alg:construct}}
  \State $L^{\ast} \leftarrow L$, $(\bm{v}^{\ast},\bm{o}^{\ast}) \leftarrow (\bm{v},\bm{o})$
  \State $L \leftarrow \lfloor (1 - r_{\mathrm{dec}})L^{\ast} \rfloor$
  \Repeat
  \State $(\bm{v},\bm{o}) \leftarrow \textnormal{GLS}(L,\bm{v},\bm{o})$ \Comment{See Algorithm~\ref{alg:guided_local_search}}
  \If{$(\bm{v},\bm{o})$ is feasible}
  \State $L^{\ast} \leftarrow L$, $(\bm{v}^{\ast},\bm{o}^{\ast}) \leftarrow (\bm{v},\bm{o})$
  \State $L \leftarrow \lfloor (1 - r_{\mathrm{dec}})L^{\ast} \rfloor$
  \State Relocate protruding pieces $P_i(o_i)$ at random position in $C(W,L)$.
  \Else
  \State $L \leftarrow \lfloor (1 + r_{\mathrm{inc}})L \rfloor$
  \If{$L \ge L^{\ast}$}
  \State $L \leftarrow \lfloor (1 - r_{\mathrm{dec}})L^{\ast} \rfloor$, $(\bm{v},\bm{o}) \leftarrow (\bm{v}^{\ast},\bm{o}^{\ast})$
  \State Relocate protruding pieces $P_i(o_i)$ at random position in $C(W,L)$.
  \EndIf
  \EndIf
  \Until{The time limit is reached}
  \State Return $(\bm{v}^{\ast},\bm{o}^{\ast})$
\end{algorithmic}
\end{small}
\end{spacing}
\end{algorithm}

\section{Intersection test for rasterized shapes via scanline representation}\label{sec:intersection}

\subsection{No-fit polygon for intersection test}\label{sec:nofit_polygon}
\begin{figure}[tb]
  \centering
  \includegraphics[height=0.25\textheight]{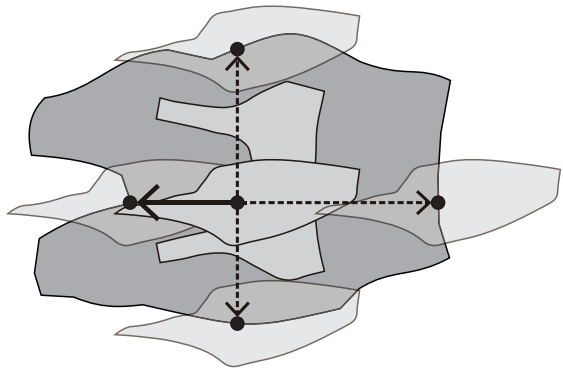}
  \caption{The no-fit polygon $\mathrm{NFP}(P_i,P_j)$ for two irregular shapes $P_i$ and $P_j$.\label{fig:nfp}}
\end{figure}

The \emph{no-fit polygon} (NFP) introduced by \citet{ArtRC1966} is a representative geometric technique used in many algorithms for the polygon packing problem.
It is also used for other applications such as image processing and robot motion planning, and is also known as the Minkowski difference and the configuration-space obstacle, respectively.
The no-fit polygon $\mathrm{NFP}(P_i,P_j)$ of an ordered pair of pieces $P_i$ and $P_j$ is defined by
\begin{equation}
\mathrm{NFP}(P_i,P_j) = P_i \oplus (-P_j) = \{ \bm{u} - \bm{w} \mid \bm{u} \in P_i, \bm{w} \in P_j \}.
\end{equation}
The NFP has an important property that $P_j \oplus \bm{v}_j$ overlaps with $P_i \oplus \bm{v}_i$ if and only if $\bm{v}_j - \bm{v}_i \in \mathrm{NFP}(P_i, P_j)$ holds.
That is, the intersection test for two irregular shapes $P_i \oplus \bm{v}_i$ and $P_j \oplus \bm{v}_j$ can be identified by testing whether a point $\bm{v}_j - \bm{v}_i$ is inside an irregular shape $\mathrm{NFP}(P_i,P_j)$ or not.
Figure~\ref{fig:nfp} shows an example of $\mathrm{NFP}(P_i,P_j)$ for two irregular shapes $P_i$ and $P_j$, where the solid arrow illustrates the minimum translation of the piece $P_j$ to separate from the piece $P_i$ within the horizontal and vertical directions.
The NFP enables us to compute the intersection test and the directed penetration depth efficiently, assuming that we compute NFPs for all pairs of pieces in advance.

\subsection{Scanline representation for rasterized shapes}\label{sec:scanline}
\begin{figure}[tb]
  \centering
  \includegraphics[width=0.95\textwidth]{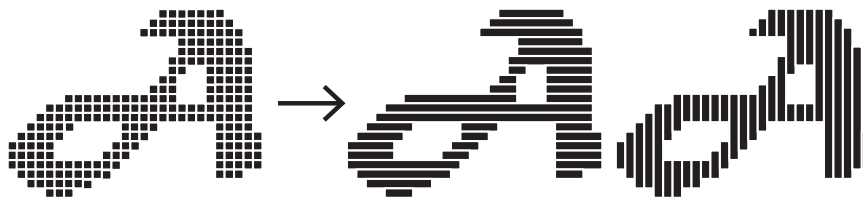}
  \caption{The double scanline representation for a rasterized shape.\label{fig:scanline}}
\end{figure}

We consider a pair of scanlines representation called the \emph{double scanline representation} that reduces the complexity of rasterized shapes by merging consecutive pixels in each row and column into strips with unit width, respectively.
Figure~\ref{fig:scanline} shows an example of the double scanline representation of a rasterized shape.
\citet{HuY2018a} developed an efficient algorithm to compute NFPs of a single (i.e., horizontal or vertical) scanline representation called the Integrate-NFP (See details of its implementation in \citet{HuY2018a}).

\begin{figure}[tb]
  \centering
  \includegraphics[width=0.85\textwidth]{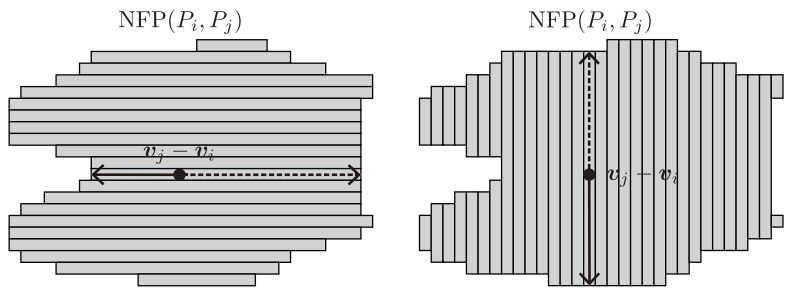}
  \caption{Computing the horizontal and vertical penetration depths from the scanline representations of NFP.\label{fig:penetration_depth}}
\end{figure}

Based on this, we compute the horizontal penetration depth $\delta(P_i \oplus \bm{v}_i,P_j \oplus \bm{v}_j,(1,0))$ efficiently.
Figure~\ref{fig:penetration_depth} shows an example of computing the horizontal (and vertical) penetration depth from NFP of the scanline representation.
Let $\mathcal{S}_{ij} = \{ S_{ij1}, S_{ij2}, \dots, S_{ijm_{ij}} \}$ be the set of horizontal strips representing $\mathrm{NFP}(P_i,P_j)$, where $m_{ij}$ is the number of strips representing $\mathrm{NFP}(P_i, P_j)$.
When the relative position $\bm{v}_j - \bm{v}_i$ of the piece $P_j$ is placed in a strip $S_{ijk}$, the horizontal penetration depth $\delta(P_i \oplus \bm{v}_i,P_j \oplus \bm{v}_j,(1,0))$ then takes the minimum length from $\bm{v}_j - \bm{v}_i$ to left and right side of the strip $S_{ijk}$.
We note that the horizontal penetration depth can be computed in $\mathrm{O}(1)$ time when $\mathrm{NFP}(P_i, P_j)$ is $y$-monotone, where a rasterized shape is $y$-monotone if it can be represented by a set of strips such that there is exactly one strip for each row and strips in adjacent rows are contiguous.
We also compute the vertical penetration depth $\delta(P_i \oplus \bm{v}_i,P_j \oplus \bm{v}_j,(0,1))$ efficiently utilizing the vertical scanline representation of $\mathrm{NFP}(P_i, P_j)$ in the same fashion.
In Figure~\ref{fig:penetration_depth}, the solid arrows illustrate the minimum translation of the piece $P_j$ to separate from the piece $P_i$ in the horizontal and vertical directions, and their lengths are the horizontal and vertical penetration depths, respectively.

\section{Coordinate descent heuristics for the overlap minimization problem}\label{sec:coordinate_descent}

\subsection{Outline of the coordinate descent heuristics}\label{sec:outline_coordinate_descent}
We develop an improvement algorithm for OMP called the \emph{coordinate descent heuristics} (CDH), which start from an initial solution and repeatedly apply the \emph{line search} that minimizes the objective function along a coordinate direction until no better solution is found in any coordinate directions.

\begin{figure}[tb]
  \centering
  \includegraphics[width=0.95\textwidth]{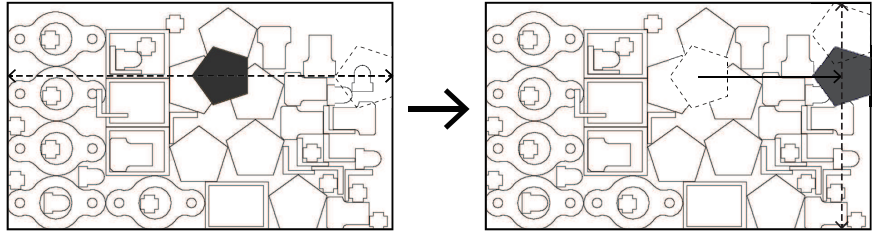}
  \caption{The neighborhood search of CDH for OMP.}\label{fig:coordinate_descent}
\end{figure}

We first explain the neighborhood of the CDH for OMP.
Let $(\bm{v},\bm{o})$ be the current solution.
The neighborhood $\mathrm{NB}(\bm{v},\bm{o})$ is defined as the set of neighbor solutions, where a neighbor solution $(\bm{v}^{\prime},\bm{o}^{\prime}) \in \mathrm{NB}(\bm{v},\bm{o})$ is obtained by setting a new orientation $o_k^{\prime} \in O_k$ of a piece $P_k$ ($1 \le k \le n$) and iteratively applying the line search to find a new position $\bm{v}_k^{\prime}$.
The quality of a solution $(\bm{v},\bm{o})$ is measured by the following weighted overlap penalty function
\begin{equation}
\widetilde{F}(\bm{v},\bm{o}) = \sum_{1 \le i < j \le n} \alpha_{ij} \cdot f_{ij}(\bm{v}_i,\bm{v}_j,o_i,o_j),
\end{equation}
where $\alpha_{ij} > 0$ are the penalty weights and $f_{ij}$ are the overlap penalties defined in (\ref{eq:coordinate_penetration_depth}) for a pair of pieces $P_i(o_i) \oplus \bm{v}_i$ and $P_j(o_j) \oplus \bm{v}_j$.
The penalty weights $\alpha_{ij}$ are adaptively controlled by GLS to be explained in Section~\ref{sec:guided_local_search}.
For a piece $P_k(o_k^{\prime})$, the neighborhood search finds a new position $\bm{v}_k^{\prime}$ in the container $C(W,L)$ such that the following weighted overlap penalty function
\begin{equation}
\label{eq:overlap_penalty_piece}
\widetilde{F}_k(\bm{v}_k^{\prime},o_k^{\prime}) = \sum_{1 \le j \le n, j \not= k} \alpha_{kj} \cdot f_{kj}(\bm{v}_k^{\prime},\bm{v}_j,o_k^{\prime},o_j)
\end{equation}
is minimized.
For this, the neighborhood search repeatedly moves the piece $P_k(o_k^{\prime})$ in the horizontal and vertical directions alternately until no better position is found in either direction.
Figure~\ref{fig:coordinate_descent} shows how the neighborhood search proceeds.
For each move of the piece $P_k(o_k^{\prime})$ in a specified direction $\bm{d} \in \{ (1,0), (0,1) \}$, let
\begin{equation}
N = \{ t \mid P(o_k^{\prime}) \oplus (\bm{v}_k + t \bm{d}) \subseteq C(W,L), t \in \mathbb{Z} \}
\end{equation}
be the set of valid $t$ such that the piece $P_k(o_k^{\prime})$ placed on the line $\bm{v}_k + t \bm{d}$ ($t \in \mathbb{Z}$) is contained in the container $C(W,L)$, where $\mathbb{Z}$ is the set of integer values.
The line search (to be explained in Section~\ref{sec:line_search}) finds a new valid position $\bm{v}_k^{\prime} = \bm{v}_k + t \bm{d}$ ($t \in N$) that minimizes the weighted overlap penalty function $\widetilde{F}_k(\bm{v}_k + t\bm{d},o_k^{\prime})$.
The neighborhood search is formally described in Algorithm~\ref{alg:neighborhood_search}.
\begin{algorithm}
  \caption{{\sc NeighborSearch}($L,\bm{v},\bm{o},P_k,o_k^{\prime}$)}\label{alg:neighborhood_search}
  \begin{spacing}{1.0}
    \begin{small}
\begin{algorithmic}[1]
  \Statex
  \State $\bm{v}^{\prime} \leftarrow \bm{v}$, $o_j^{\prime} \leftarrow o_j$ ($j \not=k$), $\bm{d} \leftarrow (1,0)$
  \State Find $t \in N$ to minimize $\widetilde{F}_k(\bm{v}_k^{\prime} + t \bm{d}, o_k^{\prime})$
  \If{$\widetilde{F}_k(\bm{v}_k^{\prime} + t\bm{d}, o_k^{\prime}) < \widetilde{F}_k(\bm{v}_k^{\prime}, o_k^{\prime})$}
  \State $\bm{v}_k^{\prime} \leftarrow \bm{v}_k^{\prime} + t \bm{d}$
  \EndIf
  \State $\bm{d} \leftarrow (0,1)$
  \Repeat
  \State Find $t \in N$ to minimize $\widetilde{F}_k(\bm{v}_k^{\prime} + t \bm{d}, o_k^{\prime})$
  \If{$\widetilde{F}_k(\bm{v}_k^{\prime} + t\bm{d}, o_k^{\prime}) < \widetilde{F}_k(\bm{v}_k^{\prime}, o_k^{\prime})$}
  \State $\bm{v}_k^{\prime} \leftarrow \bm{v}_k^{\prime} + t \bm{d}$
  \EndIf
  \If{$\bm{d} = (1,0)$}
  \State $\bm{d} \leftarrow (0,1)$
  \Else
  \State $\bm{d} \leftarrow (1,0)$
  \EndIf
  \Until{$\widetilde{F}_k(\bm{v}_k^{\prime} + t\bm{d}, o_k^{\prime}) = \widetilde{F}_k(\bm{v}_k^{\prime}, o_k^{\prime})$}
  \State Return $(\bm{v}^{\prime},\bm{o}^{\prime})$
\end{algorithmic}
\end{small}
\end{spacing}
\end{algorithm}

We now describe the outline of CDH for OMP.
Starting from an initial solution $(\bm{v},\bm{o})$ with some overlapping pieces, CDH repeatedly replaces the current solution $(\bm{v},\bm{o})$ with the first improved solution $(\bm{v}^{\prime},\bm{o}^{\prime}) \in \mathrm{NB}(\bm{v},\bm{o})$ obtained by the neighborhood search.
That is, CDH searches $\mathrm{NB}(\bm{v},\bm{o})$ in random order, and if it finds an improved solution $(\bm{v}^{\prime},\bm{o}^{\prime}) \in \mathrm{NB}(\bm{v},\bm{o})$ satisfying $\widetilde{F}(\bm{v}^{\prime},\bm{o}^{\prime}) < \widetilde{F}(\bm{v},\bm{o})$, then it immediately replaces the current solution $(\bm{v},\bm{o})$ with $(\bm{v}^{\prime},\bm{o}^{\prime})$.
If no overlapping piece exists in the current solution $(\bm{v},\bm{o})$ (i.e., $\widetilde{F}(\bm{v},\bm{o}) = 0$) or no better solution found in $\mathrm{NB}(\bm{v},\bm{o})$, then CDH outputs $(\bm{v},\bm{o})$ as a locally optimal solution and the best solution $(\bm{v}^{\ast},\bm{o}^{\ast})$ obtained so far, measured by the original penalty function $F$, and halts.

We also incorporate the \emph{fast local search} (FLS) strategy~\citep{VoudourisC1999} to improve the efficiency of CDH.
The strategy decomposes the neighborhood into a number of subneighborhoods, which are labeled active or inactive depending on whether they are being searched or not, i.e., it skips evaluating all neighbor solutions in inactive subneighborhoods.
We define the subneighborhood $\mathrm{NB}_k(\bm{v},\bm{o})$ ($1 \le k \le n$) of the current solution $(\bm{v},\bm{o})$ as the set of solutions obtainable by setting orientation $o_k^{\prime} \in O_k$ and applying the neighborhood search to the piece $P_k$, i.e., the neighborhood $\mathrm{NB}(\bm{v},\bm{o})$ is partitioned with respect to the pieces $P_k$ ($1 \le k \le n$).
The CDH first sets all subneighborhoods $\mathrm{NB}_k(\bm{v},\bm{o})$ ($1 \le k \le n$) to be active, and it searches active subneighborhoods $\mathrm{NB}_k(\bm{v},\bm{o})$ in random order.
If no improvement has been made in an active subneighborhood $\mathrm{NB}_k(\bm{v},\bm{o})$, then CDH inactivates it.
If CDH finds an improved solution $(\bm{v}^{\prime},\bm{o}^{\prime})$ satisfying $\widetilde{F}(\bm{v}^{\prime},\bm{o}^{\prime}) < \widetilde{F}(\bm{v},\bm{o})$ in an active subneighborhood $\mathrm{NB}_k(\bm{v},\bm{o})$, then it activates all subneighborhood $\mathrm{NB}_j(\bm{v},\bm{o})$ corresponding to the pieces $P_j$ overlapping with the piece $P_k$ before and after its move.
The CDH is formally described in Algorithm~\ref{alg:coordinate_descent}, where $A$ denotes the set of indices $k$ ($1 \le k \le n$) corresponding to the active subneighborhood $\mathrm{NB}_k(\bm{v},\bm{o})$, and $(\bm{v}^{\ast},\bm{o}^{\ast})$ and $(\tilde{\bm{v}}, \tilde{\bm{o}})$ denote the best solutions of the original overlap penalty function $F$ and the weighted overlap penalty function $\widetilde{F}$, respectively.
\begin{algorithm}
  \caption{{\sc CDH}($L,\bm{v},\bm{o}$)}\label{alg:coordinate_descent}
  \begin{spacing}{1.0}
    \begin{small}
\begin{algorithmic}[1]
  \Statex
  \State $(\tilde{\bm{v}},\tilde{\bm{o}}) \leftarrow (\bm{v},\bm{o})$, $(\bm{v}^{\ast},\bm{o}^{\ast}) \leftarrow (\bm{v},\bm{o})$, $A \leftarrow \{ 1,\dots,n\}$
  \While{$A \not= \emptyset$}
  \State Randomly select $k \in A$
  \State $O \leftarrow O_k$
  \While{$O \not= \emptyset$}
  \State Randomly select $o_k^{\prime} \in O$
  \State $(\bm{v}^{\prime},\bm{o}^{\prime}) \leftarrow \textnormal{{\sc NeighborSearch}}(L,\bm{v},\bm{o},P_k,o_k^{\prime})$ \Comment{See Algorithm~\ref{alg:neighborhood_search}}
  \If{$F(\bm{v}^{\prime},\bm{o}^{\prime}) < F(\bm{v}^{\ast},\bm{o}^{\ast})$}
  \State $(\bm{v}^{\ast},\bm{o}^{\ast}) \leftarrow (\bm{v}^{\prime},\bm{o}^{\prime})$
  \If{$F(\bm{v}^{\ast},\bm{o}^{\ast}) = 0$}
  \State Return $(\bm{v}^{\ast},\bm{o}^{\ast})$, $(\tilde{\bm{v}},\tilde{\bm{o}})$
  \EndIf
  \EndIf
  \If{$\widetilde{F}(\bm{v}^{\prime},\bm{o}^{\prime}) < \widetilde{F}(\tilde{\bm{v}},\tilde{\bm{o}})$}
  \State $(\tilde{\bm{v}},\tilde{\bm{o}}) \leftarrow (\bm{v}^{\prime},\bm{o}^{\prime})$
  \For{$P_j$ overlapping with $P_k$ before and after the move}
  \State $A \leftarrow A \cup \{ j \}$
  \EndFor
  \EndIf
  \State $O \leftarrow O \setminus \{ o_k^{\prime} \}$
  \EndWhile
  \State $A \leftarrow A \setminus \{ k \}$
  \EndWhile
  \State Return $(\bm{v}^{\ast},\bm{o}^{\ast})$, $(\tilde{\bm{v}},\tilde{\bm{o}})$
\end{algorithmic}
    \end{small}
  \end{spacing}
\end{algorithm}

\subsection{Efficient implementation of the line search}\label{sec:line_search}
\begin{figure}[tb]
  \centering
  \includegraphics[height=0.2\textheight]{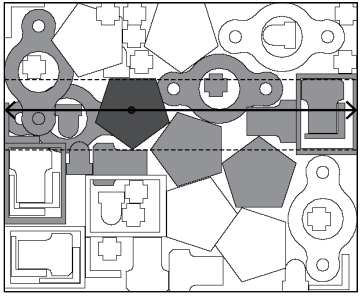}
  \caption{Detecting the overlapping pieces in the line search.}\label{fig:detect_overlap}
\end{figure}

We develop an efficient line search algorithm for the neighborhood search, which finds a new position when a piece $P_k(o_k^{\prime})$ moves in the horizontal or vertical direction $\bm{d} \in \{ (1,0), (0,1) \}$.
Recall that the line search finds the new position $\bm{v}_k^{\prime} = \bm{v}_k + t \bm{d}$ to minimize the weighted overlap penalty function $\widetilde{F}_k(\bm{v}_k + t \bm{d}, o_k^{\prime})$ while the piece $P_k(o_k^{\prime})$ is contained in the container $C(W,L)$.
We consider below the case when $P_k(o_k^{\prime})$ moves in the horizontal direction (i.e., $\bm{d} = (1,0)$).
The case of the vertical direction (i.e., $\bm{d} = (0,1)$) is almost the same and is omitted.
The weighted overlap penalty function $\widetilde{F}_k(\bm{v}_k + t \bm{d}, o_k^{\prime})$ is decomposed into $f_{kj}(\bm{v}_k + t \bm{d},\bm{v}_j,o_k^{\prime},o_j)$ for pairs of pieces $P_k(o_k^{\prime})$ and $P_j(o_j)$ ($1 \le j \le n, j \not= k$) by definition~(\ref{eq:overlap_penalty_piece}).
Let
\begin{equation}
I_{kj} = \{ t \mid \bm{v}_k + t \bm{d} \in \mathrm{NFP}(P_j(o_j),P_k(o_k^{\prime})), t \in N \}
\end{equation}
be the set of positions of the piece $P_k(o_k^{\prime})$ in terms of $t \in N$ such that it overlaps with the other piece $P_j(o_j)$.
If $I_{kj} = \emptyset$ holds, then the overlap penalty $f_{kj}(\bm{v}_k + t \bm{d},\bm{v}_j,o_k^{\prime},o_j)$ always takes zero for all $t \in N$.
The line search algorithm first detects whether $I_{kj} = \emptyset$ or not by checking the overlap between projections of pieces $P_k(o_k^{\prime}) \oplus \bm{v}_k$ and $P_j(o_j) \oplus \bm{v}_j$ onto the $y$-axis.
That is, $I_{kj} = \emptyset$ holds if and only if the intersection of their projections satisfies $[y_k^{\min}, y_k^{\max}] \cap [y_j^{\min},y_j^{\max}] = \emptyset$, where $y_i^{\min}$ and $y_i^{\max}$ ($1 \le i \le n$) of a piece $P_i(o_i) \oplus \bm{v}_i$ are defined by
\begin{align}
  y_i^{\min} &= \min\{ y \mid (x,y) \in P_i(o_i) \oplus \bm{v}_i \},\\
  y_i^{\max} &= \max\{ y \mid (x,y) \in P_i(o_i) \oplus \bm{v}_i \},
\end{align}
respectively.
Figure~\ref{fig:detect_overlap} shows an example of detecting the overlapping pieces $P_j(o_j)$ when the piece $P_k(o_k^{\prime})$ moves in the horizontal direction.
Let $N^+ = \bigcup_{1 \le j \le n, j \not= k} I_{kj}$ be the set of $t$ ($\in N$) inducing overlap with other pieces, and $N^- = N \setminus N^+$ be its complement.
If $N^- \not= \emptyset$ holds, then the line search algorithm finds the minimum $t \in N^-$ (i.e., the left most feasible position); otherwise, it finds the position $t \in N^+$ that minimizes the overlap penalty function $\widetilde{F}_k(\bm{v}_k + t \bm{d}, o_k^{\prime})$.

\begin{figure}[tb]
  \centering
  \includegraphics[width=0.95\textwidth]{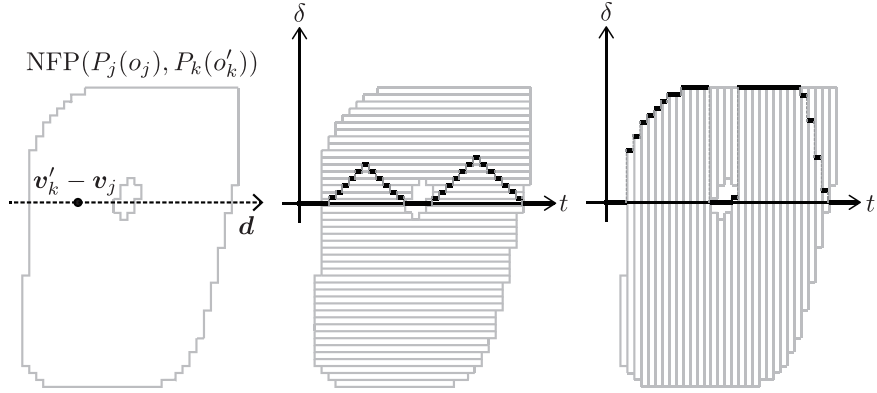}
  \caption{Computing the horizontal and vertical penetration depth in the line search.}\label{fig:line_search}
\end{figure}

We now consider how to compute the overlap penalty function $f_{kj}(\bm{v}_k + t \bm{d}, \bm{v}_j, o_k^{\prime}, o_j)$ defined in (\ref{eq:coordinate_penetration_depth}) for a given $t \in N$, where $I_{kj} \not= \emptyset$ is assumed.
Let $\bm{v}_k^{\prime} = \bm{v}_k + t \bm{d}$ be the position of the piece $P_k(o_k^{\prime})$ for a given $t \in N$.
If $t \not\in I_{kj}$ holds, then $f_{kj}(\bm{v}_k^{\prime}, \bm{v}_j, o_k^{\prime}, o_j) = 0$; otherwise, the overlap penalty $f_{kj}(\bm{v}_k^{\prime}, \bm{v}_j, o_k^{\prime}, o_j)$ is decomposed into the horizontal and vertical penetration depths.
We denote the horizontal and vertical penetration depths for a given $t \in N$ by
\begin{equation}
  \delta(P_k(o_k^{\prime}) \oplus \bm{v}_k^{\prime}, P_j(o_j) \oplus \bm{v}_j, \bm{d}) = \min \{ |s| \mid \bm{v}_k^{\prime} - \bm{v}_j + s \bm{d} \not\in \mathrm{NFP}(P_j(o_j),P_k(o_k^{\prime}) \}, 
\end{equation}
for $\bm{d} \in \{ (1,0), (0,1) \}$, respectively.
Figure~\ref{fig:line_search} shows an example of computing the horizontal and vertical penetration depths in the line search from NFP of the double scanline representation.
Recall the horizontal and vertical scanline representations of $\mathrm{NFP}(P_j(o_j),P_k(o_k^{\prime}))$ in Figure~\ref{fig:penetration_depth}.
When the relative position $\bm{v}_k^{\prime} - \bm{v}_j$ of the piece $P_k(o_k^{\prime})$ is placed in a strip, the horizontal (resp., vertical) penetration depth then takes the minimum distance $|s|$ from $\bm{v}_k^{\prime} - \bm{v}_j$ to left and right (resp., bottom and top) side of the strip.

\begin{figure}[tb]
  \centering
  \includegraphics[width=0.95\textwidth]{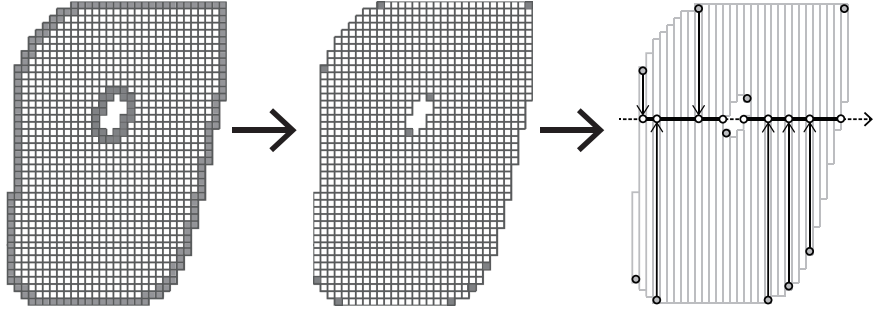}
  \caption{The corner detection algorithm for the NFP.}\label{fig:corner_detection}
\end{figure}

The line search algorithm is very time consuming when $|N^+|$ becomes larger according to high-resolution of the raster representation as long as it computes the overlap penalty $\widetilde{F}_k(\bm{v}_k+t\bm{d},o_k^{\prime})$ for all $t \in N^+$.
As shown in Figure~\ref{fig:line_search}, the horizontal penetration depth forms a regular stepwise with steady changes, and it possibly takes the minimum value only at the left and right end of the strips on the horizontal line $\bm{v}_k + t \bm{d}$ ($t \in N$).
On the other hand, the vertical penetration depth forms an irregular stepwise function that sometimes changes rapidly; however, it may occur only at the ``corners'' of $\mathrm{NFP}(P_j(o_j),P_k(o_k^{\prime}))$.
We accordingly introduce a corner detection technique to restrict the search space of the line search only having rapid changes of the vertical penetration depth.
We first detect the contour of $\mathrm{NFP}(P_j(o_j),P_k(o_k^{\prime}))$ and then apply a fast corner detection algorithm called FAST \citep{RostenE2006} to the contour as shown in Figure~\ref{fig:corner_detection}.
Let $\overline{\mathrm{NFP}}(P_j(o_j),P_k(o_k^{\prime}))$ be the set of detected corners in $\mathrm{NFP}(P_j(o_j),P_k(o_k^{\prime}))$ obtained by the FAST.
We consider the set of positions $\bar{I}_{kj}$ ($\subseteq I_{kj}$) of the piece $P_k(o_k^{\prime})$ in terms of $t \in N$ that contains (i)~the left and right end of the horizontal strips on the horizontal line $\bm{v}_k + t \bm{d}$ ($t \in N^+$), and (ii)~the projection of $\overline{\mathrm{NFP}}(P_j(o_j),P_k(o_k^{\prime}))$ onto the horizontal line $\bm{v}_k + t \bm{d}$ ($t \in N^+$).
Let $\overline{N}^+ = \bigcup_{1 \le j \le n, j \not= k} \bar{I}_{kj}$ ($\subseteq N^+$) be the reduced search space of the line search; i.e., the line search algorithm computes the weighted overlap penalty $\widetilde{F}_k(\bm{v}_k+t\bm{d},o_k^{\prime})$ only for $t \in \overline{N}^+$ (instead of $N^+$) when $N^- = \emptyset$ holds.
In Figure~\ref{fig:corner_detection} (right), the grey nodes show the detected corners in $\overline{\mathrm{NFP}}(P_j(o_j),P_k(o_k^{\prime}))$ by FAST, and the white nodes show the positions in $\bar{I}_{kj}$ to be evaluated by the line search algorithm.

\subsection{Guided local search}\label{sec:guided_local_search}
It is often the case that CDH alone may not attain a sufficiently good solution.
To improve its performance, we incorporate it into one of the representative metaheuristics called the \emph{guided local search} (GLS)~\citep{VoudourisC1999}.
The GLS repeats CDH while updating the penalty weights of the objective function adaptively to resume the search from the previous locally optimal solution.
The GLS starts from an initial solution $(\bm{v},\bm{o})$ with some overlapping pieces, where the penalty weights $\alpha_{ij}$ are initialized to $1.0$.
Whenever CDH stops at a locally optimal solution $(\bm{v},\bm{o})$, GLS updates the penalty weights $\alpha_{ij}$ by the following formula:
\begin{equation}
\label{eq:update_weight}
\alpha_{ij} \leftarrow \alpha_{ij} + \displaystyle\frac{f_{ij}(\bm{v}_i,\bm{v}_j,o_i,o_j)}{\displaystyle\max_{1\le k < l \le n} f_{kl}(\bm{v}_k,\bm{v}_l,o_k,o_l)}, \quad 1 \le i < j \le n.
\end{equation}
By updating the penalty weights $\alpha_{ij}$ ($1 \le i < j \le n$) repeatedly, the current solution $(\bm{v},\bm{o})$ becomes no longer locally optimal under the updated weighted overlap penalty function $\widetilde{F}$, and GLS resumes the search from the current solution.
The GLS stops these operations if it fails to improve the best solution $(\bm{v}^{\ast},\bm{o}^{\ast})$ of the original penalty function $F$ after a specified number $k_{\max}$ of consecutive calls to CDH.
The GLS is formally described in Algorithm~\ref{alg:guided_local_search}.
\begin{algorithm}
  \caption{GLS($L,\bm{v},\bm{o}$)}\label{alg:guided_local_search}
  \begin{spacing}{1.0}
    \begin{small}
\begin{algorithmic}[1]
\Statex
\State $(\tilde{\bm{v}},\tilde{\bm{o}}) \leftarrow (\bm{v},\bm{o})$, $(\bm{v}^{\ast},\bm{o}^{\ast}) \leftarrow (\bm{v},\bm{o})$, $\alpha_{ij} \leftarrow 1.0$ ($1 \le i < j \le n$), $k \leftarrow 0$
\While{$k < k_{\max}$}
\State $(\bm{v}^{\prime},\bm{o}^{\prime}), (\tilde{\bm{v}},\tilde{\bm{o}}) \leftarrow \textnormal{CDH}(L,\tilde{\bm{v}},\tilde{\bm{o}})$ \Comment{See Algorithm~\ref{alg:coordinate_descent}}
\If{$F(\bm{v}^{\prime},\bm{o}^{\prime}) < F(\bm{v}^{\ast},\bm{o}^{\ast})$}
\State $(\bm{v}^{\ast},\bm{o}^{\ast}) \leftarrow (\bm{v}^{\prime},\bm{o}^{\prime})$, $k \leftarrow 0$
\If{$F(\bm{v}^{\ast},\bm{o}^{\ast}) = 0$}
\State Return $(\bm{v}^{\ast},\bm{o}^{\ast})$
\EndIf
\EndIf
\State Update the penalty weights $\alpha_{ij}$ ($1 \le i < j \le n$) of $(\tilde{\bm{v}},\tilde{\bm{o}})$ by (\ref{eq:update_weight})
\State $k \leftarrow k+1$
\EndWhile
\State Return $(\bm{v}^{\ast},\bm{o}^{\ast})$
\end{algorithmic}
    \end{small}
    \end{spacing}
\end{algorithm}

\section{Construction algorithm for the initial layout}\label{sec:construction}
\begin{figure}[tb]
  \centering
  \includegraphics[width=0.95\textwidth]{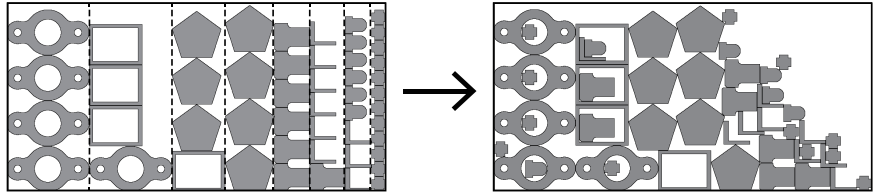}
  \caption{Constructing an initial solution for ISP.}\label{fig:construction}
\end{figure}

We generate an initial solution $(\bm{v},\bm{o})$ for ISP using the next-fit decreasing height (NFDH) algorithm.
The NFDH is a variant of the level algorithms for the rectangle packing problem~\citep{HuY2018b} that places rectangle pieces from bottom to top in columns forming levels as shown in Figure~\ref{fig:construction}.
The first level is the left side of the container, and each subsequent level is along the vertical line coinciding with the right of the longest piece placed in the previous level.
The NFDH first sorts the pieces $P_i$ ($1 \le i \le n$) in the descending order of the lengths $l_i$ of their bounding-boxes, where they are not rotated (i.e., $l_i = l_i(0)$).
Let the container length $L$ be sufficiently long, e.g., $L > \sum_{1 \le i \le n} l_i$.
The NFDH places all pieces $P_i$ ($1 \le i \le n$) one by one into the container $C$ according to the above order.
If possible, each piece $P_i$ (more precisely, its bounding-box) is placed at the bottom-most feasible position in the current level; otherwise, a new level is created to place the piece $P_i$.
We then apply a compaction algorithm (called {\sc Compact}) based on the bottom-left strategy that makes successive sliding moves to the left and the bottom alternately as long as possible.
The compaction algorithm can be implemented based on the neighborhood search (Algorithm~\ref{alg:neighborhood_search}) by replacing the condition $\widetilde{F}_k(\bm{v}_k^{\prime} + t\bm{d}, o_k^{\prime}) < \widetilde{F}_k(\bm{v}_k^{\prime}, o_k^{\prime})$ with $t < 0$ (i.e., the piece $P_k$ slides to the left or the bottom without overlap).
They are formally described in Algorithm~\ref{alg:construct}.
\begin{algorithm}
  \caption{{\sc Construct}}\label{alg:construct}
  \begin{spacing}{1.0}
    \begin{small}
\begin{algorithmic}[1]
  \Statex
  \State $L \leftarrow 0$, $\bar{l} \leftarrow 0$, $\bar{w} \leftarrow 0$
  \For{$P_i$ ($1 \le i \le n$) in the descending order of $l_i$}
  \If{$\bar{w} + w_i > W$}
  \State $L \leftarrow L + \bar{l}$, $\bar{w} \leftarrow 0$, $\bar{l} \leftarrow 0$
  \EndIf
  \State $\bm{v}_i \leftarrow (L + \lfloor \frac{l_i}{2} \rfloor, \bar{w} + \lfloor \frac{w_i}{2} \rfloor)$, $\bar{w} \leftarrow \bar{w} + w_i$, $\bar{l} \leftarrow \max\{\bar{l},l_i\}$
  \EndFor
  \For{$P_i$ ($1 \le i \le n$) in the descending order of $l_i$}
  \State $(\bm{v},\bm{o}) \leftarrow \textnormal{{\sc Compact}}(L,\bm{v},\bm{o},P_i,0)$
  \EndFor
  \State $L \leftarrow \displaystyle\max_{1 \le i \le n} \left( x_i + \lfloor \textstyle\frac{l_i}{2} \rfloor \right)$
  \State Return $L$, $(\bm{v},\bm{o})$
\end{algorithmic}
    \end{small}
    \end{spacing}
\end{algorithm}

\section{Computational results}\label{sec:computational_results}
The guided coordinate descent heuristics (GCDH) proposed in this paper were implemented with the C programming language and run on a single thread under a Mac PC with a 3.2~GHz Intel Core i7 processor and 64~GB memory.
We set the input parameters $r_{\mathrm{dec}} = 0.02$, $r_{\mathrm{inc}} = 0.005$ for GCDH (Algorithm~\ref{alg:GCDH}) and $k_{\max} = 200$ for GLS (Algorithm~\ref{alg:guided_local_search}) according to \citet{UmetaniS2009}.
The performance of GCDH was tested on two sets of instances for ISP represented in the vector model.
The first set includes 15 instances of the standard vector model without circular arcs nor holes (i.e., the simple polygon packing problem), which are available online at the web site of the working group on cutting and packing \citep{ESICUP}.
The second set includes 10 instances of an extended vector model, in which several instances incorporate circular arcs and holes into polygons~\citep{BurkeEK2010}.
Table~\ref{tab:instance} summarizes the data of the instances.
The second column ``\#shapes'' shows the number of different shapes, the third column ``\#pieces'' shows the total number of pieces, the fourth column ``avg.\#lines'' shows the average number of the line segments in an irregular shape, the fifth column ``avg.\#arcs'' shows the average number of the circular arcs in an irregular shape, the sixth column ``avg.\#holes'' shows the average number of holes in an irregular shape, and the seventh column ``degrees'' shows the permitted orientations, where we note that the permitted orientations are common to all pieces in each instance.
\begin{table}[tb]
  \caption{The test instances of ISP\label{tab:instance}}
  \medskip
\centering
      \begin{spacing}{0.8}
\begin{small}
\begin{tabular}{lrrrrrr} \hline
instance & \#shapes & \#pieces & avg.\#lines & avg.\#arcs & avg.\#holes & degrees\\ \hline
Albano & 8 & 24 & 6.83 & 0.00 & 0.00 & 0, 180 absolute\\
Dagli & 10 & 30 & 7.30 & 0.00 & 0.00 & 0, 180 absolute\\
Dighe1 & 16 & 16 & 3.88 & 0.00 & 0.00 & 0 absolute\\
Dighe2 & 10 & 10 & 4.70 & 0.00 & 0.00 & 0 absolute\\
Fu & 11 & 12 & 3.58 & 0.00 & 0.00 & 90 incremental\\
Jakobs1 & 25 & 25 & 6.00 & 0.00 & 0.00 & 90 incremental\\
Jakobs2 & 25 & 25 & 5.36 & 0.00 & 0.00 & 90 incremental\\
Mao & 9 & 20 & 8.70 & 0.00 & 0.00 & 90 incremental\\
Marques & 8 & 24 & 7.08 & 0.00 & 0.00 & 90 incremental\\
Shapes0 & 4 & 43 & 6.29 & 0.00 & 0.00 & 0 absolute\\
Shapes1 & 4 & 43 & 6.29 & 0.00 & 0.00 & 0, 180 absolute\\
Shirts & 8 & 99 & 6.05 & 0.00 & 0.00 & 0, 180 absolute\\
Swim & 10 & 48 & 20.0 & 0.00 & 0.00 & 0, 180 absolute\\
Trousers & 17 & 64 & 5.94 & 0.00 & 0.00 & 0, 180 absolute\\ \hline
Profiles1 & 8 & 32 & 4.63 & 0.63 & 0.00 & 90 incremental\\
Profiles2 & 7 & 50 & 7.54 & 1.50 & 0.38 & 90 incremental\\
Profiles3 & 6 & 46 & 7.93 & 0.65 & 0.00 & 45 incremental\\
Profiles4 & 7 & 54 & 3.83 & 0.41 & 0.00 & 90 incremental\\
Profiles5 & 5 & 50 & 7.19 & 0.00 & 0.13 & 15 incremental\\
Profiles6 & 9 & 69 & 4.60 & 1.40 & 0.00 & 90 incremental\\
Profiles7 & 9 & 9 & 4.67 & 0.00 & 0.00 & 90 incremental\\
Profiles8 & 9 & 18 & 4.67 & 0.00 & 0.00 & 90 incremental\\
Profiles9 & 16 & 57 & 26.61 & 0.00 & 0.16 & 90 incremental\\
Profiles10 & 13 & 91 & 8.23 & 0.00 & 0.00 & 0 absolute\\ \hline
\end{tabular}
\end{small}
      \end{spacing}
\end{table}

We converted these instances into the raster model with five different resolutions; i.e., we set the width $W$ of the container $C$ to 128, 256, 512, 1024 and 2048 pixels.
Tables~\ref{tab:instance_size} and \ref{tab:memory} show the instance size of the raster model for the instances (i.e., the number of pixels for representing all the irregular shapes) and the memory usage when running GCDH, respectively.
\begin{table}[tb]
  \caption{The instance size of the raster model (in pixels)\label{tab:instance_size}}
    \medskip
\centering
\begin{spacing}{0.8}
  \begin{small}
\begin{tabular}{lrrrrr} \hline
instance & \multicolumn{1}{c}{$W=128\mathrm{px}$} & \multicolumn{1}{c}{$W=256\mathrm{px}$} & \multicolumn{1}{c}{$W=512\mathrm{px}$} & \multicolumn{1}{c}{$W=1024\mathrm{px}$} & \multicolumn{1}{c}{$W=2048\mathrm{px}$}\\ \hline
Albano & $3.02 \times 10^4$ & $1.19 \times 10^5$ & $4.70 \times 10^5$ & $1.87 \times 10^6$ & $7.47 \times 10^6$ \\
Dagli & $1.46 \times 10^4$ & $5.69 \times 10^4$ & $2.25 \times 10^5$ & $8.94 \times 10^5$ & $3.56 \times 10^6$ \\
Dighe1 & $1.74 \times 10^4$ & $6.76 \times 10^4$ & $2.66 \times 10^5$ & $1.06 \times 10^6$ & $4.21 \times 10^6$ \\
Dighe2 & $1.70 \times 10^4$ & $6.71 \times 10^4$ & $2.64 \times 10^5$ & $1.05 \times 10^6$ & $4.20 \times 10^6$ \\
Fu & $1.25 \times 10^4$ & $4.94 \times 10^4$ & $1.98 \times 10^5$ & $7.88 \times 10^5$ & $3.15 \times 10^6$ \\
Jakobs1 & $4.20 \times 10^3$ & $1.64 \times 10^4$ & $6.53 \times 10^4$ & $2.59 \times 10^5$ & $1.03 \times 10^6$ \\
Jakobs2 & $4.89 \times 10^3$ & $1.89 \times 10^4$ & $7.41 \times 10^4$ & $2.92 \times 10^5$ & $1.16 \times 10^6$ \\
Mao & $1.00 \times 10^4$ & $3.94 \times 10^4$ & $1.55 \times 10^5$ & $6.12 \times 10^5$ & $2.44 \times 10^6$ \\
Marques & $1.15 \times 10^4$ & $4.49 \times 10^4$ & $1.77 \times 10^5$ & $7.02 \times 10^5$ & $2.80 \times 10^6$ \\
Shapes0 & $1.74 \times 10^4$ & $6.80 \times 10^4$ & $2.68 \times 10^5$ & $1.06 \times 10^6$ & $4.20 \times 10^6$ \\
Shapes1 & $1.74 \times 10^4$ & $6.80 \times 10^4$ & $2.68 \times 10^5$ & $1.06 \times 10^6$ & $4.20 \times 10^6$ \\
Shirts & $2.32 \times 10^4$ & $9.12 \times 10^4$ & $3.61 \times 10^5$ & $1.43 \times 10^6$ & $5.68 \times 10^6$ \\
Swim & $1.45 \times 10^4$ & $5.43 \times 10^4$ & $2.09 \times 10^5$ & $8.24 \times 10^5$ & $3.26 \times 10^6$ \\
Trousers & $4.68 \times 10^4$ & $1.83 \times 10^5$ & $7.30 \times 10^5$ & $2.90 \times 10^6$ & $1.16 \times 10^7$ \\ \hline
Profiles1 & $1.01 \times 10^4$ & $3.90 \times 10^4$ & $1.52 \times 10^5$ & $6.03 \times 10^5$ & $2.40 \times 10^6$ \\
Profiles2 & $1.71 \times 10^4$ & $6.44 \times 10^4$ & $2.54 \times 10^5$ & $1.01 \times 10^6$ & $4.00 \times 10^6$ \\
Profiles3 & $3.96 \times 10^4$ & $1.55 \times 10^5$ & $6.04 \times 10^5$ & $2.39 \times 10^6$ & $9.52 \times 10^6$ \\
Profiles4 & $7.22 \times 10^4$ & $2.84 \times 10^5$ & $1.12 \times 10^6$ & $4.46 \times 10^6$ & $1.78 \times 10^7$ \\
Profiles5 & $1.17 \times 10^4$ & $4.51 \times 10^4$ & $1.76 \times 10^5$ & $6.91 \times 10^5$ & $2.74 \times 10^6$ \\
Profiles6 & $8.75 \times 10^3$ & $3.39 \times 10^4$ & $1.29 \times 10^5$ & $5.04 \times 10^5$ & $1.99 \times 10^6$ \\
Profiles7 & $3.40 \times 10^4$ & $1.33 \times 10^5$ & $5.29 \times 10^5$ & $2.11 \times 10^6$ & $8.41 \times 10^6$ \\
Profiles8 & $1.76 \times 10^4$ & $6.79 \times 10^4$ & $2.66 \times 10^5$ & $1.06 \times 10^6$ & $4.21 \times 10^6$ \\
Profiles9 & $9.68 \times 10^3$ & $3.42 \times 10^4$ & $1.29 \times 10^5$ & $4.97 \times 10^5$ & $1.95 \times 10^6$ \\
Profiles10 & $4.42 \times 10^4$ & $1.69 \times 10^5$ & $6.59 \times 10^5$ & $2.62 \times 10^6$ & $1.04 \times 10^7$ \\ \hline
avg.(all) & $2.11 \times 10^4$ & $8.21 \times 10^4$ & $3.23 \times 10^5$ & $1.28 \times 10^6$ & $5.10 \times 10^6$ \\ \hline
\end{tabular}
\end{small}
\end{spacing}
\end{table}

\begin{table}[tb]
  \caption{The memory usage when running GCDH (in MB)\label{tab:memory}}
    \medskip
\centering
\begin{spacing}{0.8}
\begin{small}
\begin{tabular}{lrrrrr} \hline
instance & \multicolumn{1}{c}{$W=128\mathrm{px}$} & \multicolumn{1}{c}{$W=256\mathrm{px}$} & \multicolumn{1}{c}{$W=512\mathrm{px}$} & \multicolumn{1}{c}{$W=1024\mathrm{px}$} & \multicolumn{1}{c}{$W=2048\mathrm{px}$}\\ \hline
Albano & 31 & 136 & 409 & 1656 & 2216 \\
Dagli & 10 & 34 & 165 & 561 & 2061 \\
Dighe1 & 25 & 111 & 420 & 1586 & 2789 \\
Dighe2 & 28 & 99 & 326 & 1689 & 2139 \\
Fu & 30 & 77 & 267 & 848 & 2480 \\
Jakobs1 & 23 & 71 & 170 & 530 & 1624 \\
Jakobs2 & 25 & 72 & 185 & 503 & 1743 \\
Mao & 22 & 60 & 383 & 1219 & 2704 \\
Marques & 18 & 67 & 267 & 979 & 2650 \\
Shapes0 & 4 & 9 & 63 & 397 & 1170 \\
Shapes1 & 6 & 14 & 62 & 483 & 1689 \\
Shirts & 7 & 20 & 67 & 479 & 1316 \\
Swim & 6 & 33 & 222 & 700 & 1920 \\
Trousers & 16 & 65 & 316 & 1029 & 2064 \\ \hline
Profiles1 & 12 & 52 & 174 & 501 & 2231 \\
Profiles2 & 14 & 68 & 197 & 541 & 2548 \\
Profiles3 & 90 & 292 & 1122 & 2130 & 3226 \\
Profiles4 & 27 & 100 & 450 & 1725 & 2964 \\
Profiles5 & 65 & 173 & 407 & 1007 & 2642 \\
Profiles6 & 5 & 17 & 49 & 246 & 985 \\
Profiles7 & 106 & 436 & 1539 & 2554 & 4501 \\
Profiles8 & 26 & 124 & 456 & 1495 & 2776 \\
Profiles9 & 31 & 114 & 381 & 1210 & 3078 \\
Profiles10 & 19 & 36 & 205 & 805 & 2401 \\ \hline
avg.(all) & 26.9 & 95.0 & 345.9 & 1036.4 & 2329.9 \\ \hline
\end{tabular}
\end{small}
\end{spacing}  
\end{table}

\begin{figure}[tb]
  \centering
  \begin{minipage}{0.32\textwidth}
    \centering
    \includegraphics[width=0.95\textwidth]{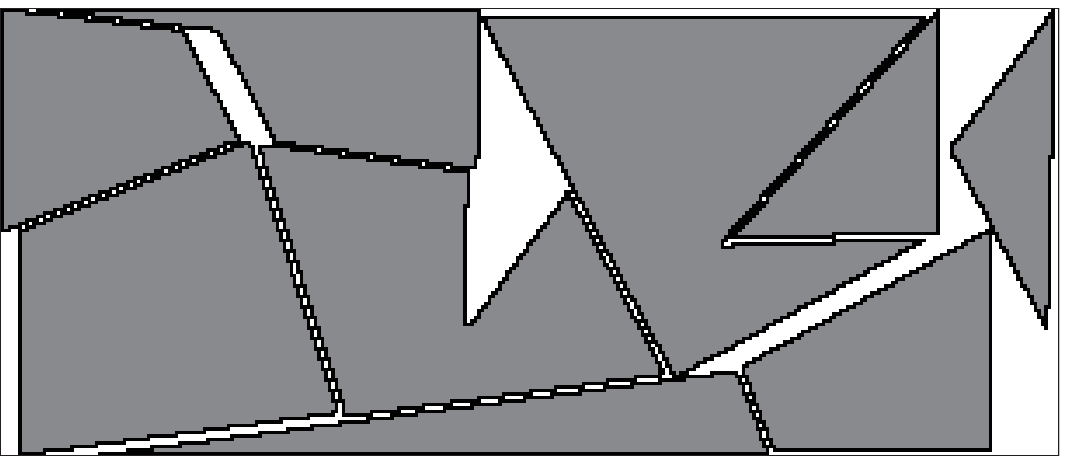}
  \end{minipage}
  \hfill
  \begin{minipage}{0.32\textwidth}
    \centering
    \includegraphics[width=0.95\textwidth]{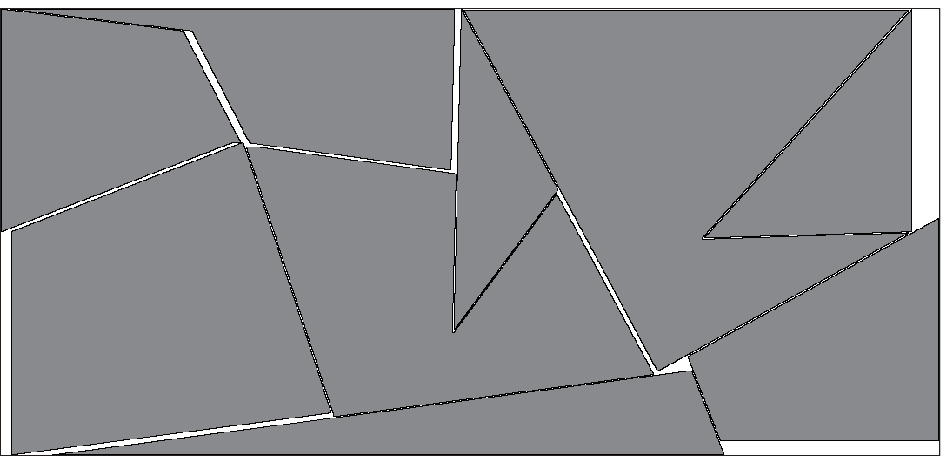}
  \end{minipage}  
  \hfill
  \begin{minipage}{0.32\textwidth}
    \centering
    \includegraphics[width=0.95\textwidth]{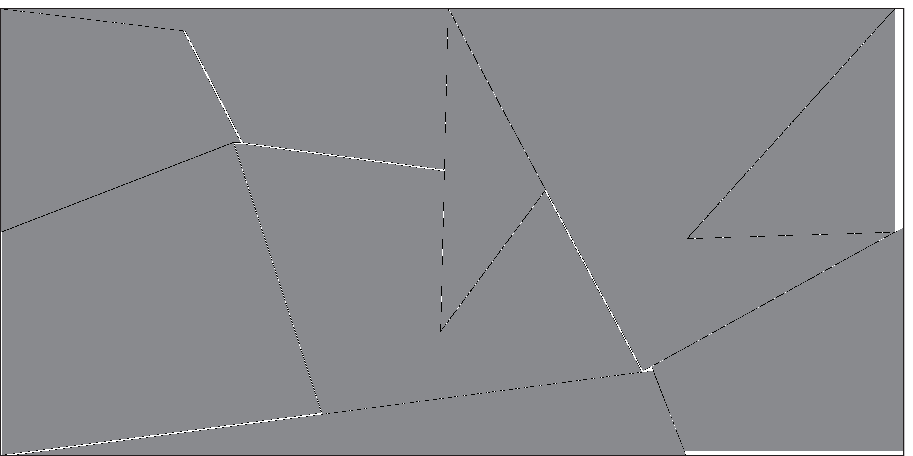}
  \end{minipage}  
  \caption{The best solutions for a test instance in different resolutions (Left: $W=128$px, Center: $W=512$px, Right: $W=2048$px) \label{fig:different_resolution}}
\end{figure}

Figure~\ref{fig:different_resolution} shows the best solutions for a test instance in different resolutions.
  We note that the rasterized shapes (especially in the low-resolution, e.g., $W=128$px) are rather different from the original shapes in the vector model, which often leads different optimal solutions.
  For example, diagonal straight lines in the vector model are often converted into jagged ones in the raster model of the low-resolution, which prevents the pieces from sticking together precisely.
  We also encounter other cases that complicated shapes in the vector model are converted into simple ones in the raster model of the low-resolution, which enables GCDH to perform efficiently and attain better solutions.
  We accordingly use the instances of the middle-resolution $W=512$px in the comparison with the other algorithms for the vector model.

Tables~\ref{tab:nfp} and \ref{tab:corner_detection} show the computation time of the preprocessing in seconds for generating NFPs for all pairs of different shapes and detecting corners for all generated NFPs.
\begin{table}[tb]
  \caption{The computation time to generate NFPs (in seconds)\label{tab:nfp}}
  \medskip
\centering
\begin{spacing}{0.8}
\begin{small}
\begin{tabular}{lrrrrr} \hline
instance & \multicolumn{1}{c}{$W=128\mathrm{px}$} & \multicolumn{1}{c}{$W=256\mathrm{px}$} & \multicolumn{1}{c}{$W=512\mathrm{px}$} & \multicolumn{1}{c}{$W=1024\mathrm{px}$} & \multicolumn{1}{c}{$W=2048\mathrm{px}$}\\ \hline
Albano & 0.03 & 0.12 & 0.48 & 1.99 & 8.15 \\
Dagli & 0.01 & 0.05 & 0.22 & 0.85 & 3.51 \\
Dighe1 & 0.05 & 0.19 & 0.77 & 3.21 & 13.16 \\
Dighe2 & 0.03 & 0.11 & 0.44 & 1.81 & 7.38 \\
Fu & 0.07 & 0.25 & 1.01 & 4.03 & 16.44 \\
Jakobs1 & 0.04 & 0.15 & 0.58 & 2.27 & 9.00 \\
Jakobs2 & 0.06 & 0.22 & 0.88 & 3.47 & 13.94 \\
Mao & 0.03 & 0.11 & 0.42 & 1.74 & 7.04 \\
Marques & 0.02 & 0.08 & 0.32 & 1.32 & 5.33 \\
Shapes0 & $<$0.01 & 0.01 & 0.03 & 0.12 & 0.48 \\
Shapes1 & $<$0.01 & 0.02 & 0.06 & 0.24 & 0.98 \\
Shirts & 0.01 & 0.02 & 0.09 & 0.37 & 1.49 \\
Swim & 0.02 & 0.06 & 0.25 & 1.01 & 4.10 \\
Trousers & 0.06 & 0.22 & 0.90 & 3.68 & 14.81 \\ \hline
Profiles1 & 0.01 & 0.06 & 0.22 & 0.85 & 3.44 \\
Profiles2 & 0.03 & 0.10 & 0.43 & 1.73 & 7.09 \\
Profiles3 & 0.19 & 0.78 & 3.19 & 13.06 & 52.76 \\
Profiles4 & 0.04 & 0.17 & 0.67 & 2.75 & 11.13 \\
Profiles5 & 0.16 & 0.62 & 2.47 & 9.95 & 39.99 \\
Profiles6 & 0.01 & 0.06 & 0.23 & 0.93 & 3.92 \\
Profiles7 & 0.29 & 1.16 & 4.75 & 19.38 & 77.34 \\
Profiles8 & 0.07 & 0.28 & 1.15 & 4.77 & 19.22 \\
Profiles9 & 0.10 & 0.42 & 1.77 & 7.32 & 30.02 \\
Profiles10 & 0.02 & 0.07 & 0.29 & 1.19 & 4.78 \\ \hline
avg.(all) & 0.06 & 0.22 & 0.90 & 3.67 & 14.81 \\ \hline
\end{tabular}
\end{small}
\end{spacing}
\end{table}

\begin{table}[tb]
  \caption{The computation time for the corner detection (in seconds)\label{tab:corner_detection}}
  \medskip
  \centering
  \begin{spacing}{0.8}
\begin{small}
\begin{tabular}{lrrrrr} \hline
  instance & \multicolumn{1}{c}{$W=128\mathrm{px}$} & \multicolumn{1}{c}{$W=256\mathrm{px}$} & \multicolumn{1}{c}{$W=512\mathrm{px}$} & \multicolumn{1}{c}{$W=1024\mathrm{px}$} & \multicolumn{1}{c}{$W=2048\mathrm{px}$} \\ \hline
Albano & 0.06 & 0.20 & 0.98 & 4.12 & 10.81 \\
Dagli & 0.03 & 0.08 & 0.29 & 1.65 & 5.83 \\
Dighe1 & 0.07 & 0.26 & 1.48 & 6.81 & 17.31 \\
Dighe2 & 0.04 & 0.14 & 0.79 & 3.71 & 9.98 \\
Fu & 0.14 & 0.43 & 1.82 & 8.98 & 29.19 \\
Jakobs1 & 0.19 & 0.44 & 1.31 & 5.27 & 28.02 \\
Jakobs2 & 0.21 & 0.53 & 1.56 & 6.67 & 35.49 \\
Mao & 0.09 & 0.26 & 1.20 & 6.05 & 16.85 \\
Marques & 0.06 & 0.17 & 0.72 & 3.73 & 12.32 \\
Shapes0 & $<$0.01 & 0.02 & 0.07 & 0.39 & 1.21 \\
Shapes1 & 0.01 & 0.03 & 0.12 & 0.73 & 2.20 \\
Shirts & 0.02 & 0.04 & 0.15 & 0.82 & 3.27 \\
Swim & 0.03 & 0.12 & 0.57 & 2.77 & 9.79 \\
Trousers & 0.08 & 0.24 & 1.01 & 4.95 & 14.13 \\ \hline
Profiles1 & 0.05 & 0.13 & 0.43 & 1.77 & 7.34 \\
Profiles2 & 0.05 & 0.14 & 0.49 & 2.30 & 9.03 \\
Profiles3 & 0.38 & 1.60 & 8.95 & 30.99 & 77.15 \\
Profiles4 & 0.08 & 0.32 & 1.59 & 5.72 & 16.52 \\
Profiles5 & 0.52 & 1.42 & 4.59 & 18.37 & 73.86 \\
Profiles6 & 0.03 & 0.08 & 0.24 & 1.03 & 5.59 \\
Profiles7 & 0.45 & 2.16 & 10.60 & 27.01 & 118.71 \\
Profiles8 & 0.14 & 0.43 & 2.15 & 10.06 & 26.75 \\
Profiles9 & 0.18 & 0.55 & 2.25 & 12.22 & 43.48 \\
Profiles10 & 0.04 & 0.12 & 0.51 & 2.69 & 8.59 \\ \hline
avg.(all) & 0.12 & 0.41 & 1.83 & 7.03 & 24.31 \\ \hline
\end{tabular}
\end{small}
\end{spacing}
\end{table}

\begin{figure}[tb]
  \centering
  \includegraphics[width=0.8\textwidth]{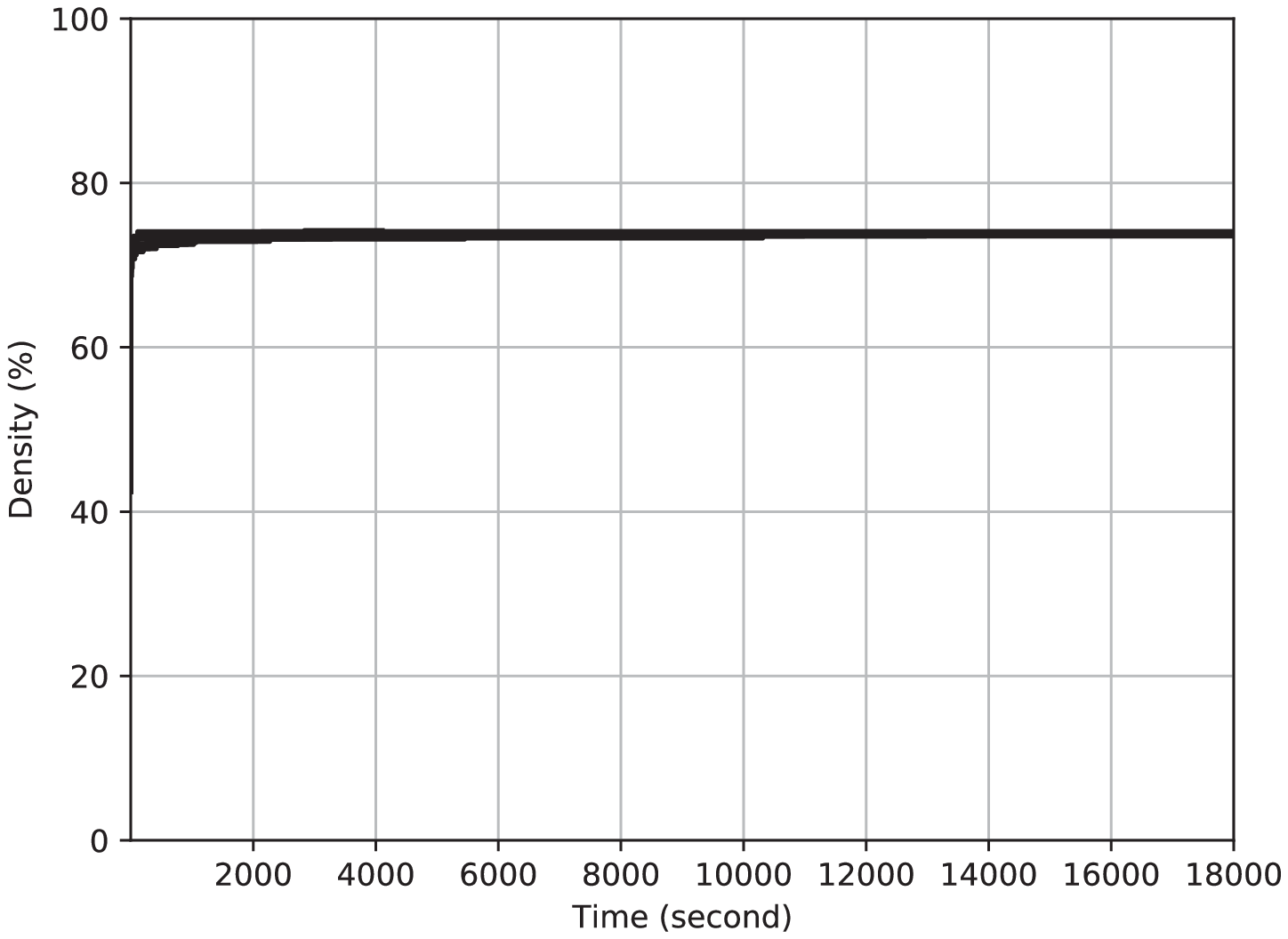}
  \caption{The evolutions of density when GCDH running 10 times on Swim instance (normal plot) \label{fig:density_normal_plot}}
\end{figure}

\begin{figure}[tb]
  \centering
  \includegraphics[width=0.8\textwidth]{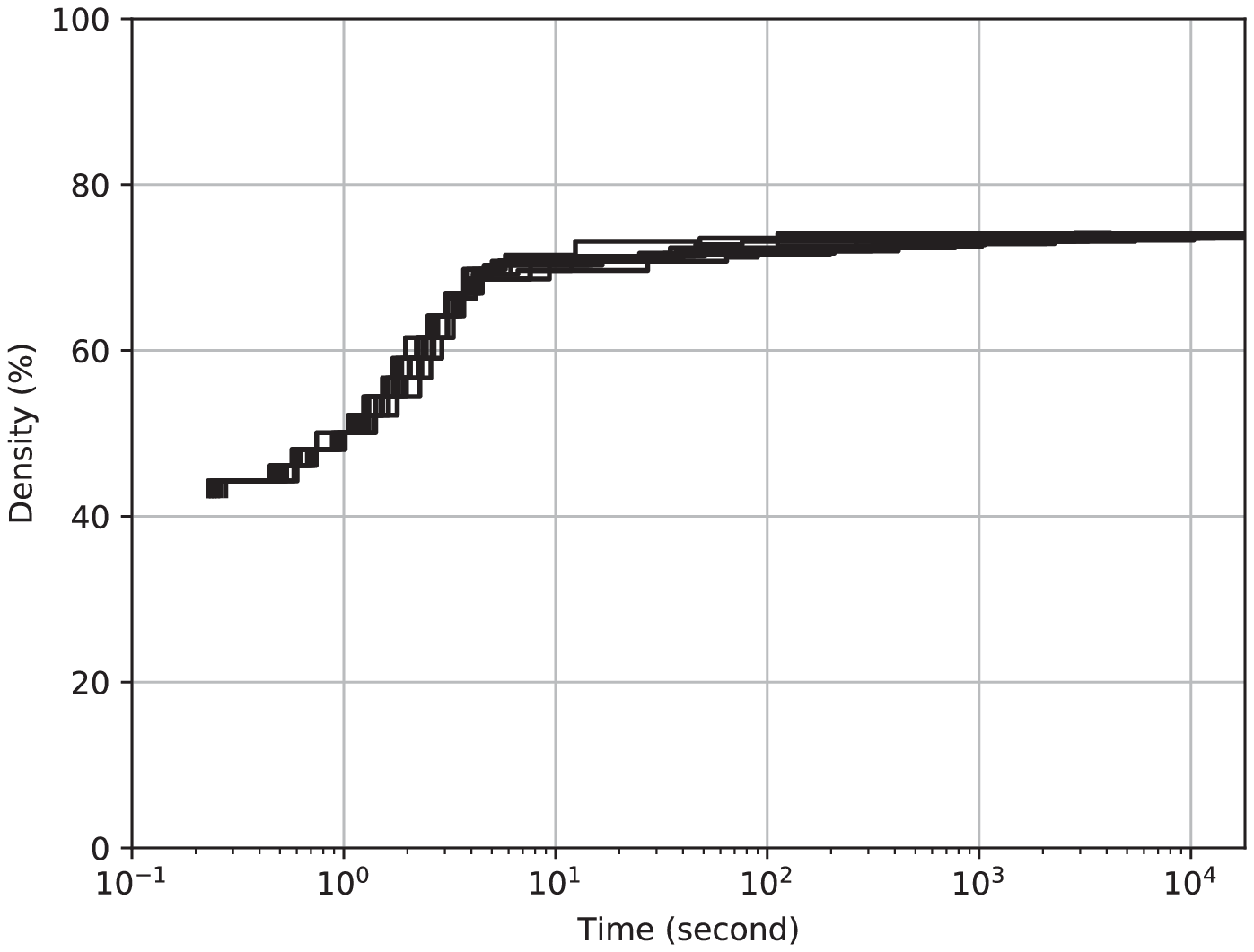}
  \caption{The evolutions of density when GCDH running 10 times on Swim instance (linear-log plot) \label{fig:density_linlog_plot}}
\end{figure}

Figures~\ref{fig:density_normal_plot} and \ref{fig:density_linlog_plot} show the evolution of density (\%) when GCDH running 10 times on Swim instance with the time limit of 18000 seconds.
The GCDH starts from an initial solution of the density 42.47\% obtained by the construction algorithm (Algorithm~\ref{alg:construct}), and then rapidly improves it within several seconds to attain the density over 70\%.
We observed that GCDH performs similarly for other instances.

\begin{figure}[tb]
  \centering
  \includegraphics[width=0.8\textwidth]{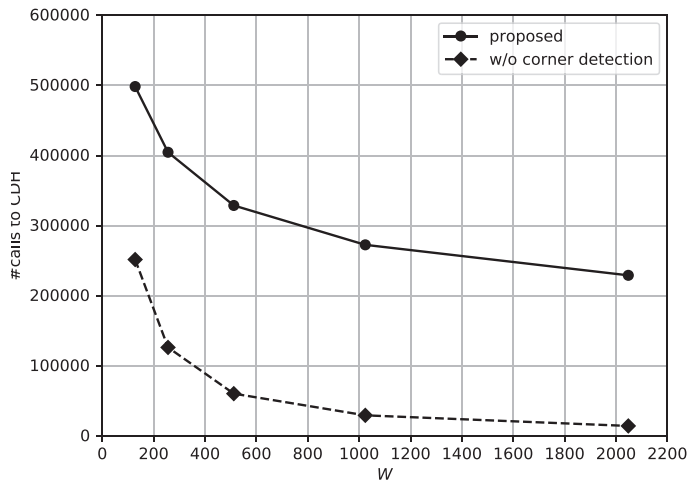}
  \caption{The number of calls to CDH when running GCDH (normal plot) \label{fig:CDH_normal_plot}}
\end{figure}

\begin{figure}[tb]
  \centering
  \includegraphics[width=0.8\textwidth]{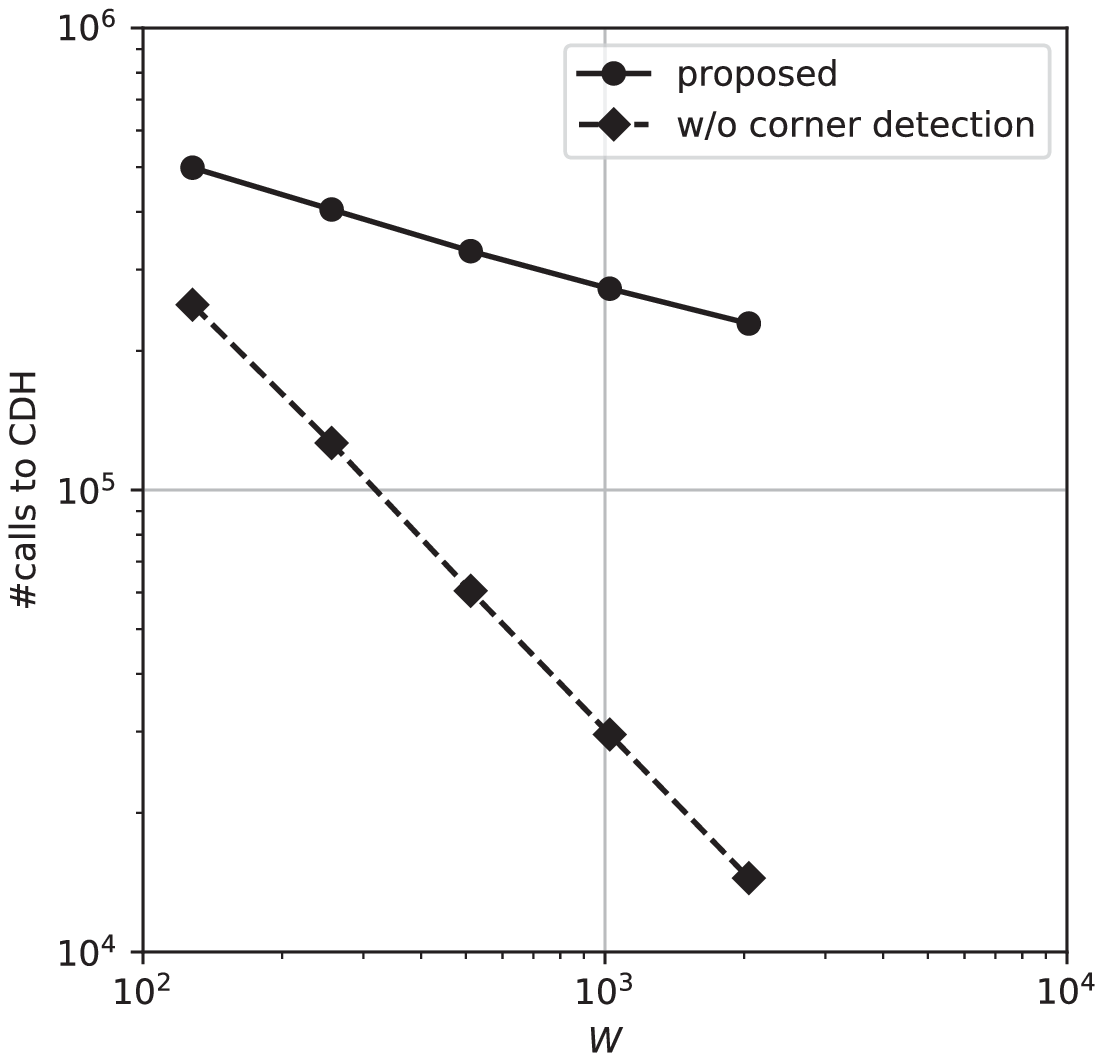}
  \caption{The number of calls to CDH when running GCDH (log-log plot) \label{fig:CDH_loglog_plot}}
\end{figure}

We first evaluated the effect of the corner detection technique on GCDH.
We tested GCDH 10 times for each instance with the time limit of 1200 seconds for each run, which is the same for all resolutions and does not include the computation time of the preprocessing for generating NFPs and detecting their corners (Tables~\ref{tab:nfp} and \ref{tab:corner_detection}).
Tables~\ref{tab:CDH} and \ref{tab:CDH_wo_corner} show the number of calls to CDH (Algorithm~\ref{alg:coordinate_descent}) with and without the corner detection, respectively.
Table~\ref{tab:speedup} shows the improvement in the computational efficiency of GCDH via the corner detection with respect to the number of calls to CDH, where that of GCDH without it is set to one.
The GCDH with the corner detection was 14.08 times faster on average than GCDH without it for the instances of high-resolution ($W=2048$px).
Figures~\ref{fig:CDH_normal_plot} and \ref{fig:CDH_loglog_plot} also show the average number of calls to CDH for all instances.
The computational efficiency has been much improved by the corner detection technique; i.e., the numbers of calls to CDH decrease in proportion to the cube root of the value of $W$ roughly, while those decrease in proportion of that of $W$ without the corner detection.

\begin{table}[tb]
  \caption{The number of calls to CDH when running GCDH with the corner detection \label{tab:CDH}}
  \medskip
\centering
\begin{spacing}{0.8}
\begin{small}
\begin{tabular}{lrrrrr} \hline
  instance & \multicolumn{1}{c}{$W=128\mathrm{px}$} & \multicolumn{1}{c}{$W=256\mathrm{px}$} & \multicolumn{1}{c}{$W=512\mathrm{px}$} & \multicolumn{1}{c}{$W=1024\mathrm{px}$} & \multicolumn{1}{c}{$W=2048\mathrm{px}$}\\ \hline
Albano & $6.95 \times 10^5$ & $5.33 \times 10^5$ & $4.55 \times 10^5$ & $4.00 \times 10^5$ & $3.65 \times 10^5$ \\
Dagli & $4.55 \times 10^5$ & $3.44 \times 10^5$ & $2.75 \times 10^5$ & $2.45 \times 10^5$ & $2.33 \times 10^5$ \\
Dighe1 & $1.17 \times 10^6$ & $8.07 \times 10^5$ & $6.11 \times 10^5$ & $4.88 \times 10^5$ & $3.93 \times 10^5$ \\
Dighe2 & $2.92 \times 10^6$ & $2.56 \times 10^6$ & $2.05 \times 10^6$ & $1.55 \times 10^6$ & $1.24 \times 10^6$ \\
Fu & $1.40 \times 10^6$ & $1.24 \times 10^6$ & $1.05 \times 10^6$ & $9.42 \times 10^5$ & $7.89 \times 10^5$ \\
Jakobs1 & $3.25 \times 10^5$ & $2.72 \times 10^5$ & $2.23 \times 10^5$ & $2.08 \times 10^5$ & $1.62 \times 10^5$ \\
Jakobs2 & $2.72 \times 10^5$ & $2.07 \times 10^5$ & $1.80 \times 10^5$ & $1.50 \times 10^5$ & $1.17 \times 10^5$ \\
Mao & $4.97 \times 10^5$ & $4.53 \times 10^5$ & $3.58 \times 10^5$ & $2.87 \times 10^5$ & $2.54 \times 10^5$ \\
Marques & $4.13 \times 10^5$ & $3.78 \times 10^5$ & $2.97 \times 10^5$ & $2.39 \times 10^5$ & $2.16 \times 10^5$ \\
Shapes0 & $4.40 \times 10^5$ & $3.44 \times 10^5$ & $3.13 \times 10^5$ & $2.98 \times 10^5$ & $2.85 \times 10^5$ \\
Shapes1 & $1.95 \times 10^5$ & $1.60 \times 10^5$ & $1.32 \times 10^5$ & $1.25 \times 10^5$ & $1.07 \times 10^5$ \\
Shirts & $1.11 \times 10^5$ & $7.98 \times 10^4$ & $6.12 \times 10^4$ & $5.08 \times 10^4$ & $4.40 \times 10^4$ \\
Swim & $1.98 \times 10^5$ & $1.22 \times 10^5$ & $8.72 \times 10^4$ & $7.02 \times 10^4$ & $5.80 \times 10^4$ \\
Trousers & $1.98 \times 10^5$ & $1.51 \times 10^5$ & $1.38 \times 10^5$ & $1.14 \times 10^5$ & $9.36 \times 10^4$ \\ \hline
Profiles1 & $2.44 \times 10^5$ & $1.77 \times 10^5$ & $1.54 \times 10^5$ & $1.43 \times 10^5$ & $1.17 \times 10^5$ \\
Profiles2 & $1.06 \times 10^5$ & $7.49 \times 10^4$ & $6.67 \times 10^4$ & $5.82 \times 10^4$ & $4.78 \times 10^4$ \\
Profiles3 & $2.87 \times 10^4$ & $2.07 \times 10^4$ & $1.71 \times 10^4$ & $1.53 \times 10^4$ & $1.18 \times 10^4$ \\
Profiles4 & $5.97 \times 10^4$ & $4.67 \times 10^4$ & $4.14 \times 10^4$ & $3.47 \times 10^4$ & $3.46 \times 10^4$ \\
Profiles5 & $2.36 \times 10^4$ & $1.40 \times 10^4$ & $9.82 \times 10^3$ & $8.61 \times 10^3$ & $6.66 \times 10^3$ \\
Profiles6 & $8.64 \times 10^4$ & $6.20 \times 10^4$ & $4.21 \times 10^4$ & $3.32 \times 10^4$ & $2.53 \times 10^4$ \\
Profiles7 & $1.50 \times 10^6$ & $1.19 \times 10^6$ & $9.60 \times 10^5$ & $7.74 \times 10^5$ & $6.66 \times 10^5$ \\
Profiles8 & $4.44 \times 10^5$ & $3.60 \times 10^5$ & $2.82 \times 10^5$ & $2.43 \times 10^5$ & $1.90 \times 10^5$ \\
Profiles9 & $1.02 \times 10^5$ & $6.60 \times 10^4$ & $4.59 \times 10^4$ & $3.31 \times 10^4$ & $2.78 \times 10^4$ \\
Profiles10 & $6.74 \times 10^4$ & $5.01 \times 10^4$ & $3.99 \times 10^4$ & $3.25 \times 10^4$ & $2.52 \times 10^4$ \\ \hline
avg.(all) & $4.98 \times 10^5$ & $4.05 \times 10^5$ & $3.29 \times 10^5$ & $2.73 \times 10^5$ & $2.29 \times 10^5$ \\ \hline
\end{tabular}
\end{small}
\end{spacing}
\end{table}

\begin{table}[tb]
  \caption{The number of calls to CDH when running GCDH without the corner detection \label{tab:CDH_wo_corner}}
  \medskip
\centering
\begin{spacing}{0.8}
\begin{small}
\begin{tabular}{lrrrrr} \hline
instance & \multicolumn{1}{c}{$W=128\mathrm{px}$} & \multicolumn{1}{c}{$W=256\mathrm{px}$} & \multicolumn{1}{c}{$W=512\mathrm{px}$} & \multicolumn{1}{c}{$W=1024\mathrm{px}$} & \multicolumn{1}{c}{$W=2048\mathrm{px}$}\\ \hline
Albano & $2.84 \times 10^5$ & $1.38 \times 10^5$ & $6.90 \times 10^4$ & $3.41 \times 10^4$ & $1.64 \times 10^4$ \\
Dagli & $2.61 \times 10^5$ & $1.26 \times 10^5$ & $6.07 \times 10^4$ & $3.02 \times 10^4$ & $1.45 \times 10^4$ \\
Dighe1 & $5.84 \times 10^5$ & $2.71 \times 10^5$ & $1.28 \times 10^5$ & $6.58 \times 10^4$ & $3.30 \times 10^4$ \\
Dighe2 & $1.30 \times 10^6$ & $6.92 \times 10^5$ & $3.35 \times 10^5$ & $1.53 \times 10^5$ & $7.15 \times 10^4$ \\
Fu & $6.15 \times 10^6$ & $3.17 \times 10^5$ & $1.49 \times 10^5$ & $7.49 \times 10^4$ & $3.66 \times 10^4$ \\
Jakobs1 & $2.37 \times 10^5$ & $1.22 \times 10^5$ & $6.20 \times 10^4$ & $3.27 \times 10^4$ & $1.51 \times 10^4$ \\
Jakobs2 & $2.20 \times 10^5$ & $9.68 \times 10^4$ & $4.89 \times 10^4$ & $2.59 \times 10^4$ & $1.30 \times 10^4$ \\
Mao & $3.73 \times 10^5$ & $2.09 \times 10^5$ & $9.88 \times 10^4$ & $4.45 \times 10^4$ & $2.28 \times 10^4$ \\
Marques & $2.56 \times 10^5$ & $1.44 \times 10^5$ & $6.90 \times 10^4$ & $3.37 \times 10^4$ & $1.64 \times 10^4$ \\
Shapes0 & $2.85 \times 10^5$ & $1.39 \times 10^5$ & $6.87 \times 10^4$ & $3.72 \times 10^4$ & $2.01 \times 10^4$ \\
Shapes1 & $1.36 \times 10^5$ & $7.16 \times 10^4$ & $3.27 \times 10^4$ & $1.80 \times 10^4$ & $8.76 \times 10^3$ \\
Shirts & $7.91 \times 10^4$ & $3.70 \times 10^4$ & $1.68 \times 10^4$ & $7.89 \times 10^3$ & $3.78 \times 10^3$ \\
Swim & $1.55 \times 10^5$ & $7.06 \times 10^4$ & $3.08 \times 10^4$ & $1.44 \times 10^4$ & $7.12 \times 10^3$ \\
Trousers & $8.84 \times 10^4$ & $3.92 \times 10^4$ & $1.90 \times 10^4$ & $9.26 \times 10^3$ & $4.54 \times 10^3$ \\ \hline
Profiles1 & $1.78 \times 10^5$ & $9.05 \times 10^4$ & $4.34 \times 10^4$ & $2.15 \times 10^4$ & $1.08 \times 10^4$ \\
Profiles2 & $7.80 \times 10^4$ & $3.75 \times 10^4$ & $1.85 \times 10^4$ & $9.08 \times 10^3$ & $4.65 \times 10^3$ \\
Profiles3 & $1.80 \times 10^4$ & $8.39 \times 10^3$ & $4.18 \times 10^3$ & $2.02 \times 10^3$ & $1.11 \times 10^3$ \\
Profiles4 & $2.01 \times 10^4$ & $1.01 \times 10^4$ & $4.95 \times 10^3$ & $2.47 \times 10^3$ & $1.32 \times 10^3$ \\
Profiles5 & $1.94 \times 10^4$ & $8.24 \times 10^3$ & $3.67 \times 10^3$ & $1.69 \times 10^3$ & $8.09 \times 10^2$ \\
Profiles6 & $6.99 \times 10^4$ & $3.49 \times 10^4$ & $1.68 \times 10^4$ & $8.29 \times 10^3$ & $3.90 \times 10^3$ \\
Profiles7 & $4.35 \times 10^7$ & $2.05 \times 10^5$ & $9.31 \times 10^4$ & $4.55 \times 10^4$ & $2.24 \times 10^4$ \\
Profiles8 & $2.27 \times 10^5$ & $1.09 \times 10^5$ & $5.29 \times 10^4$ & $2.44 \times 10^4$ & $1.21 \times 10^4$ \\
Profiles9 & $8.66 \times 10^4$ & $4.01 \times 10^4$ & $1.87 \times 10^4$ & $9.02 \times 10^3$ & $4.54 \times 10^3$ \\
Profiles10 & $3.73 \times 10^4$ & $1.65 \times 10^4$ & $7.59 \times 10^3$ & $3.75 \times 10^3$ & $1.79 \times 10^3$ \\ \hline
avg.(all) & $2.52 \times 10^5$ & $1.26 \times 10^5$ & $6.05 \times 10^4$ & $2.96 \times 10^4$ & $1.45 \times 10^4$ \\ \hline
\end{tabular}
\end{small}
\end{spacing}
\end{table}

\begin{table}[tb]
\caption{The improvement in the computational efficiency of GCDH via the corner detection\label{tab:speedup}}
\medskip
\centering
\begin{spacing}{0.8}
\begin{small}
\begin{tabular}{lrrrrr} \hline
instance & \multicolumn{1}{c}{$W=128\mathrm{px}$} & \multicolumn{1}{c}{$W=256\mathrm{px}$} & \multicolumn{1}{c}{$W=512\mathrm{px}$} & \multicolumn{1}{c}{$W=1024\mathrm{px}$} & \multicolumn{1}{c}{$W=2048\mathrm{px}$} \\ \hline
Albano & 2.45 & 3.85 & 6.60 & 11.74 & 21.72 \\
Dagli & 1.74 & 2.72 & 4.53 & 8.12 & 16.07 \\
Dighe1 & 2.01 & 2.98 & 4.76 & 7.42 & 11.89 \\
Dighe2 & 2.25 & 3.70 & 6.13 & 10.11 & 17.38 \\
Fu & 2.27 & 3.90 & 7.08 & 12.58 & 21.52 \\
Jakobs1 & 1.37 & 2.24 & 3.59 & 6.38 & 10.77 \\
Jakobs2 & 1.24 & 2.14 & 3.68 & 5.80 & 9.01 \\
Mao & 1.33 & 2.17 & 3.62 & 6.44 & 11.16 \\
Marques & 1.61 & 2.63 & 4.30 & 7.09 & 13.20 \\
Shapes0 & 1.55 & 2.47 & 4.56 & 8.02 & 14.23 \\
Shapes1 & 1.44 & 2.23 & 4.05 & 6.94 & 12.22 \\
Shirts & 1.40 & 2.16 & 3.63 & 6.44 & 11.64 \\
Swim & 1.28 & 1.73 & 2.83 & 4.87 & 8.14 \\
Trousers & 2.24 & 3.86 & 7.23 & 12.34 & 20.62 \\ \hline
Profiles1 & 1.37 & 1.96 & 3.54 & 6.66 & 10.83 \\
Profiles2 & 1.36 & 2.00 & 3.61 & 6.40 & 10.26 \\
Profiles3 & 1.59 & 2.46 & 4.09 & 7.58 & 10.61 \\
Profiles4 & 2.97 & 4.62 & 8.35 & 14.06 & 26.16 \\
Profiles5 & 1.21 & 1.70 & 2.68 & 5.09 & 8.23 \\
Profiles6 & 1.24 & 1.78 & 2.50 & 4.00 & 6.47 \\
Profiles7 & 3.46 & 5.80 & 10.31 & 17.01 & 29.71 \\
Profiles8 & 1.95 & 3.30 & 5.33 & 9.96 & 15.74 \\
Profiles9 & 1.18 & 1.64 & 2.45 & 3.67 & 6.12 \\
Profiles10 & 1.81 & 3.03 & 5.26 & 8.68 & 14.12 \\ \hline
avg.(all) & 1.76 & 2.79 & 4.78 & 8.22 & 14.08 \\ \hline
\end{tabular}
\end{small}
\end{spacing}
\end{table}

Tables~\ref{tab:result} and \ref{tab:result_wo_corner} show the computational results obtained by GCDH with and without the corner detection respectively, which are evaluated in the best and average density (\%).
  Table~\ref{tab:improvement} shows the improvement in the average density (\%) of GCDH via the corner detection.
The improvement in the computational efficiency also brought that in the quality of the obtained solutions; e.g., GCDH with the corner detection attained 2.02 point better than GCDH without it in the average density (\%) for the instances of high-resolution ($W=2048$px).
\begin{table}[t]
\caption{The density (\%) of solutions obtained by GCDH with the corner detection\label{tab:result}}
\medskip
\centering
\begin{spacing}{0.8}
\begin{small}
\begin{tabular}{lrrcrrcrrcrrcrr} \hline
  instance & \multicolumn{2}{c}{$W=128\mathrm{px}$} & & \multicolumn{2}{c}{$W=256\mathrm{px}$} & & \multicolumn{2}{c}{$W=512\mathrm{px}$} & & \multicolumn{2}{c}{$W=1024\mathrm{px}$} & & \multicolumn{2}{c}{$W=2048\mathrm{px}$}\\ \cline{2-3} \cline{5-6} \cline{8-9} \cline{11-12} \cline{14-15}
  & \multicolumn{1}{c}{best} & \multicolumn{1}{c}{avg.} & & \multicolumn{1}{c}{best} & \multicolumn{1}{c}{avg.} & & \multicolumn{1}{c}{best} & \multicolumn{1}{c}{avg.} & & \multicolumn{1}{c}{best} & \multicolumn{1}{c}{avg.} & & \multicolumn{1}{c}{best} & \multicolumn{1}{c}{avg.} \\ \hline
Albano & 89.01 & 88.74 & {} & 88.20 & 87.75 & {} & 88.20 & 87.47 & {} & 87.68 & 87.09 & {} & 87.08 & 86.80 \\
Dagli & 87.96 & 87.23 & {} & 86.84 & 86.44 & {} & 86.32 & 85.70 & {} & 86.36 & 85.71 & {} & 85.92 & 85.19 \\
Dighe1 & 92.60 & 91.07 & {} & 96.07 & 95.82 & {} & 97.60 & 97.55 & {} & 98.90 & 98.80 & {} & 99.33 & 99.27 \\
Dighe2 & 93.01 & 93.01 & {} & 95.01 & 94.77 & {} & 97.28 & 97.17 & {} & 98.79 & 98.74 & {} & 99.28 & 99.20 \\
Fu & 92.31 & 91.11 & {} & 90.99 & 90.31 & {} & 90.75 & 89.62 & {} & 91.90 & 90.16 & {} & 89.87 & 89.02 \\
Jakobs1 & 88.58 & 84.94 & {} & 84.32 & 84.32 & {} & 88.01 & 86.54 & {} & 88.48 & 86.71 & {} & 87.98 & 86.68 \\
Jakobs2 & 81.33 & 81.33 & {} & 81.91 & 78.77 & {} & 79.51 & 78.86 & {} & 79.47 & 78.44 & {} & 80.66 & 78.68 \\
Mao & 87.17 & 86.80 & {} & 85.12 & 84.84 & {} & 84.84 & 83.62 & {} & 84.03 & 83.45 & {} & 84.10 & 83.28 \\
Marques & 91.39 & 91.39 & {} & 89.92 & 89.74 & {} & 90.43 & 89.33 & {} & 89.36 & 88.73 & {} & 88.75 & 88.37 \\
Shapes0 & 68.49 & 67.51 & {} & 66.74 & 66.31 & {} & 66.13 & 65.33 & {} & 65.14 & 64.56 & {} & 64.81 & 64.31 \\
Shapes1 & 75.30 & 74.24 & {} & 73.38 & 72.62 & {} & 72.35 & 71.86 & {} & 71.88 & 71.23 & {} & 71.46 & 71.09 \\
Shirts & 86.72 & 86.39 & {} & 85.82 & 85.06 & {} & 84.33 & 84.14 & {} & 84.43 & 83.68 & {} & 84.24 & 83.50 \\
Swim & 78.87 & 77.27 & {} & 74.74 & 73.99 & {} & 74.08 & 72.91 & {} & 73.60 & 72.52 & {} & 72.99 & 72.21 \\
Trousers & 89.19 & 88.43 & {} & 88.41 & 87.40 & {} & 88.29 & 87.09 & {} & 88.02 & 86.76 & {} & 88.01 & 86.99 \\ \hline
Profiles1 & 84.75 & 84.65 & {} & 86.03 & 84.79 & {} & 86.93 & 85.08 & {} & 87.07 & 84.17 & {} & 85.73 & 84.09 \\
Profiles2 & 79.43 & 79.15 & {} & 77.66 & 76.97 & {} & 77.28 & 76.36 & {} & 76.39 & 76.02 & {} & 76.57 & 75.53 \\
Profiles3 & 76.32 & 75.10 & {} & 75.52 & 73.38 & {} & 73.87 & 72.91 & {} & 73.51 & 72.37 & {} & 72.67 & 72.02 \\
Profiles4 & 86.04 & 85.21 & {} & 85.92 & 85.19 & {} & 85.63 & 84.71 & {} & 85.31 & 84.43 & {} & 84.64 & 84.09 \\
Profiles5 & 83.88 & 83.18 & {} & 82.15 & 81.73 & {} & 82.47 & 81.42 & {} & 82.10 & 80.92 & {} & 82.44 & 81.45 \\
Profiles6 & 81.40 & 81.03 & {} & 81.14 & 80.45 & {} & 79.03 & 78.71 & {} & 77.50 & 76.85 & {} & 77.94 & 76.92 \\
Profiles7 & 87.25 & 87.25 & {} & 92.60 & 92.40 & {} & 95.80 & 95.65 & {} & 98.27 & 98.18 & {} & 98.92 & 98.83 \\
Profiles8 & 87.34 & 85.29 & {} & 86.40 & 84.51 & {} & 87.61 & 85.08 & {} & 86.65 & 85.36 & {} & 88.50 & 86.07 \\
Profiles9 & 66.95 & 66.07 & {} & 61.08 & 60.18 & {} & 57.86 & 57.39 & {} & 56.64 & 55.65 & {} & 55.92 & 54.91 \\
Profiles10 & 70.90 & 70.38 & {} & 68.69 & 68.43 & {} & 68.20 & 67.47 & {} & 67.91 & 66.98 & {} & 67.23 & 66.60 \\ \hline
avg.(1st) & 85.85 & 84.96 & {} & 84.81 & 84.15 & {} & 84.87 & 84.08 & {} & 84.86 & 84.06 & {} & 84.60 & 83.90 \\
avg.(2nd) & 80.42 & 79.73 & {} & 79.74 & 78.84 & {} & 79.38 & 78.42 & {} & 79.22 & 78.09 & {} & 79.06 & 78.05 \\
avg.(all) & 83.59 & 82.78 & {} & 82.70 & 81.94 & {} & 82.58 & 81.72 & {} & 82.51 & 81.57 & {} & 82.29 & 81.46 \\ \hline
\end{tabular}
\end{small}
\end{spacing}
\end{table}

\begin{table}[t]
\caption{The density (\%) of solutions obtained by GCDH without the corner detection\label{tab:result_wo_corner}}
\medskip
\centering
\begin{spacing}{0.8}
\begin{small}
\begin{tabular}{lrrcrrcrrcrrcrr} \hline
  instance & \multicolumn{2}{c}{$W=128\mathrm{px}$} & & \multicolumn{2}{c}{$W=256\mathrm{px}$} & & \multicolumn{2}{c}{$W=512\mathrm{px}$} & & \multicolumn{2}{c}{$W=1024\mathrm{px}$} & & \multicolumn{2}{c}{$W=2048\mathrm{px}$}\\ \cline{2-3} \cline{5-6} \cline{8-9} \cline{11-12} \cline{14-15}
  & \multicolumn{1}{c}{best} & \multicolumn{1}{c}{avg.} & & \multicolumn{1}{c}{best} & \multicolumn{1}{c}{avg.} & & \multicolumn{1}{c}{best} & \multicolumn{1}{c}{avg.} & & \multicolumn{1}{c}{best} & \multicolumn{1}{c}{avg.} & & \multicolumn{1}{c}{best} & \multicolumn{1}{c}{avg.} \\ \hline
Albano & 89.01 & 88.68 & {} & 88.54 & 87.80 & {} & 87.61 & 86.79 & {} & 87.89 & 86.38 & {} & 86.54 & 85.71 \\
Dagli & 87.96 & 87.09 & {} & 86.84 & 86.07 & {} & 86.66 & 85.34 & {} & 84.76 & 84.18 & {} & 86.22 & 84.33 \\
Dighe1 & 91.98 & 87.77 & {} & 96.07 & 93.60 & {} & 97.42 & 94.05 & {} & 92.03 & 88.65 & {} & 95.95 & 86.35 \\
Dighe2 & 93.01 & 93.01 & {} & 95.01 & 94.84 & {} & 97.28 & 97.16 & {} & 98.79 & 98.74 & {} & 99.28 & 99.21 \\
Fu & 90.60 & 90.27 & {} & 90.14 & 89.56 & {} & 89.28 & 88.44 & {} & 91.25 & 88.67 & {} & 89.14 & 87.42 \\
Jakobs1 & 84.03 & 84.03 & {} & 89.00 & 85.26 & {} & 86.82 & 84.42 & {} & 85.20 & 83.36 & {} & 85.15 & 82.94 \\
Jakobs2 & 81.33 & 80.67 & {} & 78.42 & 78.42 & {} & 79.94 & 78.65 & {} & 78.59 & 77.38 & {} & 77.90 & 76.47 \\
Mao & 87.17 & 86.06 & {} & 85.59 & 84.75 & {} & 85.08 & 83.42 & {} & 83.91 & 82.98 & {} & 83.10 & 82.21 \\
Marques & 91.39 & 91.30 & {} & 91.33 & 89.70 & {} & 89.49 & 88.85 & {} & 88.55 & 88.05 & {} & 88.40 & 87.55 \\
Shapes0 & 68.15 & 67.55 & {} & 66.91 & 66.03 & {} & 66.30 & 64.88 & {} & 64.90 & 63.85 & {} & 64.06 & 63.18 \\
Shapes1 & 74.89 & 73.88 & {} & 72.78 & 72.03 & {} & 72.05 & 71.03 & {} & 71.33 & 70.11 & {} & 70.65 & 69.90 \\
Shirts & 87.55 & 86.67 & {} & 85.61 & 85.04 & {} & 84.53 & 83.84 & {} & 83.90 & 83.19 & {} & 83.96 & 83.00 \\
Swim & 77.26 & 76.79 & {} & 74.22 & 73.81 & {} & 73.28 & 72.38 & {} & 72.21 & 71.49 & {} & 71.74 & 70.88 \\
Trousers & 89.19 & 88.16 & {} & 88.30 & 87.00 & {} & 87.32 & 86.44 & {} & 86.54 & 85.57 & {} & 87.15 & 85.66 \\ \hline
Profiles1 & 84.75 & 84.75 & {} & 84.59 & 83.90 & {} & 84.70 & 83.73 & {} & 85.18 & 82.63 & {} & 84.32 & 82.55 \\
Profiles2 & 79.91 & 79.20 & {} & 77.18 & 76.55 & {} & 76.33 & 75.61 & {} & 76.69 & 74.98 & {} & 75.10 & 73.66 \\
Profiles3 & 75.32 & 74.74 & {} & 73.99 & 73.29 & {} & 72.39 & 71.37 & {} & 72.67 & 70.87 & {} & 71.56 & 69.83 \\
Profiles4 & 86.30 & 84.95 & {} & 85.19 & 84.31 & {} & 84.18 & 83.53 & {} & 84.35 & 83.00 & {} & 83.28 & 81.74 \\
Profiles5 & 84.03 & 83.22 & {} & 82.97 & 82.08 & {} & 81.36 & 80.62 & {} & 80.77 & 80.12 & {} & 81.27 & 79.29 \\
Profiles6 & 81.40 & 80.66 & {} & 82.66 & 80.70 & {} & 78.78 & 78.24 & {} & 77.14 & 75.92 & {} & 77.01 & 75.78 \\
Profiles7 & 87.54 & 87.34 & {} & 92.60 & 92.50 & {} & 95.80 & 95.66 & {} & 98.22 & 97.85 & {} & 98.78 & 98.30 \\
Profiles8 & 86.79 & 84.86 & {} & 85.56 & 83.89 & {} & 85.31 & 83.09 & {} & 85.29 & 82.90 & {} & 83.95 & 82.08 \\
Profiles9 & 66.36 & 65.84 & {} & 60.53 & 60.02 & {} & 57.46 & 56.80 & {} & 56.05 & 55.10 & {} & 54.39 & 53.68 \\
Profiles10 & 70.76 & 69.92 & {} & 69.33 & 68.19 & {} & 67.52 & 66.78 & {} & 66.60 & 66.02 & {} & 65.56 & 64.88 \\ \hline
avg.(1st) & 85.25 & 84.42 & {} & 84.91 & 83.85 & {} & 84.50 & 83.26 & {} & 83.56 & 82.33 & {} & 83.52 & 81.77 \\
avg.(2nd) & 80.31 & 79.55 & {} & 79.46 & 78.54 & {} & 78.38 & 77.54 & {} & 78.30 & 76.94 & {} & 77.52 & 76.18 \\
avg.(all) & 83.19 & 82.39 & {} & 82.64 & 81.64 & {} & 81.95 & 80.88 & {} & 81.37 & 80.08 & {} & 81.02 & 79.44 \\ \hline
\end{tabular}
\end{small}
\end{spacing}
\end{table}

\begin{table}[tb]
\caption{The improvement in the average density via the corner detection\label{tab:improvement}}
\medskip
\centering
\begin{spacing}{0.8}
\begin{small}
\begin{tabular}{lrrrrr} \hline
instance & \multicolumn{1}{c}{$W=128\mathrm{px}$} & \multicolumn{1}{c}{$W=256\mathrm{px}$} & \multicolumn{1}{c}{$W=512\mathrm{px}$} & \multicolumn{1}{c}{$W=1024\mathrm{px}$} & \multicolumn{1}{c}{$W=2048\mathrm{px}$} \\ \hline
Albano & 0.07 & -0.05 & 0.68 & 0.70 & 1.09 \\
Dagli & 0.13 & 0.37 & 0.36 & 1.53 & 0.86 \\
Dighe1 & 3.30 & 2.22 & 3.49 & 10.15 & 12.92 \\
Dighe2 & 0.00 & -0.07 & 0.02 & 0.00 & -0.01 \\
Fu & 0.84 & 0.76 & 1.18 & 1.49 & 1.60 \\
Jakobs1 & 0.91 & -0.94 & 2.12 & 3.35 & 3.75 \\
Jakobs2 & 0.66 & 0.35 & 0.21 & 1.06 & 2.21 \\
Mao & 0.74 & 0.09 & 0.21 & 0.46 & 1.07 \\
Marques & 0.09 & 0.04 & 0.48 & 0.67 & 0.82 \\
Shapes0 & -0.03 & 0.28 & 0.44 & 0.71 & 1.12 \\
Shapes1 & 0.36 & 0.59 & 0.82 & 1.12 & 1.19 \\
Shirts & -0.29 & 0.02 & 0.30 & 0.49 & 0.50 \\
Swim & 0.47 & 0.18 & 0.53 & 1.03 & 1.33 \\
Trousers & 0.28 & 0.40 & 0.65 & 1.42 & 1.33 \\ \hline
Profiles1 & -0.10 & 0.89 & 1.35 & 1.54 & 1.54 \\
Profiles2 & -0.05 & 0.42 & 0.75 & 1.04 & 1.87 \\
Profiles3 & 0.36 & 0.49 & 1.54 & 1.50 & 2.19 \\
Profiles4 & 0.26 & 0.88 & 0.56 & 1.43 & 2.35 \\
Profiles5 & -0.04 & -0.35 & 0.80 & 0.80 & 2.16 \\
Profiles6 & 0.37 & -0.25 & 0.47 & 0.93 & 1.14 \\
Profiles7 & -0.09 & -0.10 & -0.01 & 0.33 & 0.53 \\
Profiles8 & 0.42 & 0.61 & 1.99 & 2.47 & 3.99 \\
Profiles9 & 0.23 & 0.16 & 0.58 & 0.54 & 1.23 \\
Profiles10 & 0.47 & 0.24 & 0.69 & 0.97 & 1.72 \\ \hline
avg.(all) & 0.39 & 0.30 & 0.84 & 1.49 & 2.02 \\ \hline
\end{tabular}
\end{small}
\end{spacing}
\end{table}

We next compared the computational results of GCDH ($W=512\mathrm{px}$) with those reported by \citet{UmetaniS2009} (denoted as ``FITS''), \citet{ElkeranA2013} (denoted as ``GCS''), \citet{SatoAK2019} (denoted as ``ROMA''), and \citet{BurkeEK2010} (denoted as ``BLF'').
The FITS, GCS and ROMA are improvement algorithms for the polygon packing problem, where ROMA constrained the search space to place the pieces on a grid as same as \citet{MundimLR2017}, and computed the penetration depth efficiently via a variant of discretized Voronoi diagrams called the raster penetration map.
\citet{BurkeEK2010} considered an extended vector model incorporating circular arcs into polygons and proposed a robust orbital sliding method to compute NFPs.
They developed a BLF algorithm that restricted the search space on vertical lines with sufficiently small gaps between them and incorporated it with local search and tabu search (TS) algorithms to find a good order of given pieces.
Table~\ref{tab:algorithm_density} shows their computational results, which are evaluated in the best and average density (\%).
We note that it is not appropriate to directly compare the density of these algorithms, because GCDH was tested on the instances of the raster model while the other algorithms were tested those of the vector model.
Table~\ref{tab:algorithm_time} shows the computational environment and the computation time (in seconds) of the algorithms.
\citet{BurkeEK2010} tested four variations of their algorithm, and each were run 10 times.
Their results in Table~\ref{tab:algorithm_density} are the best results of 40 runs.
They did not use time limit but stopped their algorithm by other criteria, and their computation time in Table~\ref{tab:algorithm_time} is the time spent to find the best solution reported in Table~\ref{tab:algorithm_density} in the run that found it; i.e., the time only one run is reported.
\citet{UmetaniS2009}, \citet{ElkeranA2013} and \citet{SatoAK2019} tested their algorithms 10, 20 and 30 times for each instance, respectively, where they set the time limits for each run as shown in Table~\ref{tab:algorithm_time}.

\begin{figure}[tb]
  \centering
  \begin{minipage}{0.15\textwidth}
    \centering
    \scalebox{0.2}{\includegraphics{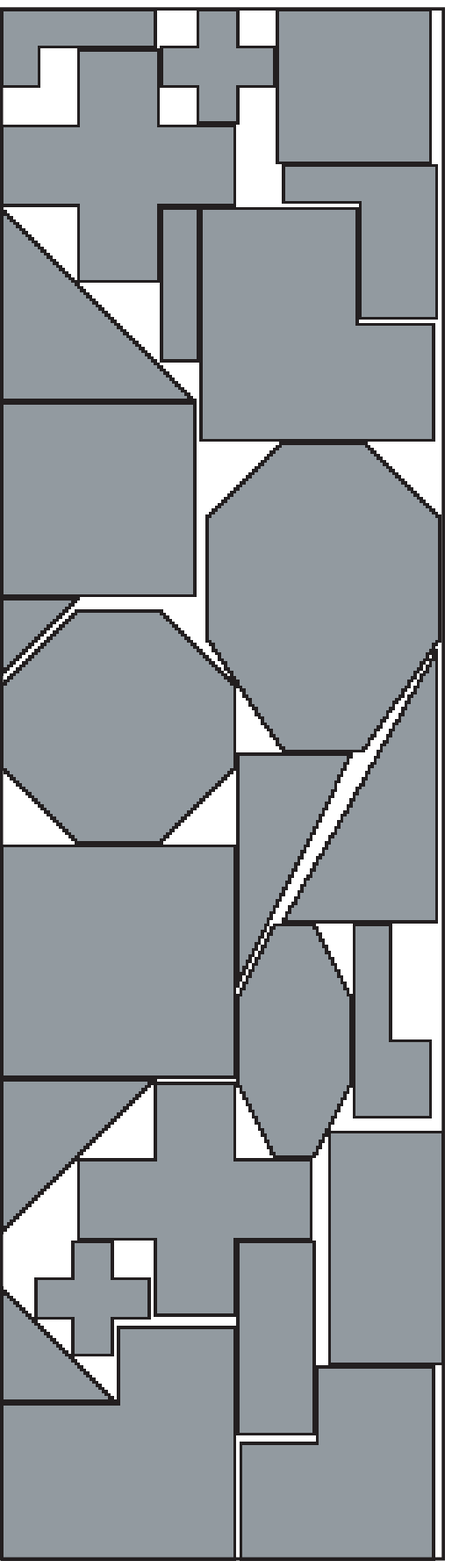}}\\
             {\small Jakobs1 (88.01\%)}
  \end{minipage}
  \hfill
  \begin{minipage}{0.15\textwidth}
    \centering
    \scalebox{0.2}{\includegraphics{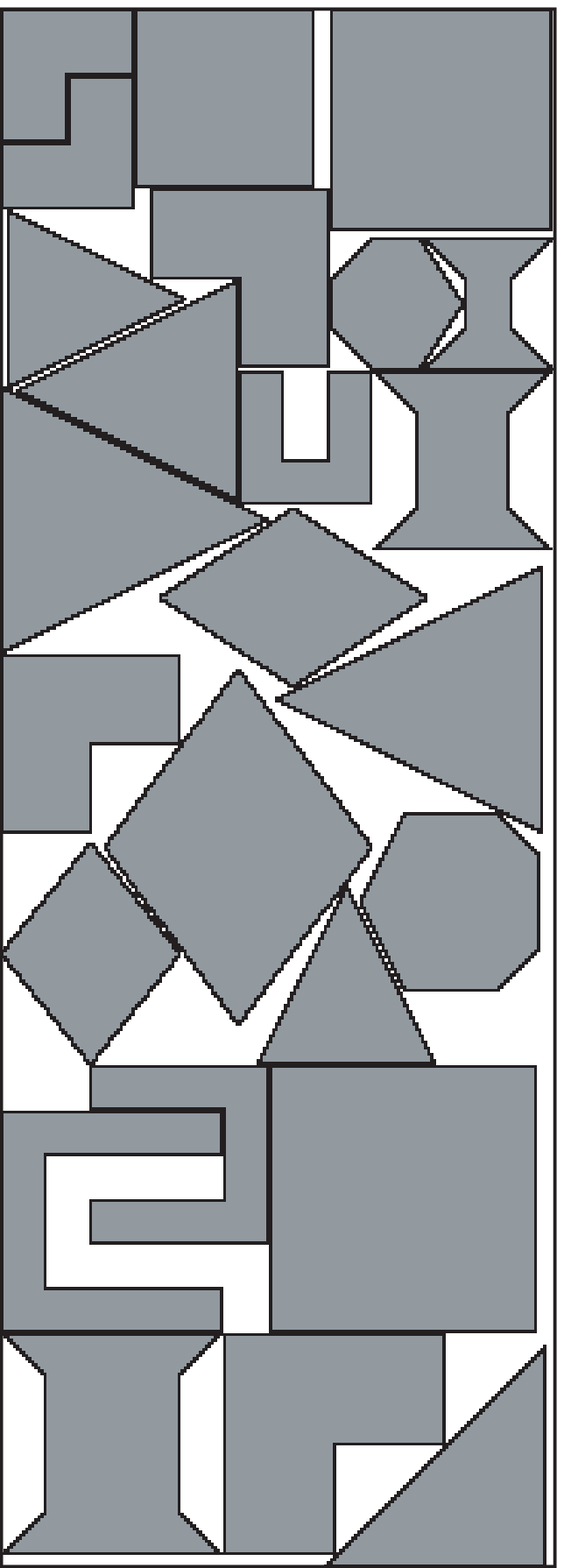}}\\
             {\small Jakobs2 (79.51\%)}
  \end{minipage}
  \hfill
  \begin{minipage}{0.30\textwidth}
    \centering
    \scalebox{0.2}{\includegraphics{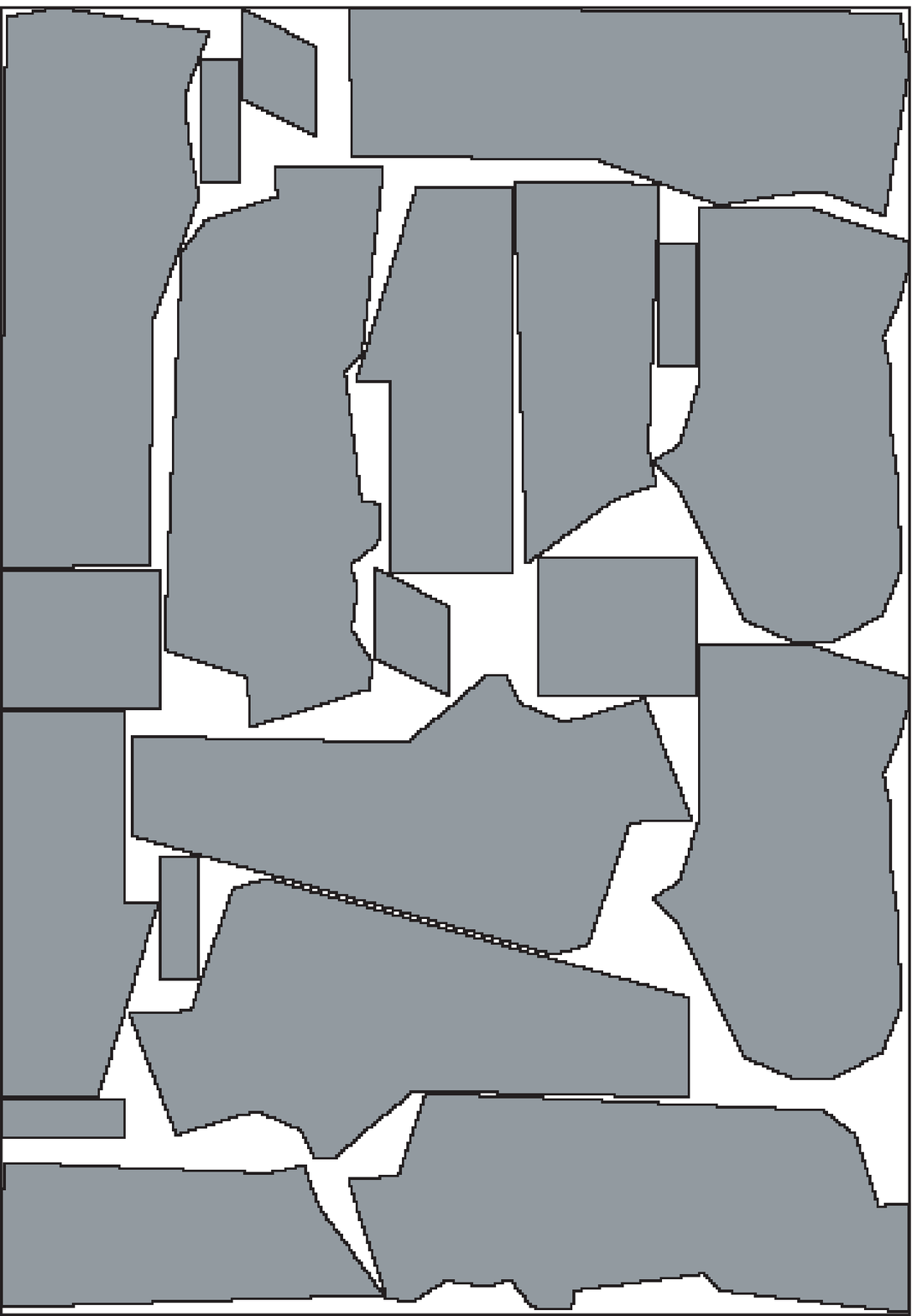}}\\
             {\small Mao (84.84\%)}
  \end{minipage}
  \hfill
  \begin{minipage}{0.30\textwidth}
    \centering
    \scalebox{0.2}{\includegraphics{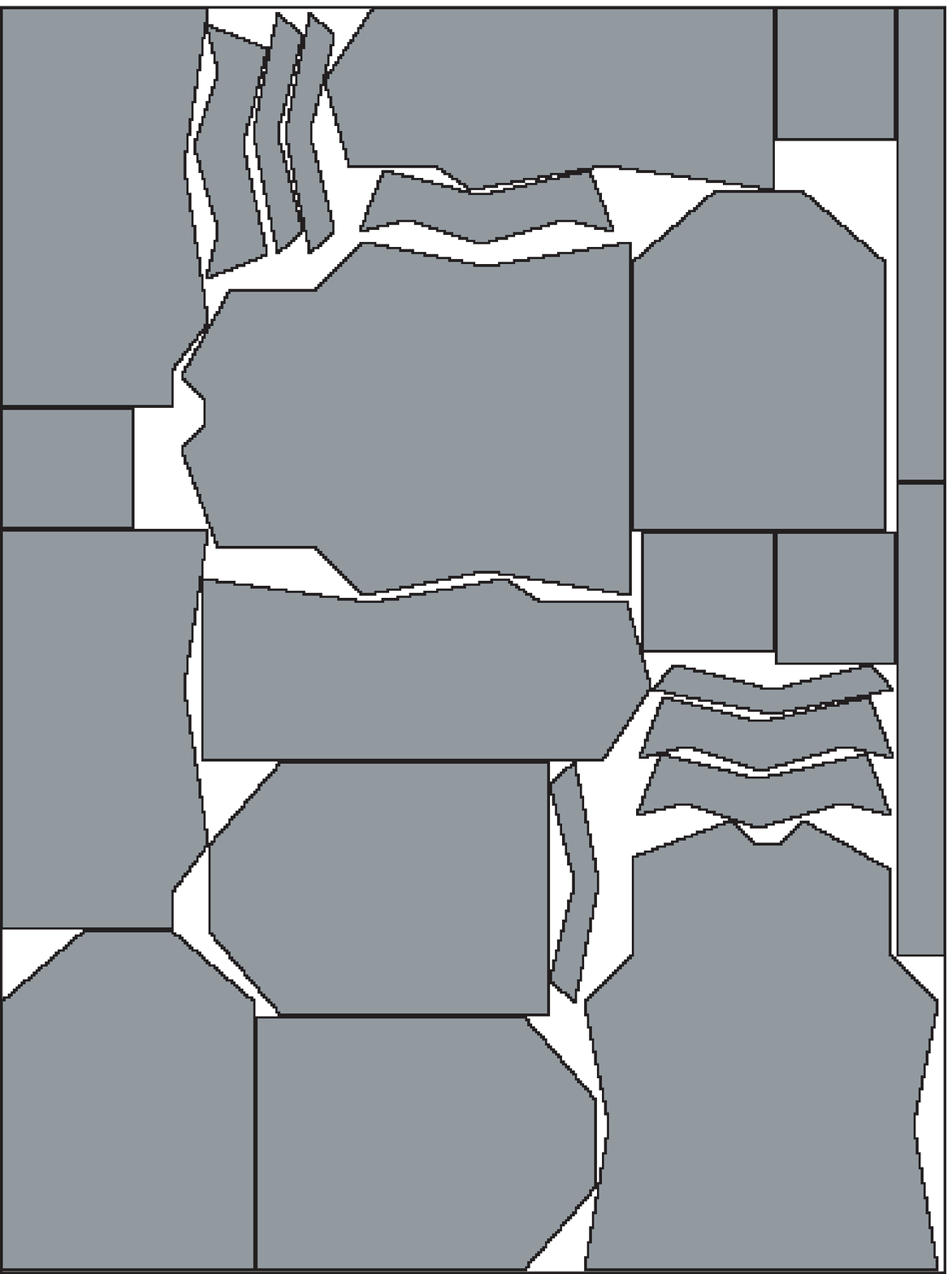}}\\
             {\small Marques (90.43\%)}
  \end{minipage}
  \bigskip
  \begin{minipage}{0.24\textwidth}
    \centering
    \scalebox{0.12}{\includegraphics{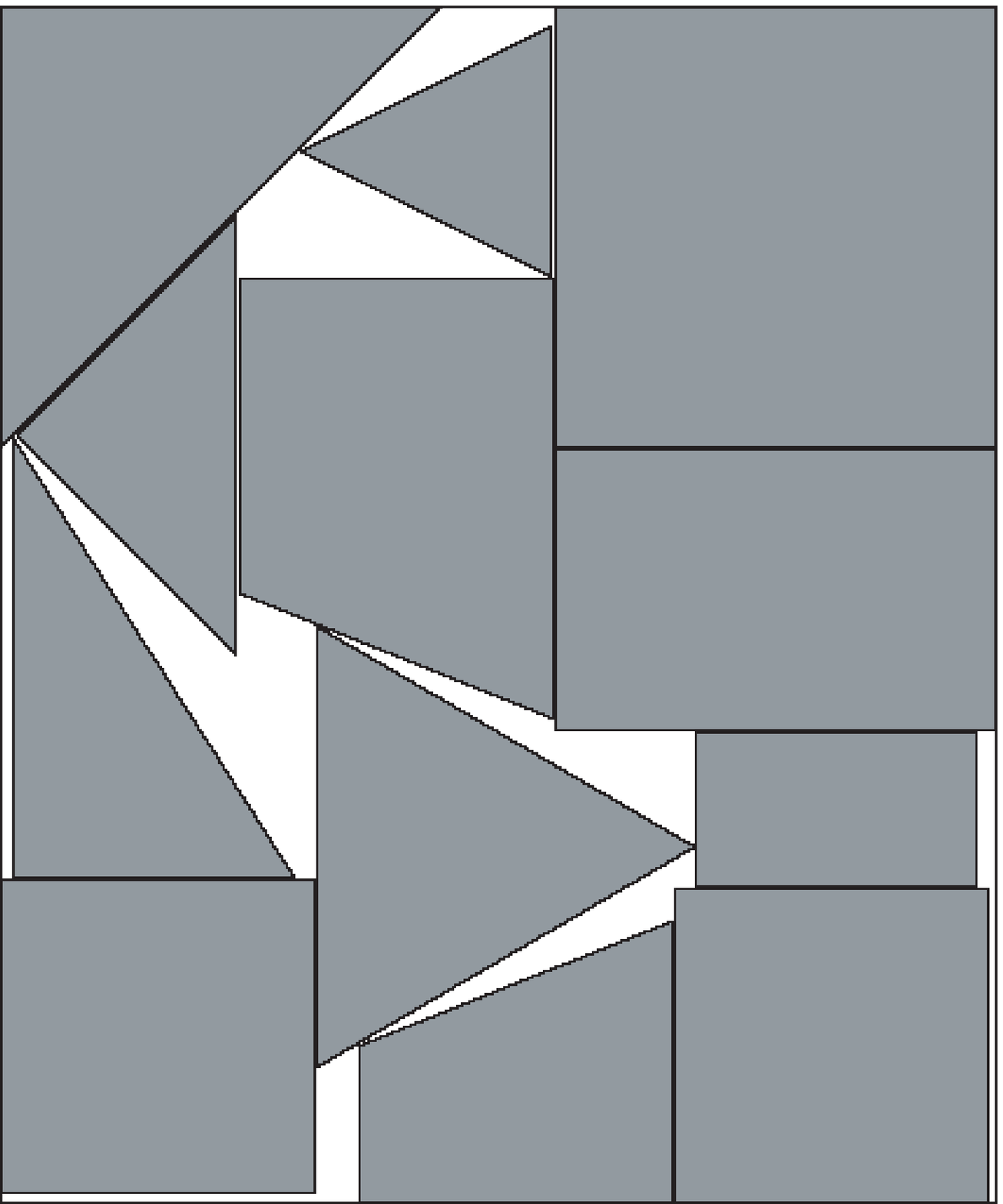}}\\
           {\small Fu (90.75\%)}
  \end{minipage}
  \hfill
  \begin{minipage}{0.24\textwidth}
    \centering
    \scalebox{0.12}{\includegraphics{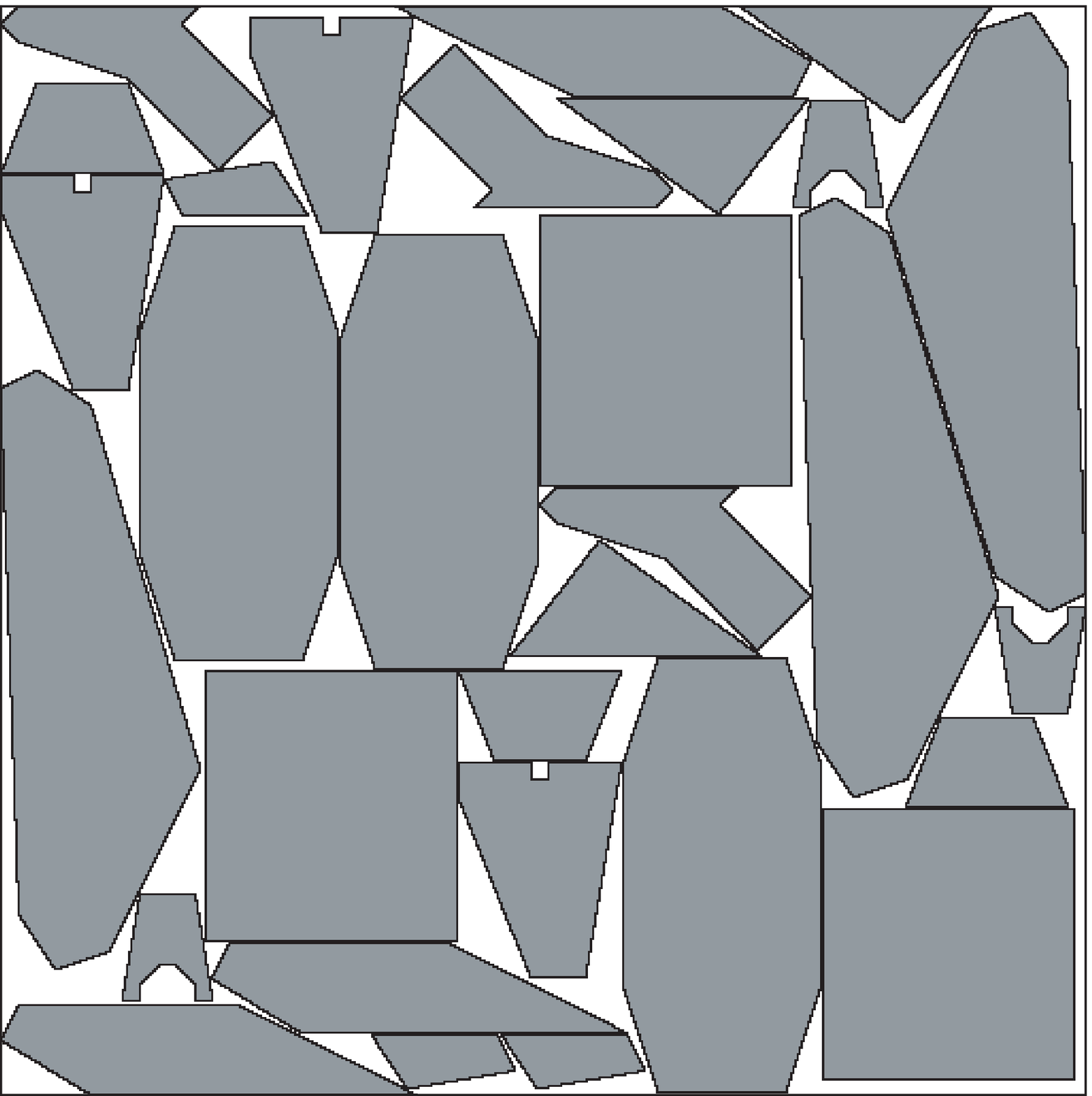}}\\
           {\small Dagli (86.32\%)}
  \end{minipage}
  \hfill
  \begin{minipage}{0.24\textwidth}
    \centering
    \scalebox{0.12}{\includegraphics{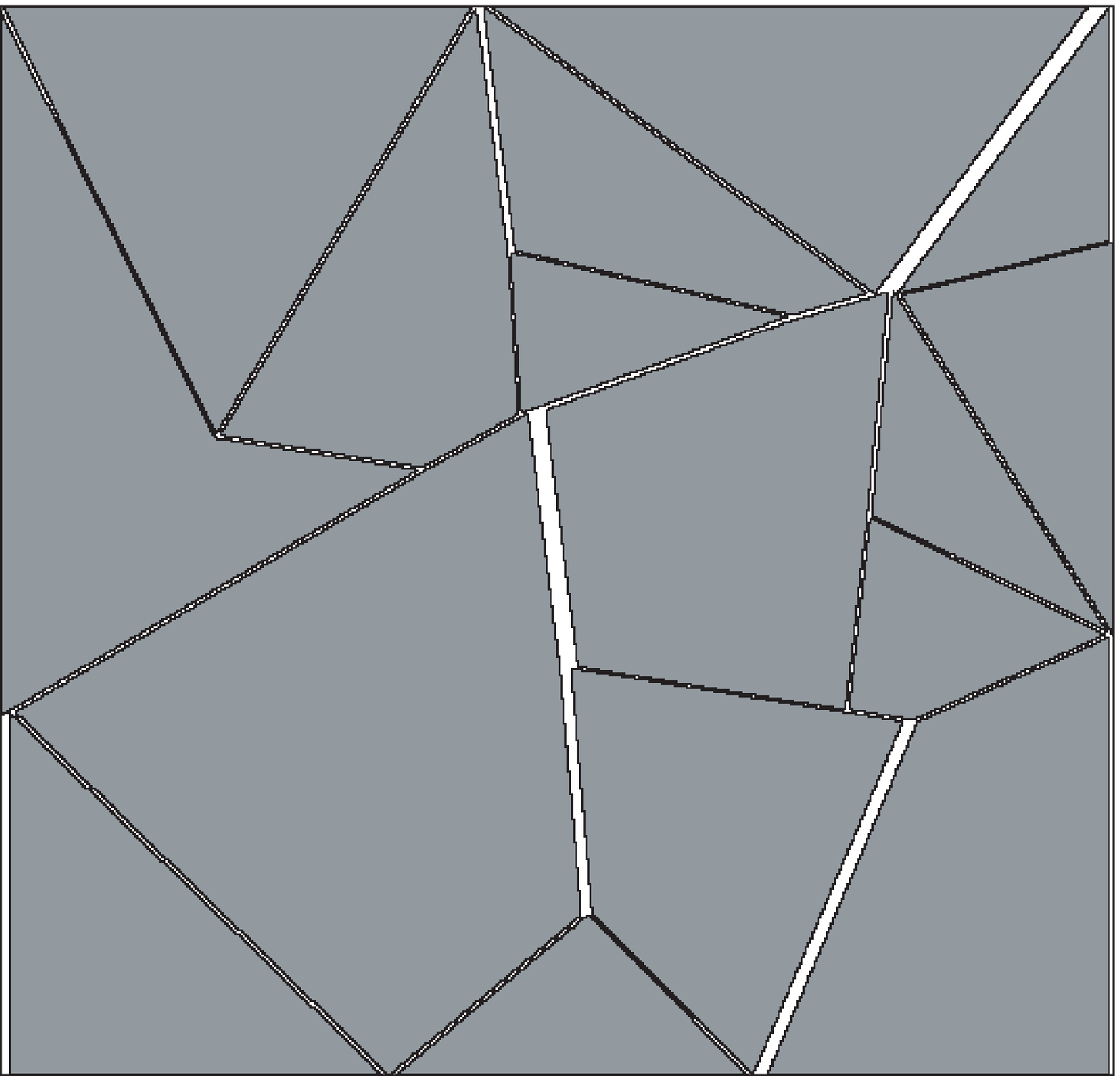}}\\
           {\small Dighe1 (97.60\%)}
  \end{minipage}
  \hfill
  \begin{minipage}{0.24\textwidth}
    \centering
    \scalebox{0.12}{\includegraphics{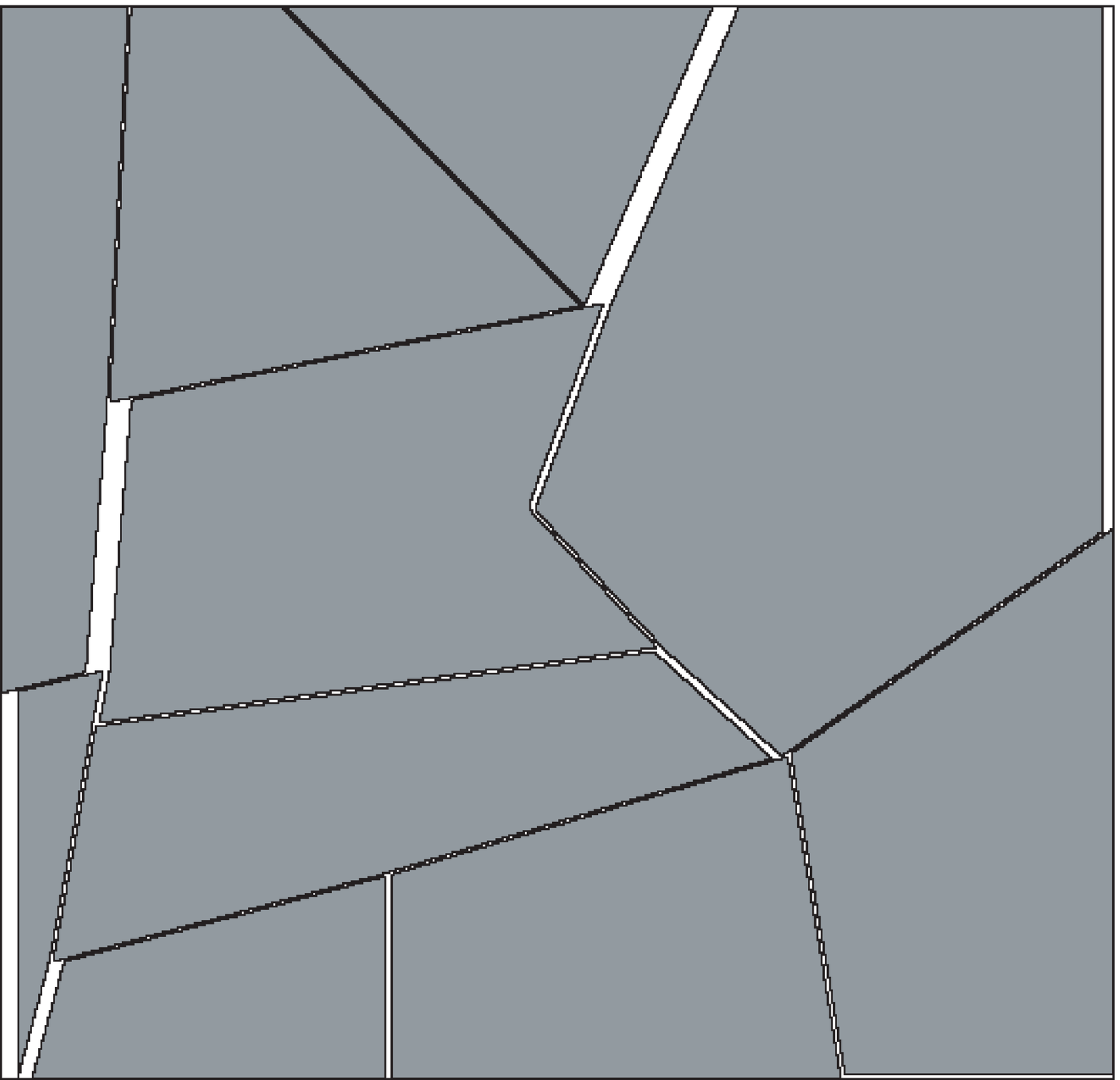}}\\
           {\small Dighe2 (97.28\%)}
  \end{minipage}
  \bigskip
  \begin{minipage}{0.48\textwidth}
    \centering
    \scalebox{0.32}{\includegraphics{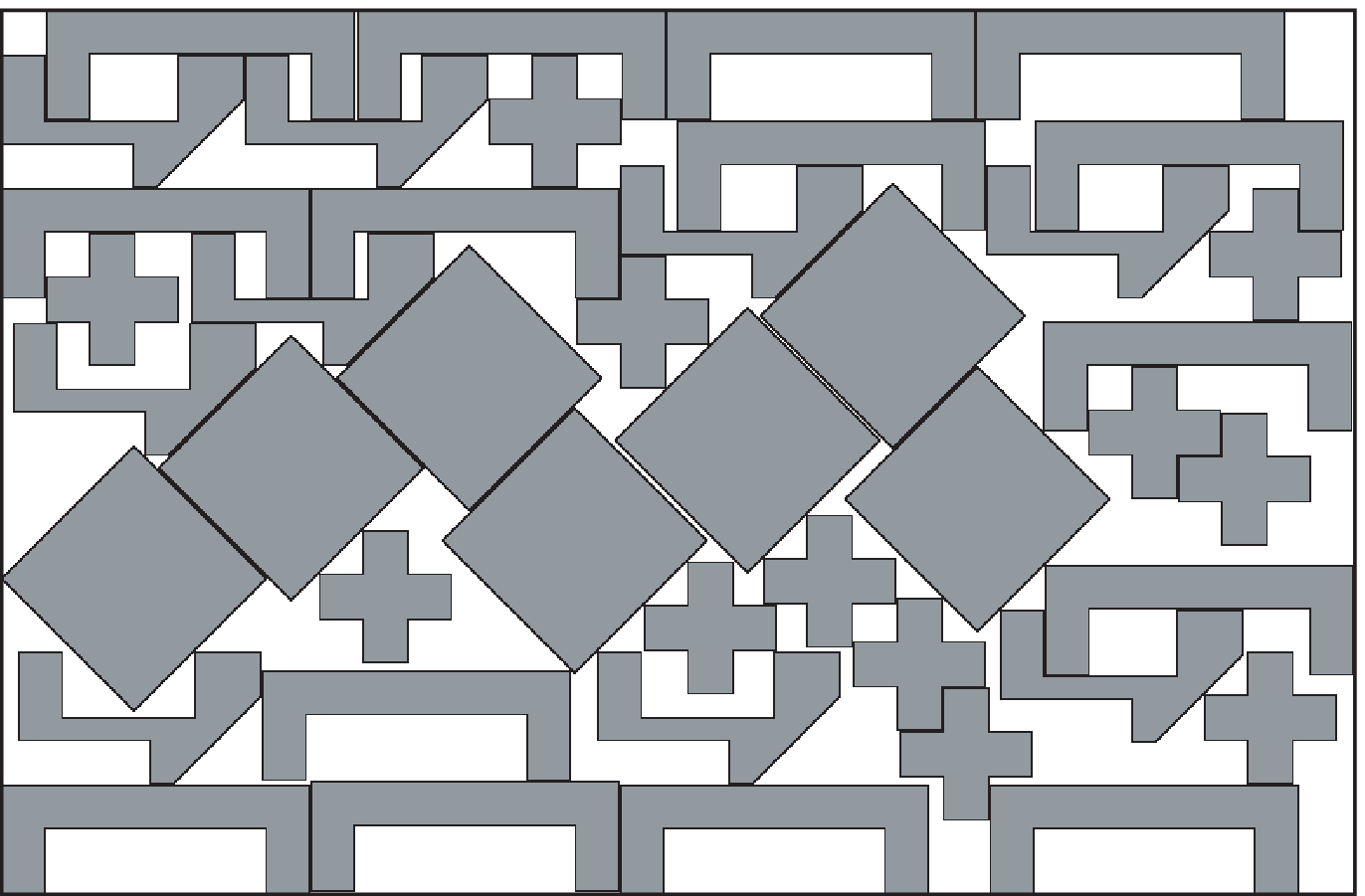}}\\
             {\small Shapes0 (66.13\%)}
  \end{minipage}
  \hfill
  \begin{minipage}{0.48\textwidth}
    \centering
    \scalebox{0.32}{\includegraphics{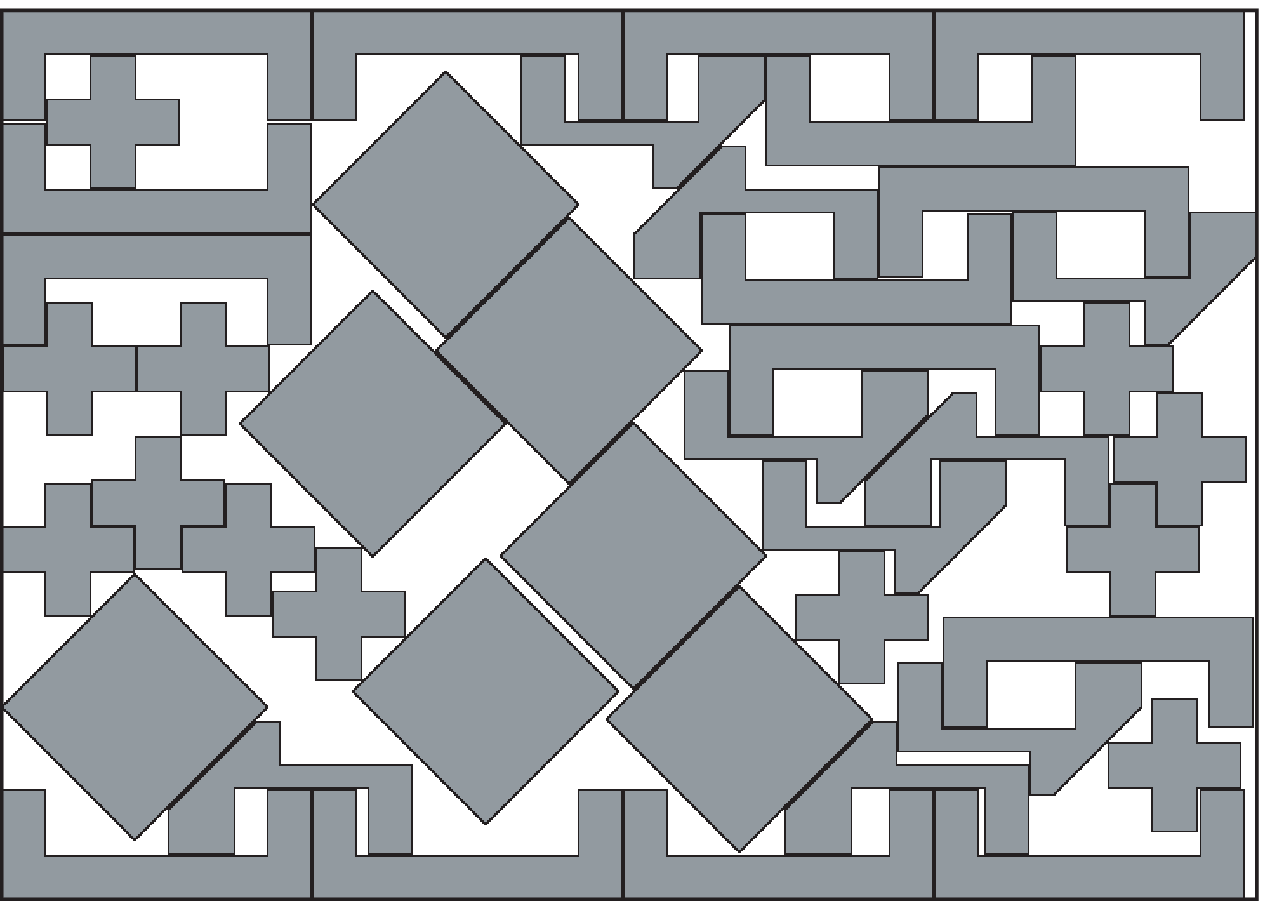}}\\
             {\small Shapes1 (72.35\%)}
  \end{minipage}
  \bigskip
  \begin{minipage}{0.38\textwidth}
    \centering
    \scalebox{0.22}{\includegraphics{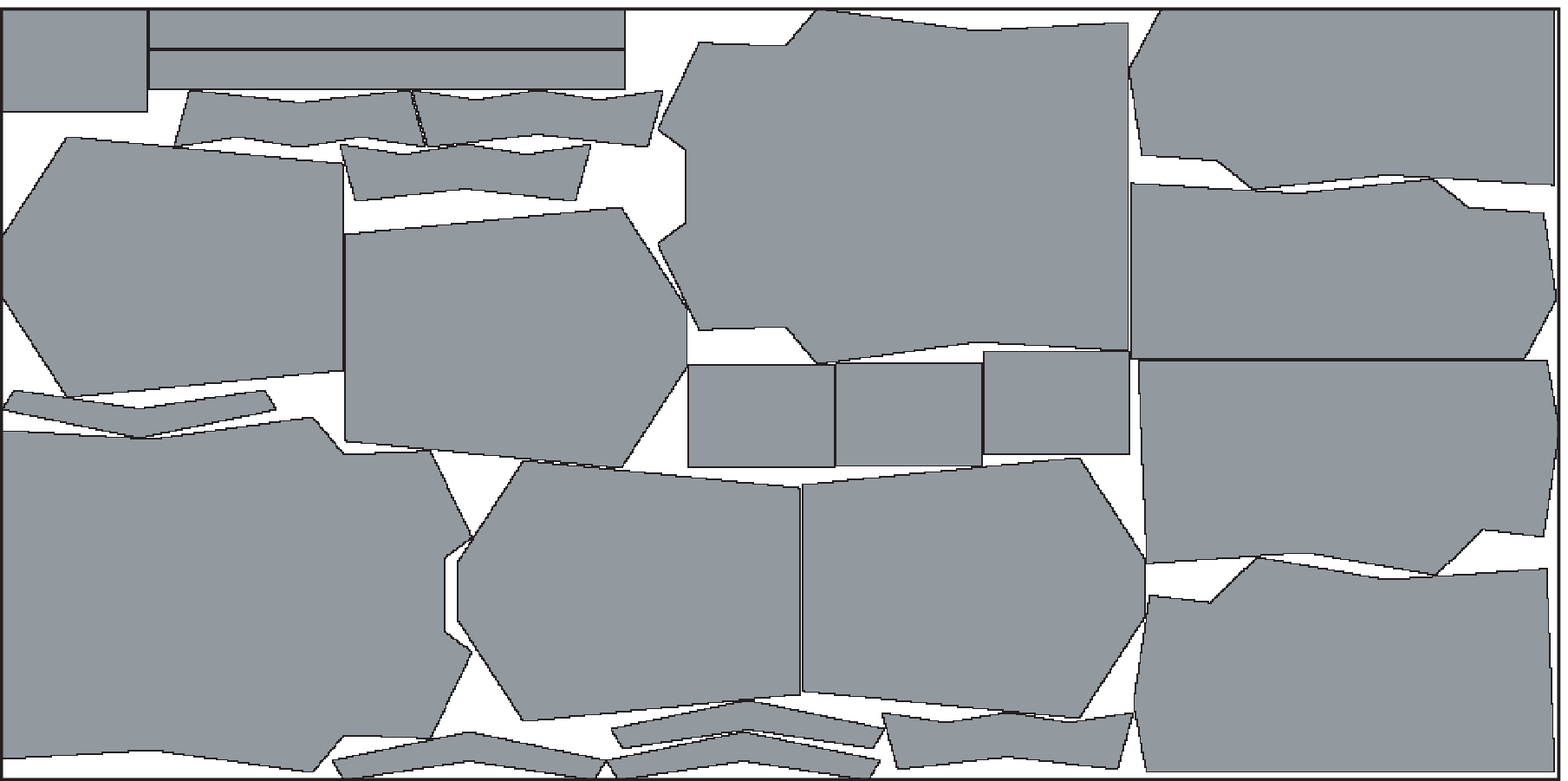}}\\
             {\small Albano (88.20\%)}
  \end{minipage}
  \hfill
  \begin{minipage}{0.58\textwidth}
    \centering
    \scalebox{0.3}{\includegraphics{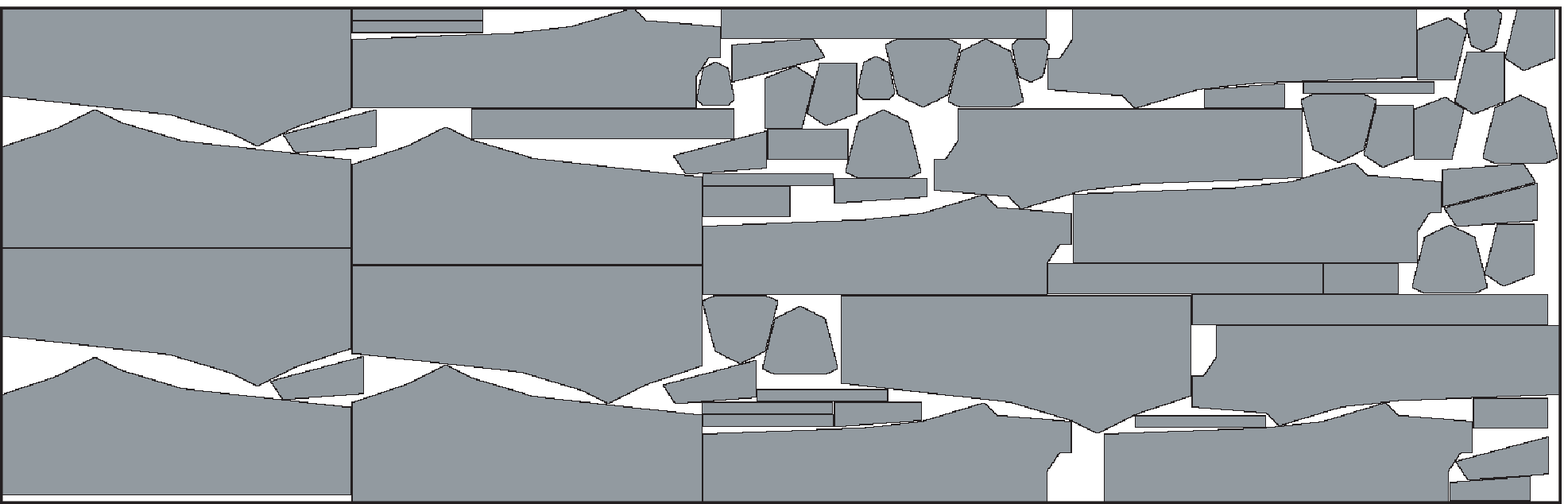}}\\
             {\small Trousers (88.29\%)}
  \end{minipage}
  \begin{minipage}{0.48\textwidth}
    \centering
    \scalebox{0.32}{\includegraphics{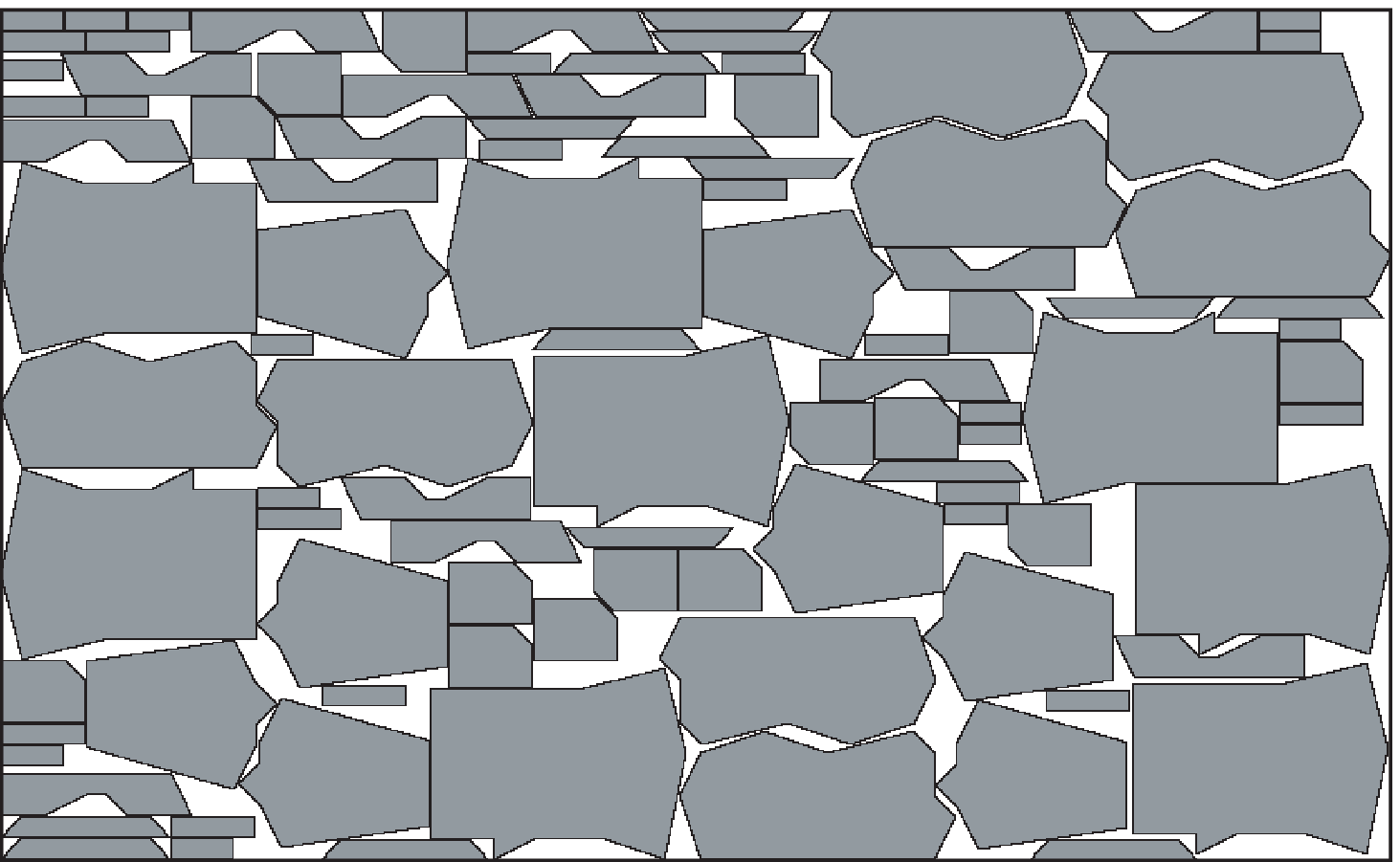}}\\
             {\small Shirts (84.33\%)}
  \end{minipage}
  \hfill
  \begin{minipage}{0.48\textwidth}
    \centering
    \scalebox{0.16}{\includegraphics{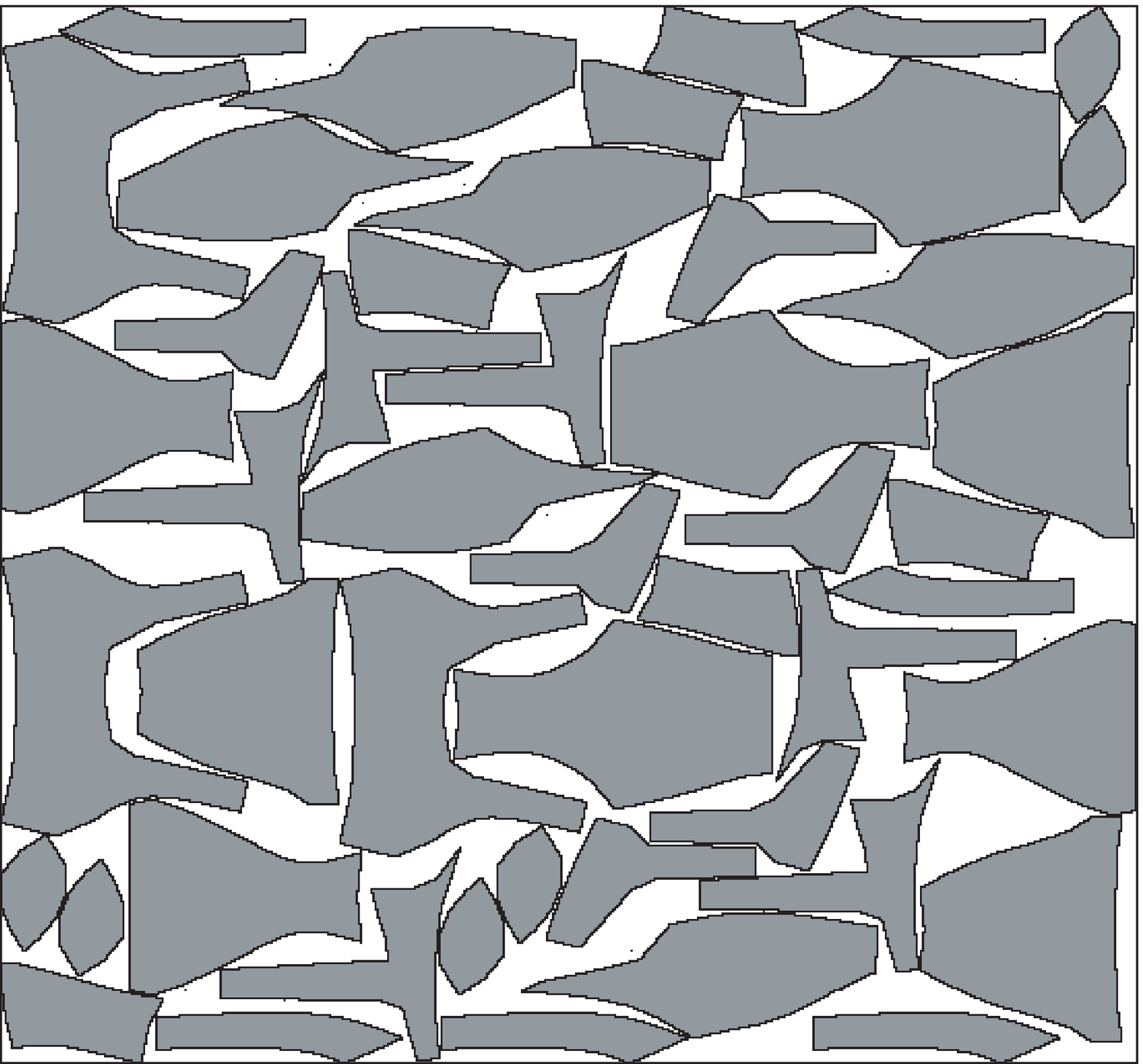}}\\
             {\small Swim (74.08\%)}
  \end{minipage}
  \caption{The best solutions for the test instances obtained by GCDH ($W=512\mathrm{px}$).}\label{fig:polygon_all}
\end{figure}

\begin{figure}[tb]
  \centering
  \begin{minipage}{0.24\textwidth}
    \centering
    \scalebox{0.16}{\includegraphics{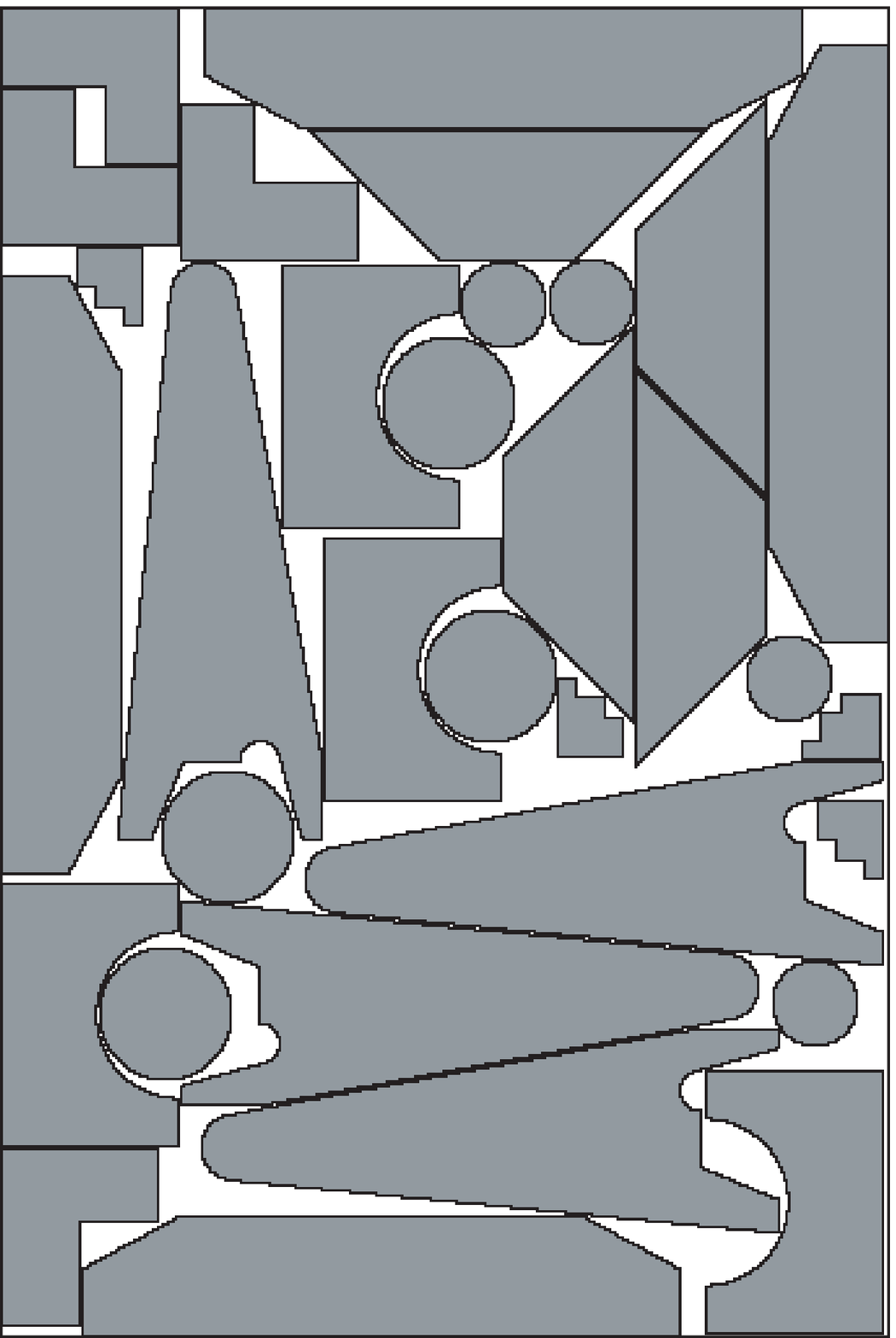}}\\
             {\small Profiles1 (86.93\%)}
  \end{minipage}
  \hfill
  \begin{minipage}{0.24\textwidth}
    \centering
    \scalebox{0.16}{\includegraphics{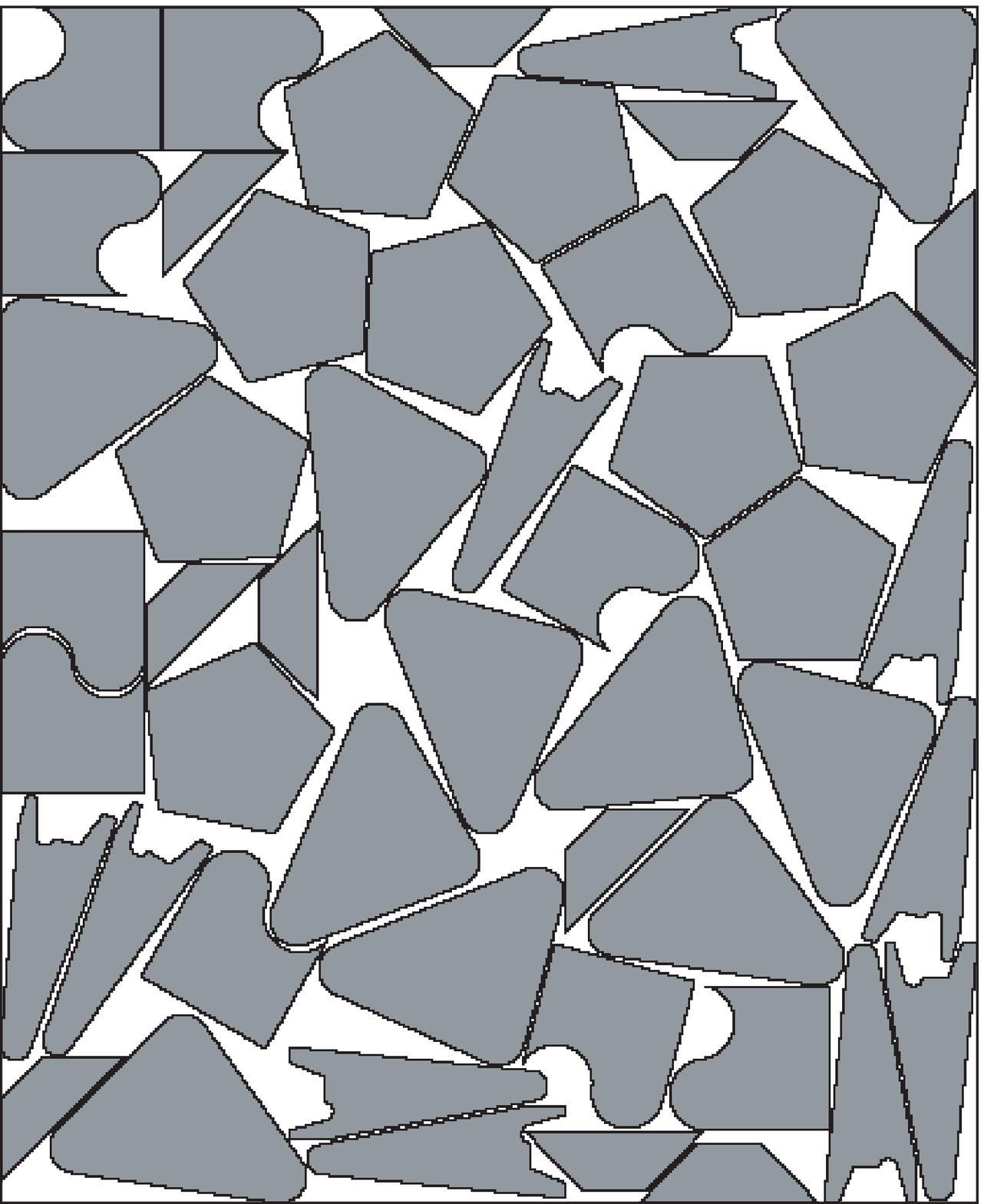}}\\
             {\small Profiles5 (82.47\%)}
  \end{minipage}
  \hfill
  \begin{minipage}{0.24\textwidth}
    \centering
    \scalebox{0.16}{\includegraphics{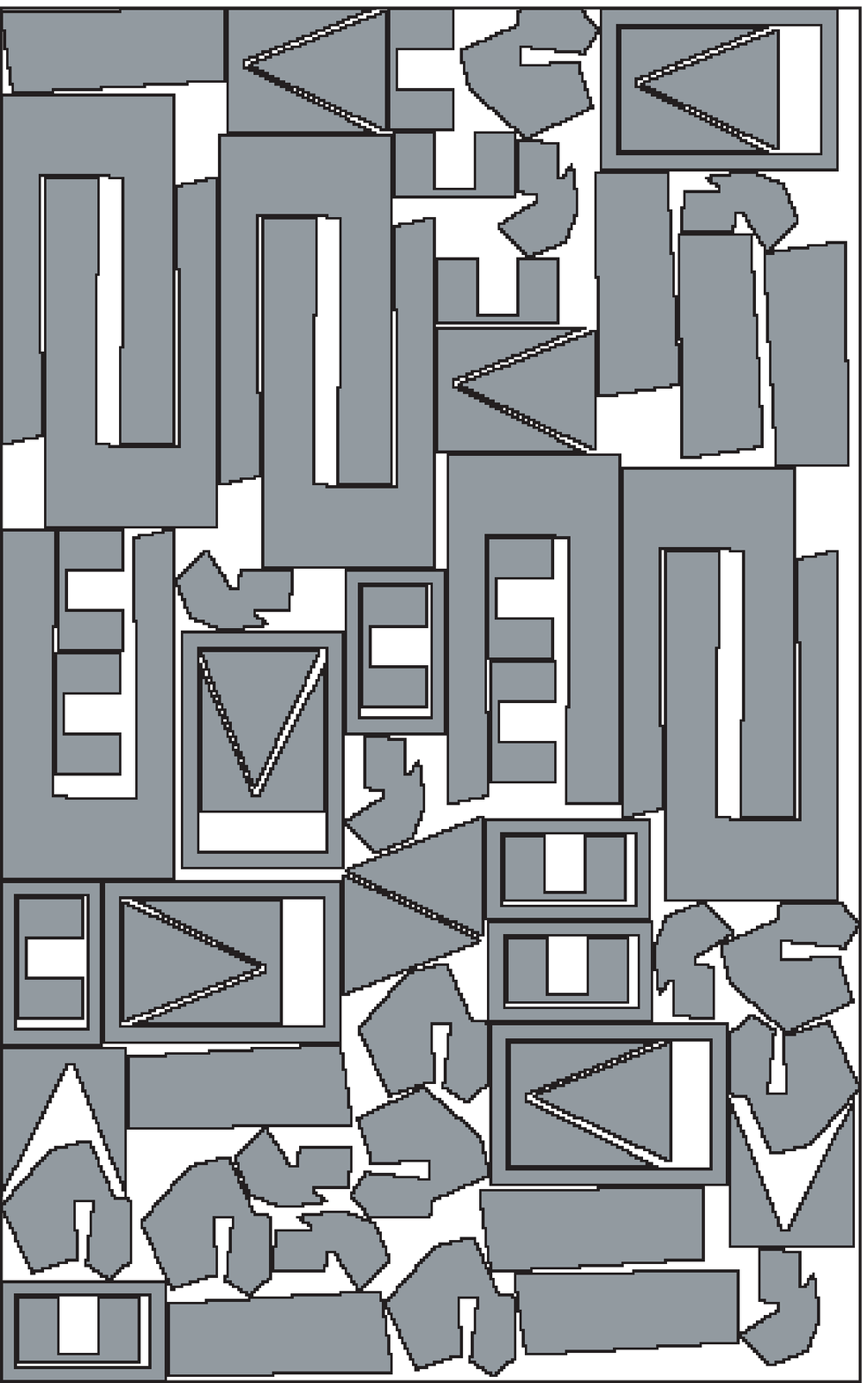}}\\
             {\small Profiles6 (79.03\%)}
  \end{minipage}
  \begin{minipage}{0.24\textwidth}
    \centering
    \scalebox{0.16}{\includegraphics{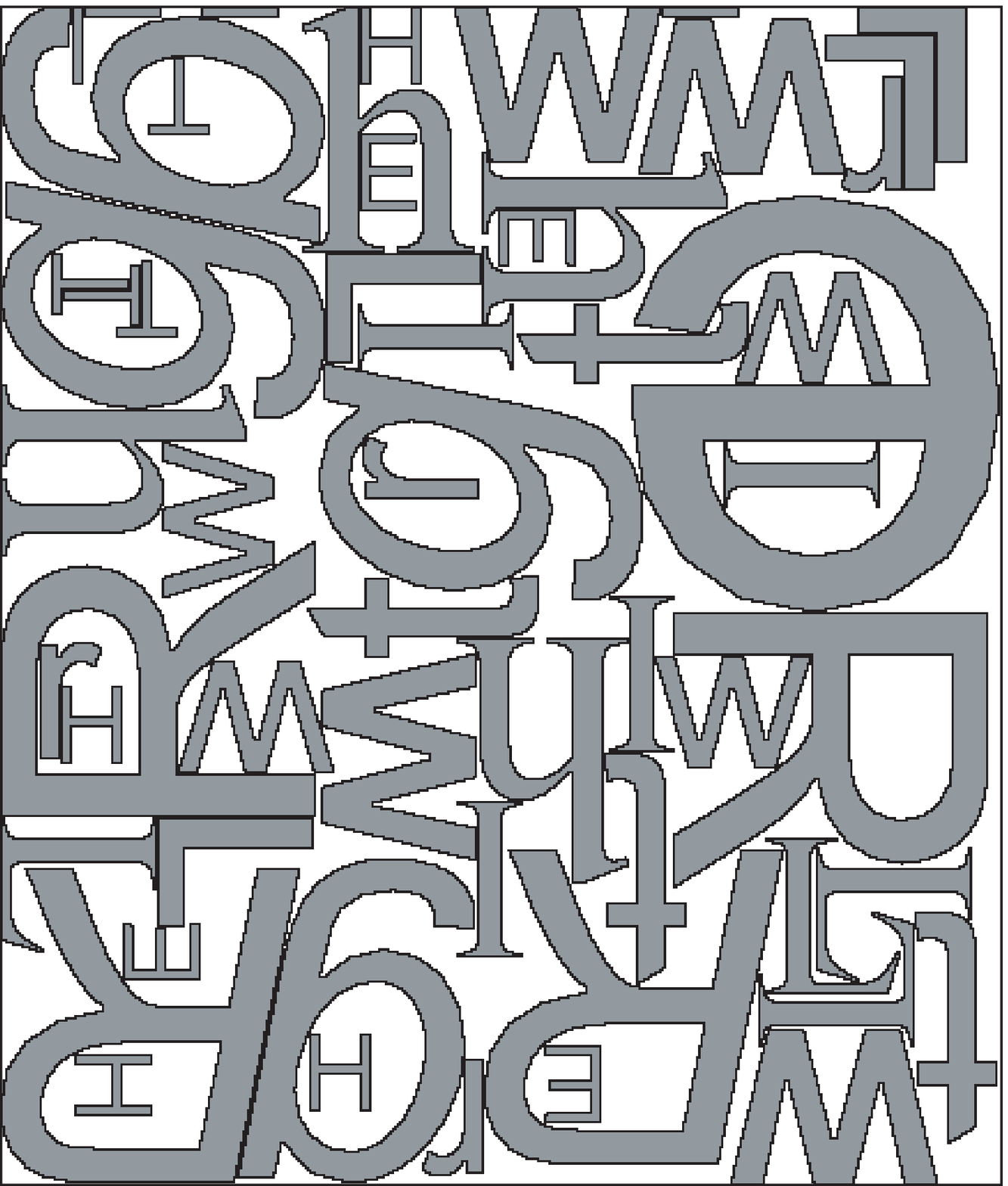}}\\
             {\small Profiles9 (57.86\%)}
  \end{minipage}\\
  \bigskip
  \begin{minipage}{0.30\textwidth}
    \centering
    \scalebox{0.25}{\includegraphics{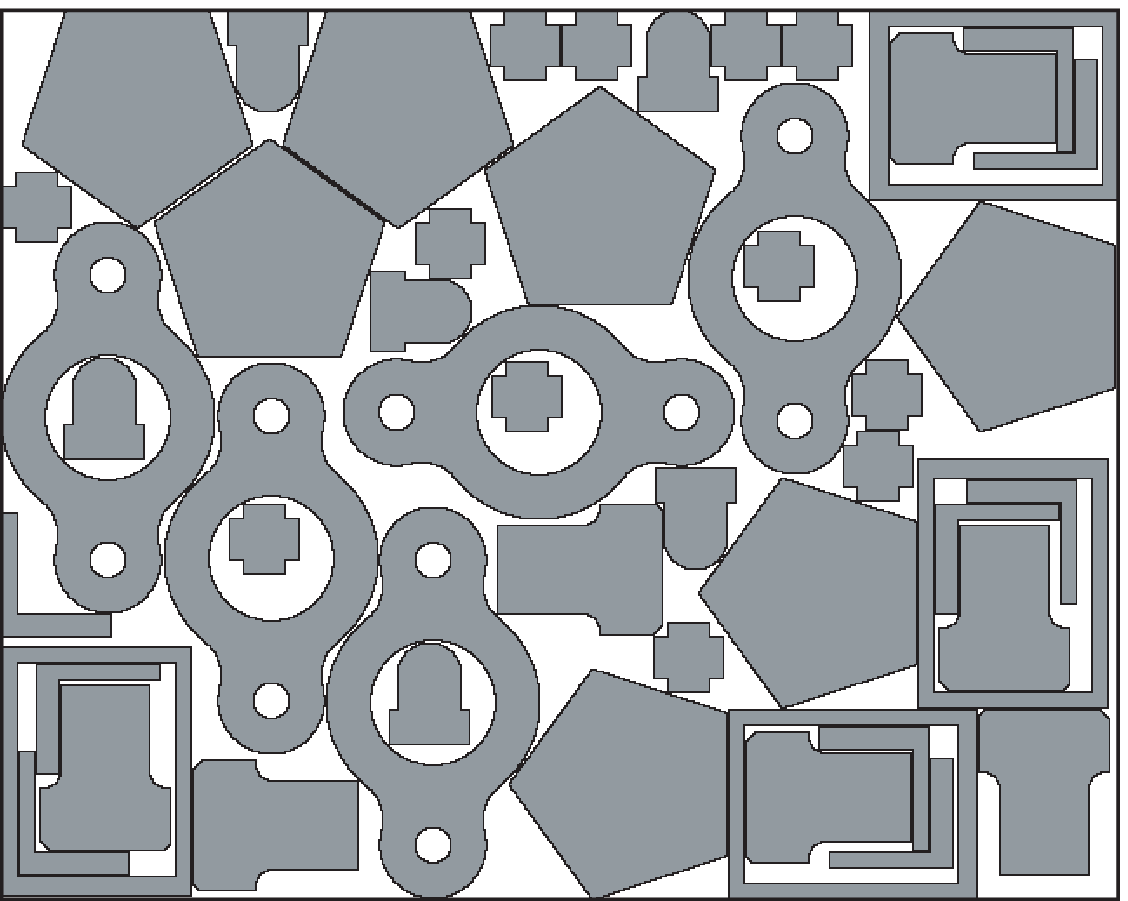}}\\
             {\small Profiles2 (77.28\%)}
  \end{minipage}
  \hfill
  \begin{minipage}{0.66\textwidth}
    \centering
    \scalebox{0.35}{\includegraphics{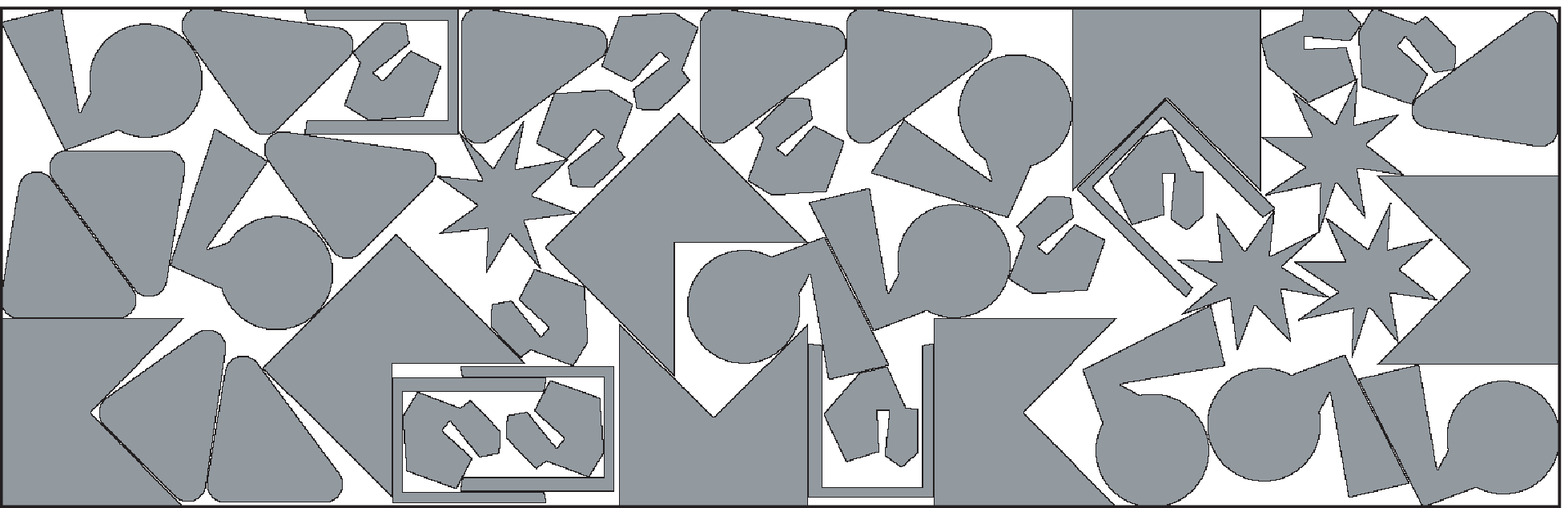}}\\
             {\small Profiles3 (73.87\%)}
  \end{minipage}\\
  \bigskip
  \begin{minipage}{0.55\textwidth}
    \centering
    \scalebox{0.28}{\includegraphics{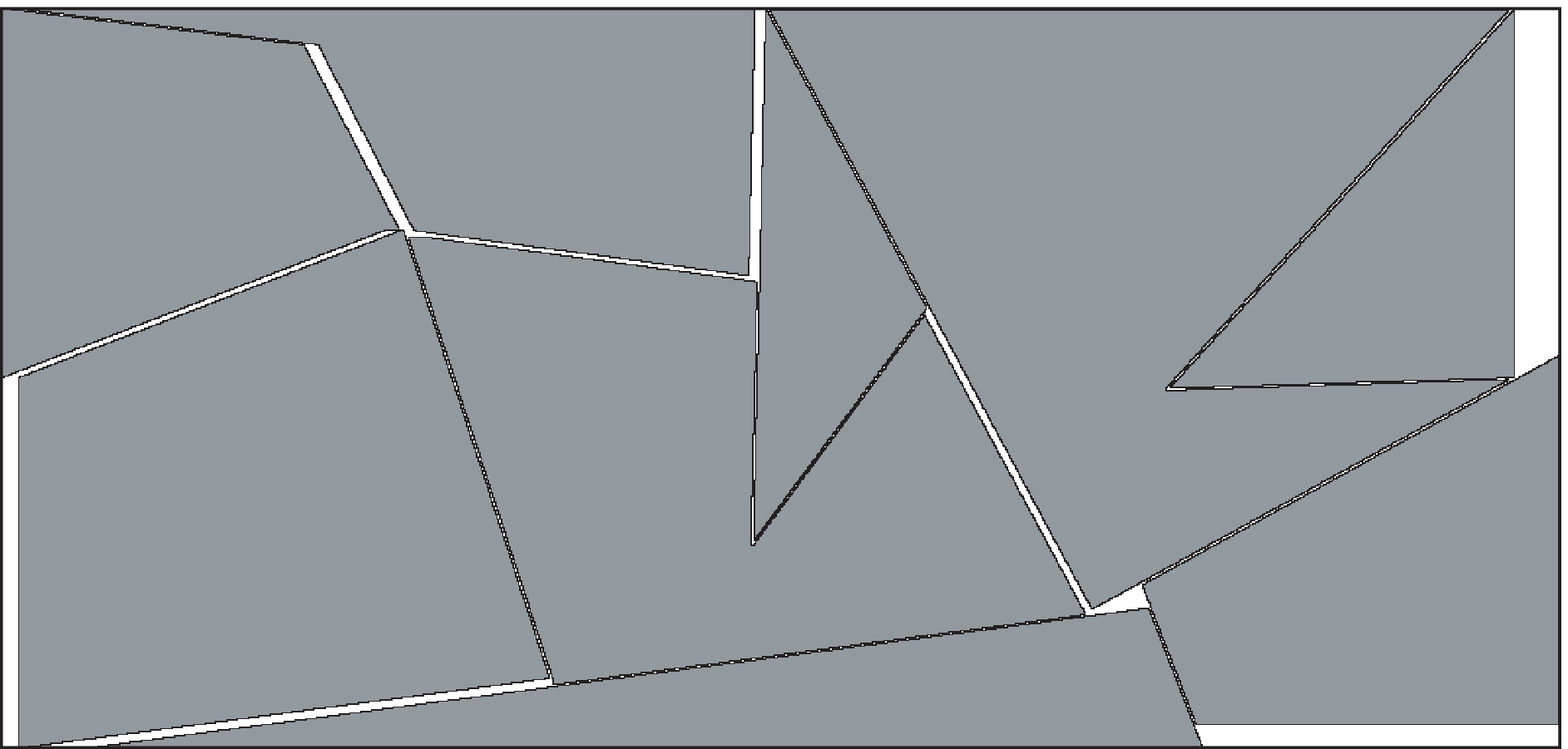}}\\
             {\small Profiles7 (95.80\%)}
  \end{minipage}
  \hfill
  \begin{minipage}{0.4\textwidth}
    \centering
    \scalebox{0.28}{\includegraphics{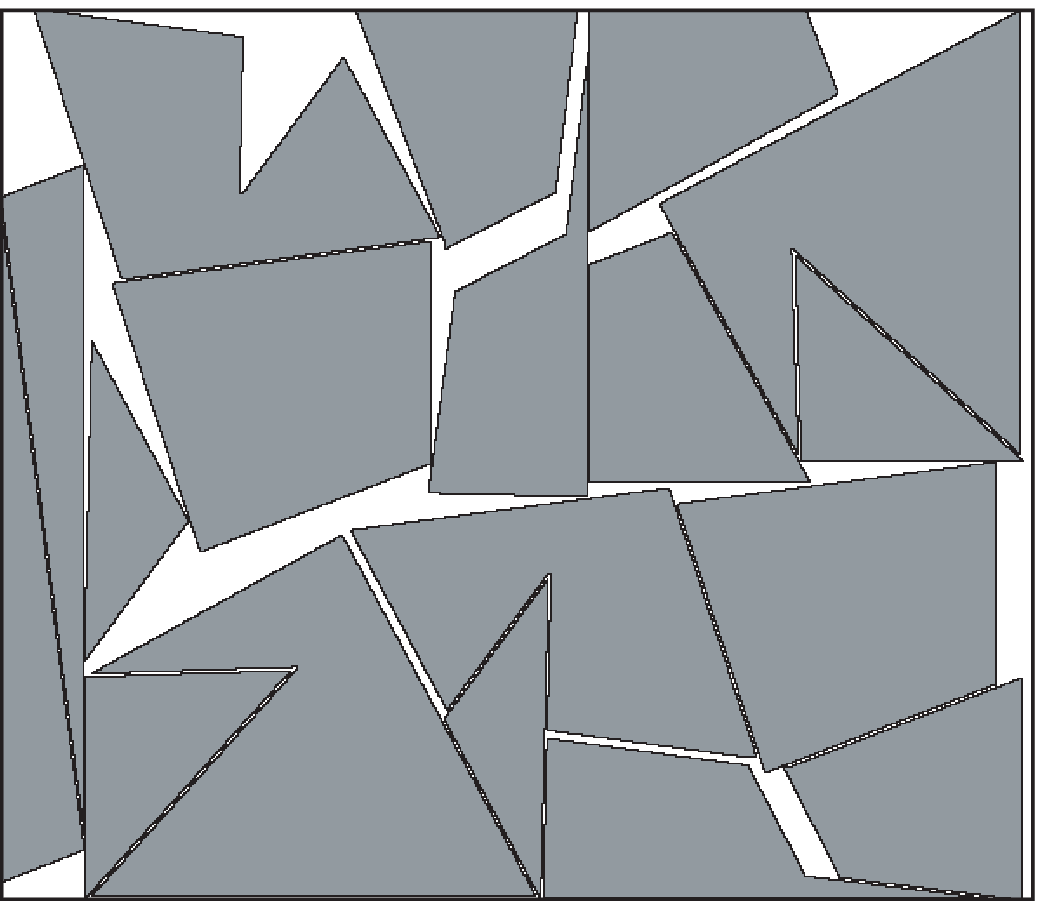}}\\
             {\small Profiles8 (87.61\%)}
  \end{minipage}\\
  \bigskip
  \begin{minipage}{0.96\textwidth}
    \centering
    \scalebox{0.56}{\includegraphics{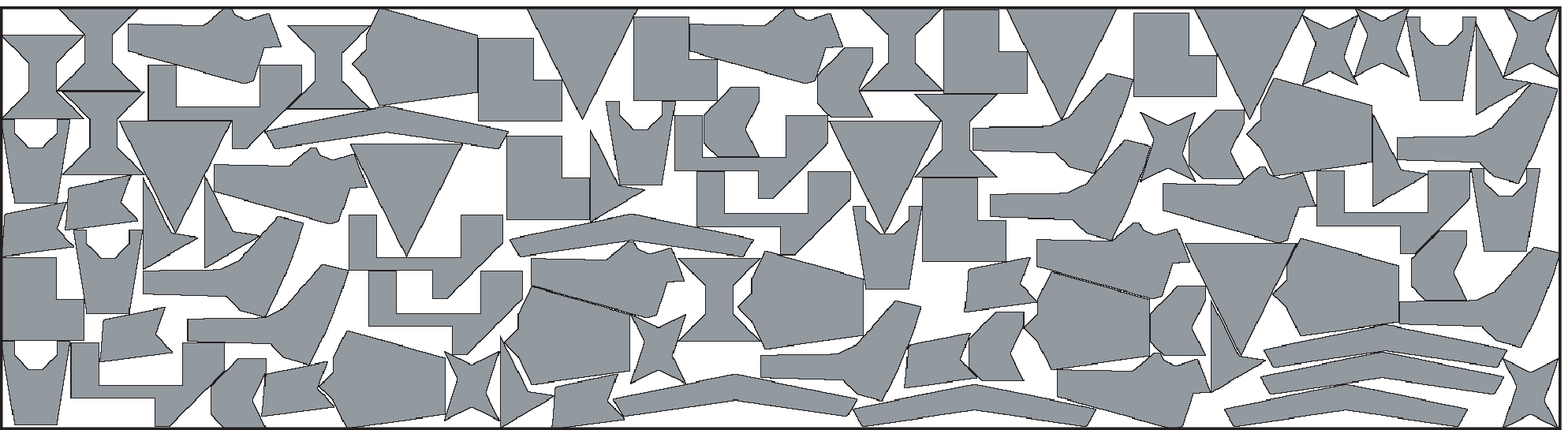}}\\
         {\small Profiles10 (68.89\%)}
  \end{minipage}\\
  \bigskip
  \begin{minipage}{0.96\textwidth}
    \centering
    \scalebox{0.4}{\includegraphics{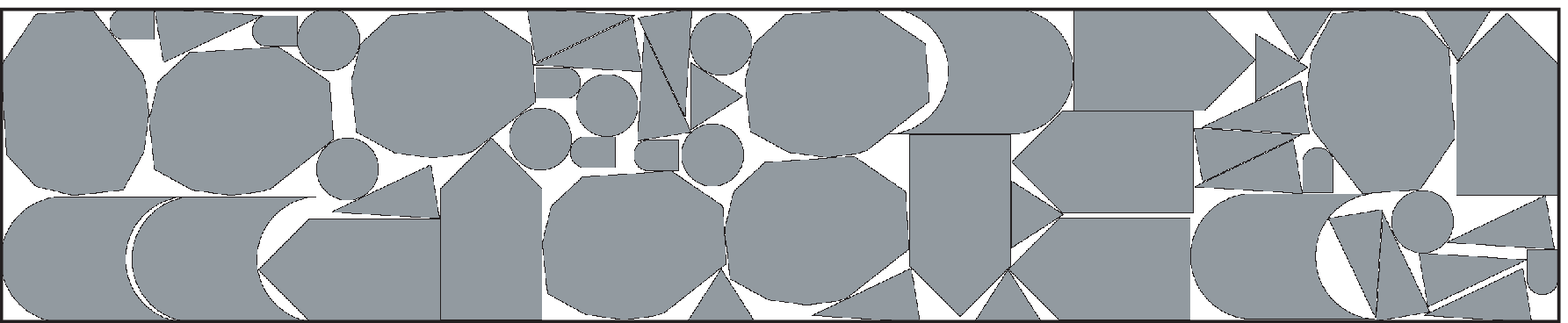}}\\
             {\small Profiles4 (85.63\%)}
  \end{minipage}
  \caption{The best solutions for the test instances obtained by GCDH ($W=512\mathrm{px}$).}\label{fig:profiles_all}
\end{figure}

The GCDH obtained comparable results to other algorithms for the first set of instances despite it was tested on the raster model of high-resolution, where we note that the test instances of the vector model are represented by a small number of line segments.
It also obtained that the better results than BLF for the second set of instances.
Figures~\ref{fig:polygon_all} and \ref{fig:profiles_all} show the best layouts obtained by GCDH for these instances ($W= 512\mathrm{px}$).
These computational results illustrate that GCDH attained a good performance for a wide variety of ISPs including circular arcs and holes.
\begin{table}[t]
  \caption{The density (\%) of FITS, GCS, ROMA, BLF and GCDH ($W=512\mathrm{px}$) on the test instances\label{tab:algorithm_density} ($\dagger~$We have corrected the density (\%) based on their lengths $L$ due to numerical errors in the literature.)}
\medskip
\centering
\begin{spacing}{0.8}
\begin{small}
  \begin{tabular}{lrrcrrcrrcrcrr} \hline
    & \multicolumn{2}{c}{FITS} & & \multicolumn{2}{c}{GCS} & & \multicolumn{2}{c}{ROMA} & & \multicolumn{1}{c}{BLF$^\dagger$} & & \multicolumn{2}{c}{GCDH}\\
    & \multicolumn{2}{c}{(vector)} & & \multicolumn{2}{c}{(vector)} & & \multicolumn{2}{c}{(vector)} & & \multicolumn{1}{c}{(vector)} & & \multicolumn{2}{c}{(raster)}\\ \cline{2-3} \cline{5-6} \cline{8-9} \cline{11-11} \cline{13-14}
  instance & \multicolumn{1}{c}{best} & \multicolumn{1}{c}{avg.} & & \multicolumn{1}{c}{best} & \multicolumn{1}{c}{avg.} & & \multicolumn{1}{c}{best} & \multicolumn{1}{c}{avg.} & & \multicolumn{1}{c}{best} & & \multicolumn{1}{c}{best} & \multicolumn{1}{c}{avg.}  \\ \hline
Albano & 88.86 & 88.30 & {} & 89.58 & 87.47 & {} & 85.72 & 82.66 & {} & 86.0 & {} & 88.20 & 87.47 \\
Dagli & 87.76 & 87.34 & {} & 89.51 & 87.06 & {} & 88.73 & 87.25 & {} & 82.2 & {} & 86.32 & 85.70 \\
Dighe1 & 99.90 & 99.89 & {} & 100.00 & 100.00 & {} & 100.00 & 100.00 & {} & 82.1 & {} & 97.60 & 97.55 \\
Dighe2 & 100.00 & 100.00 & {} & 100.00 & 100.00 & {} & 100.00 & 100.00 & {} & 84.3 & {} & 97.28 & 97.17 \\
Fu & 92.14 & 91.79 & {} & 92.41 & 90.68 & {} & 91.94 & 91.94 & {} & 89.2 & {} & 90.75 & 89.62 \\
Jakobs1 & 89.10 & 89.09 & {} & 89.10 & 88.90 & {} & 89.09 & 89.09 & {} & 82.6 & {} & 88.01 & 86.54 \\
Jakobs2 & 80.56 & 80.43 & {} & 87.73 & 81.14 & {} & 87.73 & 82.53 & {} & 75.1 & {} & 79.51 & 78.86 \\
Mao & 85.19 & 84.64 & {} & 85.44 & 82.93 & {} & 83.61 & 81.08 & {} & 78.7 & {} & 84.84 & 83.62  \\
Marques & 90.65 & 89.91 & {} & 90.59 & 89.40 & {} & 91.02 & 89.87 & {} & 86.5 & {} & 90.43 & 89.33 \\
Shapes0 & 67.56 & 66.94 & {} & 68.79 & 67.26 & {} & 68.79 & 68.72 & {} & 65.5 & {} & 66.13 & 65.33 \\
Shapes1 & 73.81 & 73.29 & {} & 76.73 & 73.79 & {} & 76.73 & 75.86 & {} & 71.5 & {} & 72.35 & 71.86 \\
Shirts & 88.08 & 87.30 & {} & 88.96 & 87.59 & {} & 88.52 & 87.29 & {} & 82.8 & {} & 84.33 & 84.14 \\
Swim & 75.70 & 75.20 & {} & 75.94 & 74.49 & {} & 73.23 & 70.37 & {} &  67.2 & {} & 74.08 & 72.91 \\
Trousers & 90.19 & 89.74 & {} & 91.00 & 89.02 & {} & 90.75 & 90.36 & {} & 86.9 & {} & 88.29 & 87.09 \\ \hline
Profiles1 & -- & -- & {} & -- & -- & {} & -- & -- & {} & 82.5 & {} & 86.93 & 85.08 \\
Profiles2 & -- & -- & {} & -- & -- & {} & -- & -- & {} & 73.8 & {} & 77.28 & 76.36 \\
Profiles3 & -- & -- & {} & -- & -- & {} & -- & -- & {} & 70.8 & {} & 73.87 & 72.91 \\
Profiles4 & -- & -- & {} & -- & -- & {} & -- & -- & {} & 86.8 & {} & 85.63 & 84.71 \\
Profiles5 & -- & -- & {} & -- & -- & {} & -- & -- & {} & 75.9 & {} & 82.47 & 81.42 \\
Profiles6 & -- & -- & {} & -- & -- & {} & -- & -- & {} & 72.1 & {} & 79.03 & 78.71 \\
Profiles7 & -- & -- & {} & -- & -- & {} & -- & -- & {} & 73.3 & {} & 95.80 & 95.65 \\
Profiles8 & -- & -- & {} & -- & --  & {} & -- & -- & {} & 78.7 & {} & 87.61 & 85.08 \\
Profiles9 & -- & -- & {} & -- & -- & {} & -- & -- & {} & 52.9 & {} & 57.86 & 57.39 \\
Profiles10 & -- & -- & {} & -- & -- & {} & -- & -- & {} & 65.0 & {} & 68.20 & 67.47 \\ \hline
avg.(1st) & 86.39 & 85.99 & {} & 87.56 & 85.70 & {} & 86.84 & 85.50 & {} & 80.0 & {} & 84.87 & 84.08 \\
avg.(2nd) & -- & -- & {} & -- & -- & {} & -- & -- & {} & 73.2 & {} & 79.38 & 78.42 \\
avg.(all) & -- & -- & {} & -- & -- & {} & -- & -- & {} & 77.2 & {} & 82.58 & 81.72 \\ \hline
  \end{tabular}
\end{small}
\end{spacing}
\end{table}

\begin{table}[t]
\caption{The computation time of FITS, GCS, ROMA, BLF and GCDH on the benchmark instances (in seconds)\label{tab:algorithm_time}}
\medskip
\centering
\begin{spacing}{0.8}
\begin{small}
  \begin{tabular}{lrrrrr} \hline
    & FITS & GCS & ROMA & BLF & GCDH\\
    & (vector) & (vector) & (vector) & (vector) & (raster)\\
    & Core i7 & Core i7 & Core i9 & Pentium4 & Core i7\\
    & 3.2GHz & 2.2GHz & 3.3GHz & 2.0GHz & 3.2GHz\\
    & 10 runs & 20 runs & 30 runs & $4\times10$ runs & 10 runs\\
    instance & limit & limit & limit & to best & limit\\ \hline
Albano & 1200 & 1200 & 1200 & 299 & 1200 \\
Dagli & 1200 & 1200 & 1200 & 252 & 1200 \\
Dighe1 & 1200 & 600 & 600 & 3 & 1200 \\
Dighe2 & 1200 & 600 & 600 & 148 & 1200 \\
Fu & 1200 & 600 & 600 & 139 & 1200 \\
Jakobs1 & 1200 & 600 & 600 & 29 & 1200 \\
Jakobs2 & 1200 & 600 & 600 & 51 & 1200 \\
Mao & 1200 & 1200 & 1200 & 152 & 1200 \\
Marques & 1200 & 1200 & 1200 & 21 & 1200 \\
Shapes0 & 1200 & 1200 & 1200 & 274 & 1200 \\
Shapes1 & 1200 & 1200 & 1200 & 239 & 1200 \\
Shirts & 1200 & 1200 & 1200 & 194 & 1200 \\
Swim & 1200 & 1200 & 1200 & 141 & 1200 \\
Trousers & 1200 & 1200 & 1200 & 253 & 1200 \\ \hline
Profiles1 & -- & -- & -- & 15 & 1200 \\
Profiles2 & -- & -- & -- & 295 & 1200 \\
Profiles3 & -- & -- & -- & 283 & 1200 \\
Profiles4 & -- & -- & -- & 256 & 1200 \\
Profiles5 & -- & -- & -- & 300 & 1200 \\
Profiles6 & -- & -- & -- & 171 & 1200 \\
Profiles7 & -- & -- & -- & 211 & 1200 \\
Profiles8 & -- & -- & -- & 279 & 1200 \\
Profiles9 & -- & -- & -- & 98 & 1200 \\
Profiles10 & -- & -- & -- & 247 & 1200 \\ \hline
\end{tabular}
\end{small}
\end{spacing}
\end{table}

\begin{figure}[tb]
  \centering
  \includegraphics[width=0.8\textwidth]{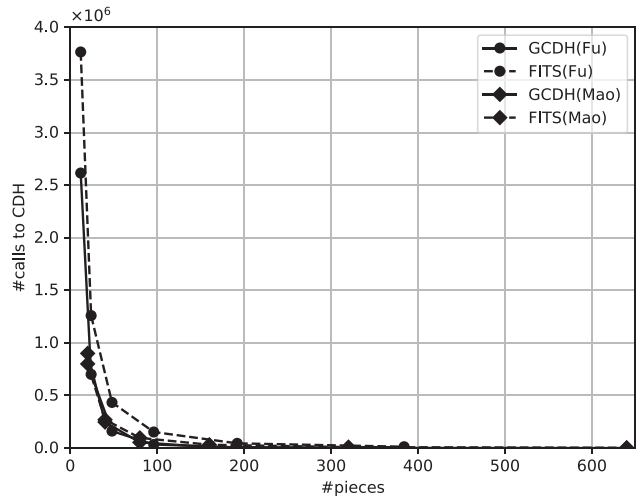}
  \caption{The number of calls to CDH on large-scale instances ($W=512$px) (normal plot) \label{fig:CDH_large_plot}}
\end{figure}

\begin{figure}[tb]
  \centering
  \includegraphics[width=0.8\textwidth]{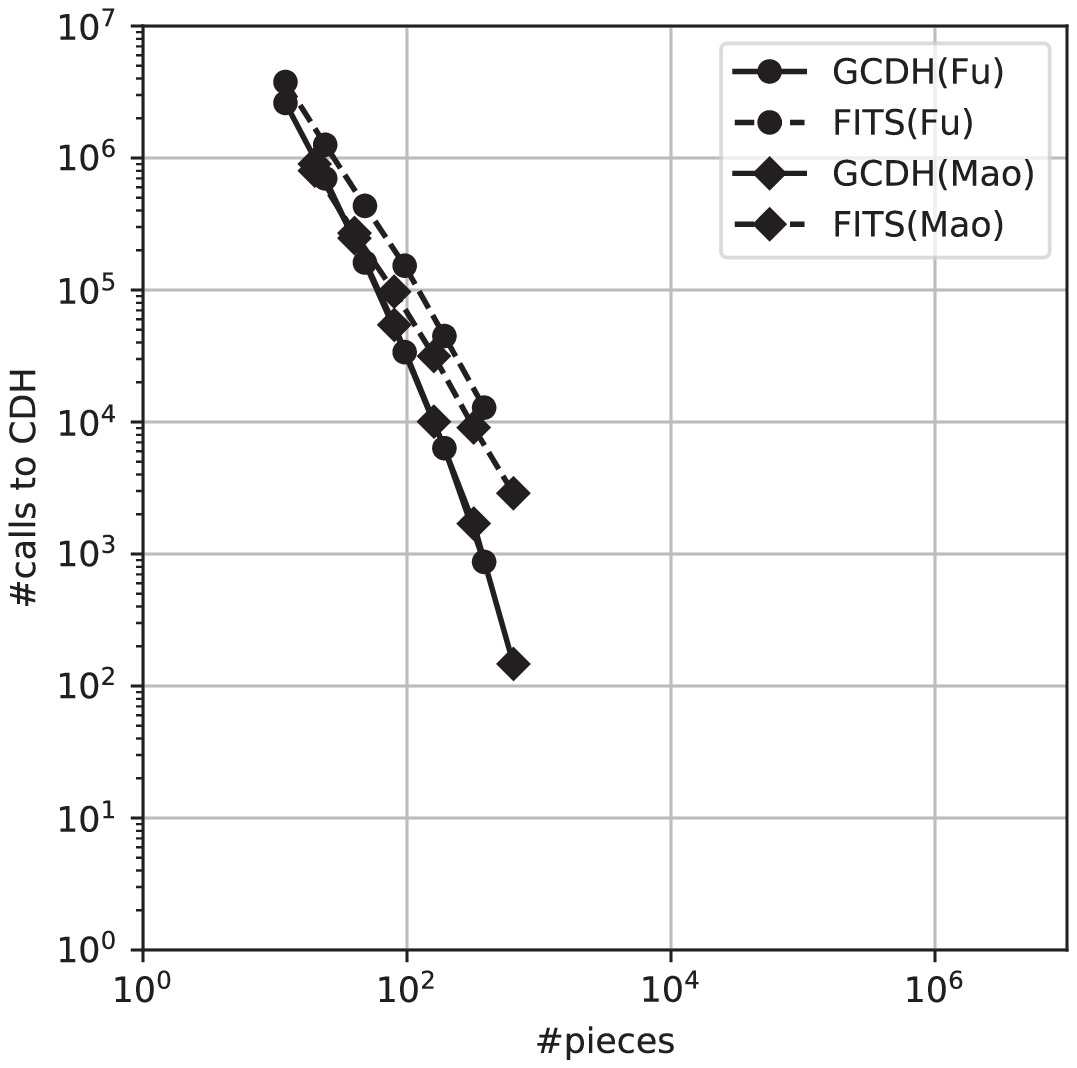}
  \caption{The number of calls to CDH on large-scale instances ($W=512$px) (log-log plot) \label{fig:CDH_large_loglog_plot}}
\end{figure}

We last evaluated the performance of GCDH ($W=512$px) and FITS on large-scale instances that were generated by copying the irregular shapes of the instances Fu and Mao.
We tested GCDH and FITS 10 times for each instance with the time limit of 3600 seconds for each run.
Table~\ref{tab:result_large} shows the computational results of GCDH and FITS on large-scale instances, which are evaluated in the average density (\%) and the number of calls to CDH.
The number after the name of each instance shows the number of copies; i.e., ``Fu2'' contains two copies of every piece in ``Fu'' and hence the number of pieces is doubled.
Figures~\ref{fig:CDH_large_plot} and \ref{fig:CDH_large_loglog_plot} also show the number of calls to CDH.
The GCDH and FITS take similar tendency in the computational efficiency for the increase of pieces, and their numbers of calls to CDH decrease in proportion to the square of the number of pieces roughly.
\begin{table}[t]
\caption{The computational results of GCDH ($W=512$px) and FITS on large-scale instances (3600 seconds)\label{tab:result_large}}
\medskip
\centering
\begin{spacing}{0.8}
\begin{small}
  \begin{tabular}{lrrcrrcrr} \hline
    & & & {} & \multicolumn{2}{c}{FITS} & {} & \multicolumn{2}{c}{GCDH} \\
    & & & {} & \multicolumn{2}{c}{(vector)} & {} & \multicolumn{2}{c}{(raster)} \\ \cline{5-6} \cline{8-9}
  instance & \multicolumn{1}{c}{\#shapes} & \multicolumn{1}{c}{\#pieces} & {} & \multicolumn{1}{c}{density(\%)} & \multicolumn{1}{c}{\#CDHs} & {} & \multicolumn{1}{c}{density(\%)} & \multicolumn{1}{c}{\#CDHs} \\ \hline
  Fu & 11 & 12 & {} & 91.98 & $3.77 \times 10^6$ & {} & 89.78 & $2.61 \times 10^6$ \\
  Fu2 & 11 & 24 & {} & 93.70 & $1.26 \times 10^6$ & {} & 90.77 & $7.01 \times 10^5$ \\
  Fu4 & 11 & 48 & {} & 93.49 & $4.34 \times 10^5$ & {} & 88.81 & $1.61 \times 10^5$ \\
  Fu8 & 11 & 96 & {} & 92.43 & $1.53 \times 10^5$ & {} & 86.81 & $3.38 \times 10^4$ \\
  Fu16 & 11 & 192 & {} & 91.78 & $4.48 \times 10^4$ & {} & 85.66 & $6.34 \times 10^3$ \\
  Fu32 & 11 & 384 & {} & 91.80 & $1.28 \times 10^4$ & {} & 82.12 & $8.72 \times 10^2$ \\ \hline
  Mao & 9 & 20 & {} & 84.90 & $8.00 \times 10^5$ & {} & 84.25 & $9.00 \times 10^5$\\
  Mao2 & 9 & 40 & {} & 84.33 & $2.70 \times 10^5$ & {} & 82.29 & $2.47 \times 10^5$ \\
  Mao4 & 9 & 80 & {} & 82.56 & $9.72 \times 10^4$ & {} & 80.14 & $5.44 \times 10^4$\\
  Mao8 & 9 & 160 & {} & 81.24 & $3.16 \times 10^4$ & {} & 79.13 & $1.00 \times 10^4$ \\
  Mao16 & 9 & 320 & {} & 80.64 & $9.08 \times 10^3$ & {} & 78.45 & $1.70 \times 10^3$ \\
  Mao32 & 9 & 640 & {} & 80.36 & $2.88 \times 10^3$ & {} & 76.20 & $1.47 \times 10^2$ \\ \hline
\end{tabular}
\end{small}
\end{spacing}
\end{table}

\section{Conclusion}\label{sec:conclusion}
We develop coordinate descent heuristics for the the irregular strip packing problem (ISP) of the rasterized shapes that repeats a line search in the horizontal and vertical directions alternately.
The rasterized shapes provide simple procedures of the intersection test without any exceptional handling due to geometric issues, while they often require much memory and computational effort in high-resolution.
To reduce the complexity of rasterized shapes, we first propose the double scanline representation by merging consecutive pixels in each row and column into strips with unit width, respectively.
Based on this, we develop an efficient line search algorithm incorporating a corner detection technique used in computer vision to reduce the search space.
Computational results for test instances show that the proposed algorithm obtains sufficiently dense layouts of rasterized shapes in high-resolution within a reasonable computation time.

\section*{Acknowledgement}
The authors would like to thank Yusuke Nakano for technical assistance with implementing the proposed algorithm GCDH especially for processing the raster model.
This research did not receive any specific grant from funding agencies in the public, commercial, or not-for-profit sectors.

\begin{spacing}{1.0}

\end{spacing}
\end{document}